\documentclass[12pt]{article}
\usepackage[margin=1in]{geometry}
\usepackage{bm}
\usepackage{eufrak}
\usepackage{float}
\usepackage{graphicx}
\usepackage{amssymb}
\usepackage{amsmath}
\usepackage{amsthm}
\usepackage{empheq}
\usepackage{epstopdf}
\usepackage{mdframed}
\usepackage{booktabs}
\usepackage{lipsum}
\usepackage{graphicx}
\usepackage{mathtools}
\usepackage{algorithm}
\usepackage{algpseudocode}
\usepackage{color}
\usepackage{psfrag}
\usepackage{pgfplots}
\usepackage{hyperref}
\usepackage{natbib}
\usepackage{bbm}
\usepackage{caption}
\usepackage{subcaption}
\mathtoolsset{showonlyrefs}
\graphicspath{ {./images/} }
\hypersetup{
  colorlinks,
  linkcolor={red!50!black},
  citecolor={blue},
  urlcolor={blue!80!black}
}
\usepackage[toc,page]{appendix}
\usepackage{multirow}

\newtheorem{theorem}{Theorem}
\newtheorem{assumption}{Assumption}

\newtheorem{lemma}{Lemma}

\newtheorem{corollary}{Corollary}
\newtheorem{proposition}{Proposition}

\newtheorem{remark}{Remark}

\algrenewcommand\algorithmicrequire{\textbf{Input:}}
\algrenewcommand\algorithmicensure{\textbf{Output:}}

\usepackage{xr}
\makeatletter
\newcommand*{\addFileDependency}[1]{% argument=file name and extension
  \typeout{(#1)}
  \@addtofilelist{#1}
  \IfFileExists{#1}{}{\typeout{No file #1.}}
}
\makeatother

\definecolor{ocre}{RGB}{243,102,25}
\definecolor{mygray}{RGB}{243,243,244}
\definecolor{deepGreen}{RGB}{26,111,0}
\definecolor{shallowGreen}{RGB}{235,255,255}
\definecolor{deepBlue}{RGB}{61,124,222}
\definecolor{shallowBlue}{RGB}{235,249,255}

\DeclareMathOperator*{\argmax}{arg\,max}

\newcommand{\bb}[1]{\mathbb{#1}}

\newcommand{\norm}[1]{\left\lVert#1\right\rVert}

\newcommand{\ti}{_{2,\infty}}
\newcommand{\eq}[1]{\begin{equation} #1\end{equation}}
\newcommand{\ip}[1]{\left\langle#1 \right\rangle}
\newcommand{\mc}[1]{\mathcal{#1}}
\newcommand{\longeq}[1]{\begin{equation}
    \begin{split}
        #1
    \end{split}
\end{equation}}

\newcommand{\A}{\mathbf A}
\newcommand{\B}{\mathbf B}
\newcommand{\C}{\mathbf C}
\newcommand{\E}{\mathbf E}
\newcommand{\e}{\mathbf e}
\newcommand{\F}{\mathbf F}

\newcommand{\I}{\mathbf I}

\newcommand{\M}{\mathbf M}
\newcommand{\N}{\mathbf N}
\def\S{\mathbf S}

\newcommand{\bP}{\mathbf P}

\newcommand{\R}{\mathbf R}

\newcommand{\U}{\mathbf U}

\newcommand{\W}{\mathbf W}
\newcommand{\V}{\mathbf V}

\newcommand{\X}{\mathbf X}
\newcommand{\x}{\mathbf x}
\newcommand{\Y}{\mathbf Y}

\newcommand{\Z}{\mathbf Z}
\newcommand{\bT}{\boldsymbol{\Theta}}
\newcommand{\bt}{\boldsymbol{\theta}}
\newcommand{\bS}{\boldsymbol{\Sigma}}

\newcommand{\bPi}{\boldsymbol{\Pi}}
\newcommand{\bpi}{\boldsymbol{\pi}}
\newcommand{\Lam}{\mathbf \Lambda}
\def\S{\mathbf S}

\newcommand{\bo}{\boldsymbol }

\def\cov{\text{Cov}}
\def\var{\text{Var}}
\def\tr{\textsf{tr}}
\allowdisplaybreaks

\newcommand{\mb}{\mathbf}
\def\hat{\widehat}
\def\tilde{\widetilde}
\def\t{^{\top}}

\usepgfplotslibrary{colorbrewer}
\pgfplotsset{width=8cm,compat=1.9}
\usepackage[margin=1in]{geometry}
\def\spacingset#1{\renewcommand{\baselinestretch}%
{#1}\small\normalsize}

\def\diag{\textsf{diag}}

\allowdisplaybreaks

\newcommand\blfootnote[1]{%
  \begingroup
  \renewcommand\thefootnote{}\footnote{#1}%
  \addtocounter{footnote}{-1}%
  \endgroup
}

\usepackage{multibib}
\newcites{one}{References}
\newcites{two}{References}

\title{Generalized Grade-of-Membership Estimation for High-dimensional Locally Dependent Data}
\author{Ling Chen$^*$ \and Chengzhu Huang$^*$ \and Yuqi Gu$^\dagger$}
\date{Department of Statistics, Columbia University}

\begin{document}

\maketitle
\blfootnote{$^*$Chen and Huang contributed equally to this work and are co-first authors.}
\blfootnote{$^\dagger$Corresponding author. Email:  \texttt{yuqi.gu@columbia.edu}. Research is partially supported by NSF grant DMS-2210796.}

\vspace{-4mm}
\begin{abstract}
This work focuses on the mixed membership models for multivariate categorical data widely used for analyzing survey responses and population genetics data. These grade of membership (GoM) models offer rich modeling power but present significant estimation challenges for high-dimensional polytomous data. Popular existing approaches, such as Bayesian MCMC inference, are not scalable and lack theoretical guarantees in high-dimensional settings. To address this, we first observe that data from this model can be reformulated as a three-way (quasi-)tensor, with many subjects responding to many items with varying numbers of categories. We introduce a novel and simple approach that flattens the three-way quasi-tensor into a ``fat'' matrix, and then perform a singular value decomposition of it to estimate parameters by exploiting the singular subspace geometry. Our fast spectral method can accommodate a broad range of data distributions with arbitrarily locally dependent noise, which we formalize as the generalized-GoM models. We establish finite-sample entrywise error bounds for the generalized-GoM model parameters. This is supported by a new sharp two-to-infinity singular subspace perturbation theory for locally dependent and flexibly distributed noise, a contribution of independent interest. Simulations and applications to data in political surveys, population genetics, and single-cell sequencing demonstrate our method's superior performance.
\end{abstract}


\spacingset{1.7} 

\section{Introduction}
Multivariate categorical data are prevalent in various fields, including social science surveys \citep{clinton2004statistical}, population genetics \citep{pritchard2000inference}, health sciences \citep{manton1994statistical},
text analysis \citep{blei2003latent}, among others. 
These data consist of high-dimensional categorical variables, with many subjects responding to a large number of items with varying numbers of categories.
Identifying interpretable latent patterns to capture heterogeneity in such data is of great interest. Among latent variable models for multivariate categorical data, the grade of membership (GoM) model \citep{woodbury1978gom, erosheva2007describing} stands out as a popular tool that assumes each subject's latent variables live in a simplex. 
The GoM model falls within the broad family of mixed membership models for individual-level mixtures \citep{airoldi2015handbook}.
Unlike traditional population-level mixture models that restrict each individual to belong to a single latent class, the GoM model more flexibly allows each individual to partially belong to multiple latent classes. In a GoM model with $K$ extreme latent classes, each individual has $K$ nonnegative membership scores that sum to one, quantifying the extent to which the individual belongs to each class. 

The flexibility of GoM models, however, brings significant challenges in estimation, particularly for high-dimensional polytomous data. 
{Due to the complex hierarchical modeling nature of GoM models, most existing studies adopt the Bayesian approach with Markov Chain Monte Carlo (MCMC) algorithms for posterior inference
\citep{erosheva2002grade,erosheva2007describing,  bhattacharya2012simplex, gu2023dimension}.}
Despite their popularity, these methods face critical challenges in high-dimensional settings. MCMC methods often suffer from high computational cost, making them hard to scale to big data. 
Also, Bayesian or likelihood-based methods are usually tailored to a specific probabilistic model, limiting their applicability across different data distributions. Extending GoM model estimation methods developed for dichotomous data \citep{robitzsch2017package, chen2024spectral} to polytomous data also requires careful new development. 
Furthermore, the theoretical guarantees for estimating the GoM model in high-dimensional settings remain unknown.

Another challenge in GoM models is the local dependence issue among items, meaning that a subject's observed multivariate responses exhibit additional dependencies that are not fully explained by the latent variables.  
Local dependence has long been known as a practical issue when applying latent variable measurement models in social science \citep{chen1997local,chen2018robust}.
For instance, in the American National Election Studies (ANES) dateset, blocks of items concerning voting participation, presidential approval, economics, and other topics may display local dependence within each topic.
In addition to social science, in genome-wide association studies (GWAS), linkage disequilibrium (LD) -- where neighboring alleles are correlated -- represents a form of local dependence \citep{uffelmann2021genome}. 
In single-cell data, gene co-expression that reflects dependencies between genes \citep{van2018single} also constitutes local dependence.
It is crucial to devise accurate estimation methods for locally dependent GoM models, especially in high-dimensional settings where such dependencies are often more pronounced. 

To address the above challenges, we propose a fast spectral method based on singular value decomposition (SVD) for high-dimensional data with potential local dependence. We first observe that the multivariate categorical data can be reformulated as a three-way (quasi-)tensor, with many subjects responding to many items with varying numbers of categories.
We employ a simple yet novel technique to flatten this quasi-tensor into a ``fat'' matrix and then perform SVD to estimate parameters by exploiting the singular subspace geometry. We further introduce the generalized-GoM models, which encompass a wide range of data distributions with potential local dependence and cover the traditional GoM model as a special case. 
We establish finite-sample entrywise estimation error bounds for generalized-GoM parameters. To achieve this, we develop a novel sharp two-to-infinity singular subspace perturbation theory for arbitrary locally dependent data from flexible distributions, which is of independent interest. 
Extensive simulation studies demonstrate that our method efficiently and accurately estimates the generalized-GoM parameters with local dependence that either results from the flattening technique or is a consequence of the intrinsic data generation mechanism. 
We apply our method to a diverse range of real-world datasets in political surveys, population genetics, and single-cell sequencing, and uncover interesting insights.

The rest of the paper is structured as follows. Section \ref{sec:model setup} explains the background and motivation for this work and introduces the model setup. Section \ref{sec:estimation} discusses model identifiability and presents a spectral method for generalized-GoM models. Section \ref{sec-error} develops a two-to-infinity singular subspace perturbation theory for locally dependent data from flexible distributions and establishes entrywise error bounds for parameter estimation. Section \ref{sec:simulation} reports the results of simulation studies and Section \ref{sec:real data} presents three real-data applications. 
Section \ref{sec:discussion} concludes the paper.
The R codes implementing the proposed method are available at \href{https://github.com/lscientific/gGoM}{https://github.com/lscientific/gGoM}.
All technical proofs of the theoretical results and additional computation details are provided in the Supplementary Material.

\section{Motivation and Model Setup}\label{sec:model setup}

\subsection{GoM for Multivariate Categorical Data and Existing Methods}
We first review the classical GoM model for multivariate categorical data \citep{woodbury1978gom, manton1994statistical, erosheva2002grade}.
Consider categorical responses from $N$ subjects to $L$ items, and denote the response of subject $i$ to item $l$ by $\tilde R_{i,l}\in\{1,\dots, C_l\}$, for $i\in[N]$ and $l\in[L]$, where $C_l\geq 2$ denotes the number of categories of item $l$. 

The GoM model uses latent mixed membership structures to model multivariate categorical data.
It has two levels of modeling: the population level and the individual level. On the population level, there are $K$ \emph{extreme latent profiles}, each representing a prototypical response pattern. For each $k\in[K]$, the extreme latent profile $k$ corresponds to an item parameter vector $\tilde {\bo{\vartheta}}_k=(\tilde\bt_{1,k}^\top,\dots, \tilde\bt_{L, k}^\top)^\top$ with $\tilde \bt_{l, k}=(\tilde\theta_{l,k,1},\dots, \tilde\theta_{l,k,C_l})^\top\in[0,1]^{C_l}$, and 
\begin{align*}
    \tilde \theta_{l, k, c} = \mathbb P(\tilde R_{i,l}=c\mid \text{subject $i$ solely belongs to the $k$-th extreme latent profile})
\end{align*}
for $c\in[C_l]$, $l\in[L]$.
Note that $\sum_{c=1}^{C_l}\tilde\theta_{l, k, c} \equiv 1$ always holds for each $l\in[L]$ and $k\in[K]$.
On the individual level, each subject $i$ is associated with a vector of mixed membership scores $\bpi^*_i=(\pi_{i1}^*,\dots,\pi_{iK}^*)$, with $\pi^*_{ik}\ge 0$ and $\sum_{k=1}^K \pi^*_{ik}=1$. Each membership score $\pi^*_{ik}$ indicates the extent to which subject $i$ partially belongs to extreme latent profile $k$. 
The item response probability for the $i$-th subject's response to the $l$-th item is a convex combination:
\begin{equation}
     \mathbb P(\tilde R_{i, l}=c) = \sum_{k=1}^K \pi_{i,k}^*\tilde\theta_{l, k, c}, \text{ for } c\in[C_l].
\end{equation}
We collect all item parameters in matrix $\tilde \bT=(\tilde {\bo{\vartheta}}_1, \dots, \tilde {\bo{\vartheta}}_K)\in[0,1]^{(\sum_{l=1}^L C_l)\times K}$.

Most existing studies on the GoM model treat the membership scores as random effects coming from some distribution on the simplex.
It was also typically assumed that a subject's responses $\tilde R_{i,1},\dots, \tilde R_{i, L}$ are conditionally independent given the membership scores $\bo\pi_i^*$, which is called the local independence assumption. Under this assumption, if $\bpi^*_i$ are considered to be i.i.d random vectors, for example, $\bpi_i^*$ follows the Dirichlet distribution $D_{\bo\alpha^*}(\cdot)$ parameterized with $\bo\alpha^*\in\mathbb{R}^K_+$, then the marginal likelihood function of a GoM model is
\begin{equation}\label{eq:mlf}
    L(\tilde\bT, \bo\alpha^* \vert \tilde\R) = \prod_{i=1}^N \int_{\Delta_{K-1}} \prod_{l=1}^{L} \prod_{c=1}^{C_l} \left(\sum_{k=1}^K \pi^*_{i,k} \tilde\theta_{l, k, c} \right)^{\mathbbm{1} (\tilde R_{i,l}=c)} d D_{\bo\alpha^*} (\bpi_i),
\end{equation}
where $\Delta_{K-1} = \{\x=(x_1,\dots, x_K): x_i\ge 0,\ \sum_{i=1}^Kx_i=1\}$ denotes the probability simplex.
The marginal maximum likelihood (MML) approach estimates the population parameters $(\tilde\bT, \bo\alpha^*)$ by maximizing the marginal likelihood function \eqref{eq:mlf}. The integrals over the membership scores are hard to evaluate, posing challenges for MML. 
Currently, Bayesian inference with MCMC algorithms is arguably the most popular estimation paradigm for the GoM model and its variants \citep[e.g.,][]{airoldi2015handbook,erosheva2002grade,erosheva2004pnas,bhattacharya2012simplex,gu2023dimension,kang2024blockwise}.
But MCMC algorithms often suffer from high computational cost and do not scale to high-dimensional data.

In addition to Bayesian MCMC methods, there exist several optimization-based estimation methods for the GoM model. 
\cite{wang2015fitting} developed a heuristic variational inference method for the GoM model.
Another optimization-based estimation approach is the joint maximum likelihood (JML) method \citep{manton1994statistical}. JML treats the mixed membership scores as fixed parameters and maximizes the joint likelihood function to simultaneously estimate $\bPi^*$ and $\tilde\bT$. 
Another study \cite{zhao2018dirichlet} proposed a moment estimation method that leverages the second and third order moments to estimate the population parameters in Dirichlet latent variable models with mixed data types.

\subsection{Mixed Membership Structures for Other Data Types}\label{subsec-otherdata}
Related mixed membership (MM) models have also been used to model multivariate data from the Binomial and Poisson distributions in some applications. 
For example, the number of minor allele of a single nucleotide polymorphism (SNP) is usually assumed to follow a Binomial distribution. Let $\bT^*=(\theta^*_{j,k})$ and $\theta^*_{j,k}\in(0,1)$ represents the binomial parameter for the $j$-th SNP and the $k$-th \emph{ancestry profile}, then the genotype $R_{i,j}$ follows $\text{Binom}(2, \sum_{k=1}^K\pi^*_{i,k}\theta^*_{j,k})$ for $i\in[N], j\in[J]$. The marginal likelihood function for a Binomial MM model is
\begin{equation*}
    L(\bT^*, \bo\alpha^* | \R) = \prod_{i=1}^N \int_{\Delta_{K-1}} \prod_{j=1}^J \binom{2}{R_{i,j}}\left(\sum_{k=1}^K\pi^*_{i,k}\theta^*_{j,k}\right)^{R_{i,j}}\left(1-\sum_{k=1}^K\pi^*_{i,k}\theta^*_{j,k}\right)^{2-R_{i,j}} d D_{\bo\alpha^*} (\bo\pi_i)
\end{equation*}
when assuming $\pi^*_{i,k}$ follows a distribution $D_{\bo\alpha^*}(\cdot)$ parametrized by $\bo\alpha^*$.
The influential MCMC algorithm STRUCTURE \citep{pritchard2000inference} imposes a Dirichlet prior on $\pi^*_{i,k}$ and estimates this \emph{admixture model} for Binomial distributed genotype data.
As an example for count data, let $R_{i,j}$ represent the expression count for the gene $j$ in the cell $i$ in single-cell sequencing data. In this case, $\theta^*_{j,k}\in\mathbb{R}_+$ denotes the Poisson rate and $R_{i,j} \stackrel{\text{ind.}}{\sim} \text{Poisson}(\sum_{k=1}^K\pi^*_{i,k}\theta^*_{j,k})$ for $i\in[N], j\in[J]$. 
The joint likelihood function for a Poisson MM model is given by
\begin{equation}\label{eq:poisson}
    L(\bT^*, \bPi^* | \R) = \prod_{i=1}^N \prod_{j=1}^J \frac{1}{R_{i,j}!}\left(\sum_{k=1}^K \pi_{i,k}^*\theta_{j,k}^*\right)^{R_{i,j}} \exp\left(-\sum_{k=1}^K \pi_{i,k}^*\theta_{j,k}^*\right).
\end{equation}
To maximize \eqref{eq:poisson}, nonnegative matrix factorization (NMF) with iterative multiplicative updates is a popular approach \citep{lee1999learning, lee2000algorithms}.
This algorithm depends on the Poisson distribution assumption and is sensitive to initializations.

\subsection{Generalized-GoM Models and Our Assumptions}
\begin{figure}[t!]
    \centering
    \includegraphics[width=\linewidth]{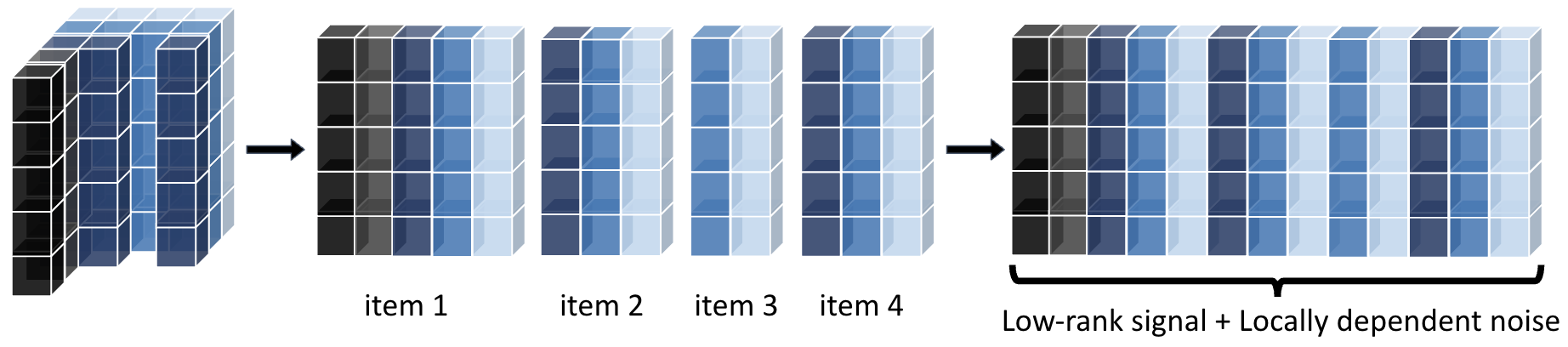}
    \caption{Visualization of the flattening technique. Cubes on the left represent the three-way quasi-tensor $\tilde\R$, and those on the right represent the flattened binary matrix $\R$.}
    \label{fig:visualization}
\end{figure}

To facilitate statistical analysis of high-dimensional polytomous data, we introduce a simple but novel approach that flattens the three-way quasi-tensor into a ``fat'' binary matrix (see Figure \ref{fig:visualization}), with columns corresponding to the response categories of all items. 
Define $J:=\sum_{l=0}^{L}C_l$ with $C_0=0$ and consider the binary matrix $\R=(R_{i,j})\in\mathbb{R}^{N\times J}$ with 
\begin{equation}\label{eq-flatten}
R_{i, \sum_{m=0}^{l-1}C_m+c}=
\mathbbm{1}(\tilde R_{i,l} = c),
\quad \text{  for } c\in[C_l].  
\end{equation}
Similarly, we define the corresponding flattened item parameter matrix $\bT^*=(\theta_{jk}^*)\in[0,1]^{J\times K}$ with
$\theta^*_{\sum_{m=0}^{l-1}C_m+c, k} = \tilde \theta_{l, k, c}$ for $c\in[C_l], l\in[L], k\in[K]$.
The columns of $\R$ consist of $L$ blocks, with the size of the $l$-th block equal to the number of categories for the $l$-th item.  Within the $l$-th block, we always have $\sum_{c=1}^{C_l} R_{i, \sum_{m=0}^{l-1}C_m+c}\equiv 1$.
Apparently, this flattening technique introduces local dependence within each block that cannot be accounted for by the latent variables. 
Figure \ref{fig:visualization} visualizes flattening a three-way quasi-tensor into a fat binary matrix. {We interpret this figure using the ANES data (see Section \ref{sec:real data}) as an illustration. Each of the five rows represents a subject that responds to four items regarding illegal immigration, climate, political participation, and taxes, respectively. The number of response categories for the four items are $5, 3, 2, 3$, as reflected as the number of columns for the cubes in the middle of the figure. For instance, item $1$ asks ``\emph{How important is the illegal immigration issue in the country today?}'' and the $5$ possible responses vary from ``Extremely important'' to ``Not at all important''. Item $3$ asks ``\emph{During the past 12 months, have you posted a message or comment online about a political issue or campaign?}'' and the possible responses are ``Yes'' and ``No''.} 
Similar concatenated data structures have been considered in the multilayer random graph models \citep{jones2020multilayer, gallagher2021spectral}, and \cite{arroyo2021inference} considers such structures in the spectral subspace. 

To accommodate flexible local dependence structures and various data types, we introduce a class of \emph{generalized-GoM} models that support general data distributions, including those with sub-exponential tails and covering the classical GoM model as a special case.
Write the (flattened) data matrix $\R$ as the sum of a mean matrix $\R^*$ and a noise matrix $\E$, 
\begin{equation}\label{eq-model}
    \R^*:=\mathbb{E}[\R]=\underbrace{\bPi^*}_{N\times K} \underbrace{\bT^{*\top}}_{K\times J },\qquad
   \E:=\R-\R^*,
\end{equation}
where $\bPi^*=(\pi^*_{i,k})\in[0,1]^{N\times K}$ includes the membership scores, and $\bT^*=(\theta_{j,k}^*)\in\mathbb{R}^{J\times K}$ collects all the item parameters.
Such expectation-noise decomposition is commonly used in the topic modeling and network modeling literature, generally assuming independent entries in the noise matrix $\E$. 
In contrast, we impose the following Assumption~\ref{assumption:block} to characterize the general block dependence structure of the noise entries, along with their tail behavior.

\begin{assumption}[Block Dependent Noise under Flexible Distributions]
\label{assumption:block}
Matrix $\E$ satisfies:
\begin{itemize}
\item[(a)] 
There exists a partition $S_1,S_2,\ldots,S_{L}$ of $[J]$ 
with $|S_l| \leq M$ for each $l\in[L]$, such that the vectors $\{\E_{i,S_l}\}_{l=1}^{L}$ are mutually independent for different $i\in[N]$ and $l\in [L]$. 
\item[(b)] Either $|E_{i,j}| \leq B$ for all $i\in[N]$, $ j\in[J]$, or there exists a random matrix $\E'=(E_{i,j}') \in\mathbb{R}^{N\times J}$ obeying the same dependence structure in (a), such that for any $i\in[N], j\in[J]$, it holds that $\norm{E'_{i,j}}_\infty\leq B$, $\mathbb E[E'_{i,j}] = 0$, $\norm{ \mathrm{Cov}(\E'_{i,:})} \lesssim \norm{\mathrm{Cov}(\E_{i,:})}$, and $\Pr(E_{i,j} = E'_{i,j}) \ge 1 - O(d^{-22})$.
\end{itemize}
\end{assumption}

The generalized-GoM models satisfying Assumption \ref{assumption:block} are suitable for handling highly asymmetric and sparse data, and are applicable across diverse fields, including social science surveys, admixture modeling in population genetics, single-cell sequencing data, among others. 
The flexible local dependence in Assumption \ref{assumption:block}(a) effectively captures the dependence in the flattened data from multivariate polytomous responses and accommodates other more complex dependence structures.
The entries in each block $S_l$ can be arbitrarily dependent on each other, which can either consist of those flattened response categories of a single item and/or concatenated responses across different intrinsically locally dependent items.
Related local dependence considerations are explored in \cite{lei2019unified} and \cite{modell2024intensity}. \cite{agterberg2022entrywise} relaxes the entrywise independence assumption using a linear-transformed entrywise-independent noise assumption, but their assumption cannot capture the dependence structure of flattened polytomous data in our motivating setting. 

{Notably, the high-probability boundedness  Assumption~\ref{assumption:block}(b) encompasses a wide range of data types, covering all sub-exponential distributions and substantially extending beyond the flattened polytomous data case in the traditional GoM model. 
For example, the mixed membership models for Binomial and Poisson data mentioned in Section \ref{subsec-otherdata} both belong to our generalized-GoM modeling framework. Specifically, one can write the data arising from these models as the sum of a low-rank signal part $\bPi^* \bT^{*\top}$ and a mean-zero Binomial/Poisson noise part $\E$, where $\bT^*$ are Binomial or Poisson parameters with their respective range.} 

Consider the SVD of the signal matrix $\R^*=\U^*\Lam^* \V^{*\top}$,
where $\Lam^*=\text{diag}(\sigma_1^*,\dots, \sigma_K^*)$ with $\sigma_1^*\ge\dots \ge \sigma_K^*> 0$ collects the $K$ singular values. Orthonormal matrices $\U^*_{N\times K}$, $\V^*_{J\times K}$ respectively consist of the left and right singular vectors.
Similarly, we consider the top-$K$ SVD for the observed matrix $\R\approx\U\Lam\V^\top$ 
where $\Lam=\text{diag}(\sigma_1,\dots, \sigma_K)$ with $\sigma_1\ge \dots, \sigma_K\ge0$, $\U\in\mathbb{R}^{N\times K}$, $\V\in\mathbb{R}^{J\times K}$, and $\U^{\top}\U=\V^{\top}\V=\mathbf{I}_K$.

Write the ratio of $J$ and $N$ as $r \coloneqq {J}/{N}$. Let $[B]=\{1,2,\ldots,B\}$ for any positive integer $B$. For a matrix $\A$, let $\lambda_k(\A)$ denote the $k$-th largest eigenvalue and $\sigma_k(\A)$ denote the $k$-th largest singular value; define its condition number $\kappa(\A):=\sigma_1(\A)/\sigma_{\min}(\A)$ with $\sigma_{\min}(\A)$ being the minimum non-zero singular value. 
Specifically, define $\kappa^*:=\sigma_1^*/\sigma_K^*$ as the condition number of $\R^*$. The $\ell_{2,\infty}$ norm of $\A$ is $\|\A\|_{2, \infty}:=\max_i\|\A_{i,:}\|_2$, which is the maximum $\ell_2$ norm of the rows of $\A$. Let $\A_{\S,:}$ denote the submatrix consisting of the rows indexed by $\S$ and similarly $\A_{:, \S}$ for columns.
Define the matrix incoherence degrees, which are standard in entrywise matrix perturbation theory:
$\mu_1 \coloneqq (N / K)\norm{\U^*}^2\ti$, and $\mu_2 \coloneqq (J / K)\norm{\V^*}^2\ti.$

Finally, we define the following notations similar to those in \cite{chen2021spectral}: (a) $f(n) = O (g(n))$ or $f(n) \lesssim g(n)$ means there is a universal constant $c > 0$ such that $|f(n)| \le c|g(n)|$ for all $n$ large enough, and $\tilde{O}(\cdot)$ differs from $O(\cdot)$ with up to poly-logarithmic factors; (b) $f(n) \asymp g(n)$ indicates that there exists universal constants $c_1, c_2> 0$ such that $c_1|g(n)| \le |f(n)| \le c_2|g(n)|$ for all $n$ large enough; (c) $f(n) = o(g(n))$ denotes that $f(n)/g(n)\stackrel{n\rightarrow \infty}{\longrightarrow} 0$; (d) $f(n) \gg g(n)$ (resp. $f(n) \ll g(n)$) is used when there exists universal constant $c > 0$ large (resp. small) enough such that $|f(n)| \ge c |g(n)|$ (resp. $|f(n)| \le c |g(n)|$).

\section{A Spectral Method for Parameter Estimation}
\label{sec:estimation}
Model identifiability is a crucial prerequisite for valid and reproducible statistical analysis. 
Conventionally, a statistical model is said to be identifiable if there is a one-to-one mapping between the population parameters and the marginal likelihood function. \cite{gu2023dimension} established such a notion of \emph{population identifiability} for a dimension-grouped variant of the GoM model, by investigating when the probability mass function (PMF) of the observed random vector (see \eqref{eq:mlf}) uniquely determines the item parameters $\bo\Theta$ and Dirichlet parameters $\bo\alpha$.
However, this notion does not concern the identifiability of the latent membership scores $\bo\pi_i$, which are also often of interest. 
To address this problem for a special case of the GoM model with binary responses, \cite{chen2024spectral} considered another simpler notion of identifiability, called \emph{expectation identifiability}. With this notion, a GoM model with parameter set $(\bPi^*, \bT^*)$ is said to be identifiable, if for any other valid parameter set $(\tilde{\bPi}, \tilde{\bT})$, the equality of expectation $\mathbb E[\R] = \tilde{\bPi}\tilde{\bT}^\top=\bPi^*\bT^{*\top}$ holds if and only if $(\bPi^*, \bT^*)$ and $(\tilde{\bPi}, \tilde{\bT})$ are identical up to a permutation of the $K$ extreme profiles. To establish expectation identifiability for generalized-GoM models, we define subject $i$ to be a pure subject for extreme profile $k$ if the only positive entry of $\bpi_i^*$ is located at index $k$; i.e., $\bpi_i^* = \mathbf e_k$.
We impose Assumption~\ref{assumption:pure} for expectation identifiability of the generalized-GoM models.

\begin{assumption}
\label{assumption:pure}
    $\bo\Pi^*$ satisfies that every extreme latent profile has at least one pure subject. In addition, rank$(\bT^*)=K$.
\end{assumption}

The pure subject condition in Assumption~\ref{assumption:pure} is analogous to the anchor word assumption in topic modeling \citep{ke2024using} and the pure node assumption in mixed membership stochastic block model for networks \citep{jin2024mixed, mao2021mm}.
Under Assumption~\ref{assumption:pure}, denote $\S = (S_1,..., S_K)$ as the index vector of one set of $K$ pure subjects such that $\bPi_{\S,:}^* = \mathbf{I}_K$. Lemma \ref{lemma:simplex} gives the model parameter expressions in terms of $\S$ and the population SVD.

\spacingset{1}
\begin{algorithm}[ht!]
\centering
\caption{Spectral Method for Estimating the GoM Model with Categorical Responses}\label{algorithm2}
\begin{algorithmic}[1]
    \Require Categorical response matrix $\tilde\R\in\mathbb{R}^{N\times L}$, number of extreme latent profiles $K$ 
    \vspace{0.3em}
    \Ensure Estimated $\hat{\S}\in \mathbb{R}^K, \hat{\bT}^{(\text{post})}\in \mathbb{R}^{J \times K}, \hat{\bPi}^{(\text{post})}\in\mathbb{R}^{N\times K}$
    \vspace{0.3em}
    \State Obtain the flattened binary matrix $\R\in\mathbb{R}^{N\times J}$ from $\tilde \R$ according to \eqref{eq-flatten}.
    \State Get the top $K$ singular value decomposition of $\R$ as $\U\Lam\V^\top$
    \State $\Y = \U$
    \For{$k \in \{1,\dots,K\}$}                    
        \State $\hat{S}_k=\argmax (\{\|\Y_{i,:}\|_2:\; i\in[N] \backslash \hat{\bP}\})$
        \State $\mathbf{u} = \Y_{\hat{S}_k,:} / \|\Y_{\hat{S}_k,:}\|_2$
        \State $\Y = \Y (\I_K - \mathbf{u}\mathbf{u}^\top)$
    \EndFor
    \State $\hat{\bPi}=\U(\U_{\hat{\S},:})^{-1}$,\quad
    % \State 
    $\hat{\bPi}^{(\text{post})}=\diag(\hat{\bPi}_+ \mathbf{1}_K)^{-1} \hat\bPi_+$
    \State $\hat{\bT}=\V \Lam \U^{\top}_{\hat\S,:}$
    \State Let $\hat{\bT}^{(\text{post})} = (\hat{\bt}_{j,k}^{(\text{post})})$ with $\hat{\theta}_{j,k}^{(\text{post})}=
    \begin{cases}
        \hat{\theta}_{j,k} & \text{if } 0 \le \hat{\theta}_{j,k}\le 1 \\
        0 & \text{if } \hat{\theta}_{j,k}<0 \\
        1 & \text{if } \hat{\theta}_{j,k}>1
    \end{cases}$
    \For{$l \in \{1, \dots, L \}$}
        \For{$k \in \{1,\dots, K\}$}
                \State ${\bt}^b = \hat \bT^{(\text{post})}_{(\sum_{t=1}^{l-1} C_t+1):(\sum_{t=1}^l C_t),k }$
                \State $\hat \bT^{(\text{post})}_{(\sum_{t=1}^{l-1} C_t+1):\sum_{t=1}^l C_t ,k } =  \bt^b / \mathbf{1}_K^\top {\bt}^b $
        \EndFor
    \EndFor
\end{algorithmic}
\end{algorithm}
\spacingset{1.7}

\begin{lemma}\label{lemma:simplex}
    Under Assumption~\ref{assumption:pure}, {the generalized-GoM model is identifiable} and 
        $\bPi^* = \U^* \U_{\S,:}^{* -1}$, $\bT^* = \V^*\Lam^* \U^{*\top}_{\S,:}$. 
\end{lemma}

The first equation in Lemma \ref{lemma:simplex} implies that the $i$-th row of the left singular matrix $\U^*$ can be written as a convex combination of $\U_{S_1}^*,\dots, \U_{S_K}^*$, with the weights being the membership scores $\bpi_{i}^*$. This geometric property motivates us to use the successive projection algorithm \citep[SPA,][]{Gillis2013} to find the indices $\mathbf S$ of the pure subjects. SPA successively finds the vertices with the largest norm in a sequentially defined orthogonal subspace. After estimating $\hat\S = (\hat S_1,\ldots,\hat S_K)$, we can estimate the parameters using Lemma \ref{lemma:simplex}.
Algorithm \ref{algorithm2} summarizes the proposed procedure for the polytomous-GoM (i.e., traditional GoM) model. The algorithm for Poisson-GoM is given in Supplementary Material S.3.2. Algorithm \ref{algorithm2} can be easily generalized to accommodate generalized-GoM models with other data types by modifying the post-processing on the item parameter estimate $\hat\bT$ (line 11) to the appropriate range. These post-processing steps truncate the estimates to valid values and can only decrease the estimation error.

Related spectral methods have been studied for network models and topic models with mixed membership structures. The geometric properties in the spectral domain have been exploited for the MM network model using the eigenvalue decomposition of the symmetric adjacency matrix \citep{jin2024mixed, ke2023special, jin2021improvements, mao2021mm, zhang2020detecting}. 
Also, SVD-based methods have been used to estimate the topic models \citep{ke2024using, wu2023sparse}. 
{Compared to these existing works, we are motivated by asymmetric mixed membership models, especially the popular GoM model for high-dimensional categorical (polytomous) responses. Data arising from this model cannot be directly written as a signal matrix plus an \emph{entrywise-independent} noise matrix, in contrast to the usual network model and topic model. For our motivating GoM models, Bayesian methods and likelihood-based optimization methods are the predominant paradigms for parameter estimation. In contrast, our spectral method offers a unifying scalable estimation approach that comes with strong theoretical guarantees under local dependence; see Section \ref{sec-error}.}

\section{Uniform Estimation Error Bounds for Parameters}\label{sec-error}
To establish entrywise estimation error bounds for our spectral estimator $\hat\bPi$ and $\hat\bT$, it is necessary to develop fine-grained perturbation theory for the empirical singular matrices $\U$ and $\V$ according to Lemma \ref{lemma:simplex}.
In Section \ref{sec:perturbation}, we establish a novel $\ell_{2,\infty}$ perturbation theory for the left and right singular subspaces for the truncated SVD under locally dependent and flexibly distributed noise. 
In Section \ref{sec:consistency}, we present uniform entrywise convergence rates for parameters using the new singular subspace perturbation theory. Finally, in Section \ref{sec:prior}, we discuss an application to GoM models with the popular Dirichlet priors.

\subsection{New $\ell_{2,\infty}$ Singular Subspace Perturbation Theory}
\label{sec:perturbation}

We parse the impact of local dependence by measuring the covariances of the noise blocks:
\begin{equation*}
    \sigma^2 \coloneqq \max_{i\in[N],j\in[J]} \var(E_{i,j}), \quad
    \tilde{\sigma}^2 \coloneqq \max_{i\in[N], \ l\in[L]}\norm{\cov(\E_{i,S_l})}.
\end{equation*}
One can verify that $\sigma^2\le \tilde\sigma^2\le M\sigma^2$. In the special case of independent entries in $\mathbf E$, $\tilde\sigma^2=\sigma^2$; in the another special case of exactly equal entries within each block, $\tilde\sigma^2= M\sigma^2$.

\begin{assumption}
 \label{assumption: signal strength}   
We assume that $K \ll \min \{N, L\}, ML \asymp J$, and
\begin{align}
    & \sigma_K(\mathbf \Pi^*) \sigma_K(\mathbf \Theta^*) \gtrsim \sigma \kappa^*(M\sqrt{N} + \sqrt{J}), \label{eq:a3} \\ 
    & MB\log^2 (N\vee J) \max\left\{ \sqrt{\frac{\mu_1 K}{N}},\sqrt{\frac{\mu_2 K}{J}} \right\} \lesssim  \sigma. \label{eq:a4}
    \end{align}
\end{assumption}
Assumption~\ref{assumption: signal strength} assumes a low-rank signal matrix and balanced block size in the dependence structure. In addition, \eqref{eq:a3} assumes a lower bound on signal strength, and \eqref{eq:a4} assumes that the incoherence degrees cannot be too large. 
Theorem \ref{theorem-bound (general r.v.)} below gives our novel $\ell_{2,\infty}$ singular subspace perturbation bounds.
% under locally dependent and general-distribution noise. 
These results apply to general low-rank matrices with locally dependent noise from flexible distributions (see Assumption~\ref{assumption:block}), and are of independent interest to other latent variable models beyond the mixed membership models. 

\begin{theorem}[Two-to-infinity Singular Subspace Perturbation Under Local Dependence]\label{theorem-bound (general r.v.)}
Assume that Assumptions~\ref{assumption:block}, 
\ref{assumption: signal strength} hold. Then with probability at least $1- O((N\vee J)^{-10})$,
\begin{align}\label{eq: general bound for U} 
    & \norm{\U \U^{\top}\U^* - \U^*}_{2,\infty} \\ \notag
    \lesssim & \frac{\tilde \sigma \sqrt{N \log(N\vee J)}}{{\sigma^*_K}} \sqrt{\frac{\mu_1 K}{N}} +  \frac{\kappa^* \sigma^2 J}{{\sigma^*_K}^2}  \sqrt{\frac{\mu_1 K}{N}} + \frac{\sigma MB\log(N \vee J) \sqrt{MN+J}}{{\sigma_K^*}^2} \sqrt{\frac{\mu_2 K}{J}} =: \xi_1,\\[3mm]
    \label{eq: general bound for V}
    & \norm{\V \V^{\top} \V^* - \V^*}_{2,\infty}  \\
    \lesssim & \frac{\sigma \sqrt{J\log(N\vee J)}}{{\sigma_K^*}} \sqrt{\frac{\mu_2 K}{J}} + \frac{\kappa^* \sigma^2 MN}{{\sigma^*_K}^2} \sqrt{\frac{\mu_2 K }{J}} + \frac{\sigma B\log(N\vee J) \sqrt{M N + J} }{{\sigma_K^*}^2} \sqrt{\frac{\mu_1 K}{N}}=:\xi_2,  \\[3mm]
    & \norm{\U \mathbf \Lambda\V^{\top} - \U^*{\mathbf \Lambda}^*{\V^*}^{\top} }_\infty \notag \\
    \lesssim & \kappa^* K \sigma \log (N \vee J) \sqrt{\mu_1\mu_2} \sqrt{\frac{M}{J} + \frac{1}{N}} + MB \log(N\vee J)\left(\frac{\mu_2 K}{J} + \frac{\mu_1 K}{MN} \right)\eqqcolon \xi_3. \label{eq: general bound for ULambdaV}
\end{align}
\end{theorem}

\begin{remark} \label{remark:1}
    By imposing additional assumptions, we can derive simpler expressions for the upper bounds presented in Theorem \ref{theorem-bound (general r.v.)}. 
    First, provided that ${MB}/{\sigma} \lesssim \log(N\vee J)$ and ${\mu_2}/{\mu_1}\lesssim {J}/{(\log (N\vee J))^3} $, then the third term in \eqref{eq: general bound for U} is dominated by the first term and
\begin{equation}\label{eq: simplified U perturbation bound}
     \norm{\U \U^{\top}\U^* - \U^*}_{2,\infty} \lesssim \left(\frac{\tilde \sigma \sqrt{N\log(N \vee J)}}{{\sigma^*_{K}}} + \frac{\kappa^* \sigma^2 J}{{\sigma^*_{K}}^2} \right)\sqrt{\frac{\mu_1 K}{N}}.
\end{equation} 
Similarly, if ${B}/{\sigma} \lesssim \log(N\vee J)$ and ${\mu_1}/{\mu_2}\lesssim {N}/{(\log (N\vee J))^3}$ hold, \eqref{eq: general bound for V} is simplified to
\begin{equation}\label{eq: simplified V perturbation bound}
    \norm{\V \V^{\top} \V^* - \V^*}_{2,\infty} \lesssim \left(\frac{\sigma \sqrt{J \log(N \vee J)}}{{\sigma_{K}^*}}+ \frac{\kappa^* \sigma^2 MN}{{\sigma^*_{K}}^2} \right)  \sqrt{\frac{\mu_2 K}{J}}.
\end{equation}
Lastly, if $\sqrt{M}B \lesssim \kappa^* \sigma \sqrt{\frac{\mu_1}{\mu_2} \wedge \frac{\mu_2}{\mu_1}} \sqrt{\frac{M^2N^2}{J} \wedge \frac{J^2}{MN}}$ holds, the entrywise bound \eqref{eq: general bound for ULambdaV} becomes
\begin{equation}\label{eq: simplified inf bound}
    \norm{\U\mathbf \Lambda \V^{\top} - \U^* \mathbf\Lambda^*{\V^*}^{\top}}_\infty \lesssim \kappa^* K \sigma\sqrt{\mu_1\mu_2} \log(N \vee J) \sqrt{\frac{M}{J} + \frac{1}{N}}.
\end{equation}
\end{remark}

\begin{remark}
We give an interpretation of \eqref{eq: simplified V perturbation bound} in Remark \ref{remark:1}, and similar interpretation holds for \eqref{eq: simplified U perturbation bound}. We employ the expansion formula in \cite[Theorem 1]{xia2021normal}:
    \begin{align}
        & \mathbf V \mathbf V^{\top} \mathbf V^* - \mathbf V^* \nonumber \\
        = & \big(\mathbf I_J - \mathbf V^* { \mathbf V^* }^{\top} \big) \underbrace{\big( \E^\top\R^* + \E^\top\E + \R^{*\top} \E \big)}_{\R^\top \R - \R^{*\top} \R^*} \mathbf V^* { \mathbf \Lambda^*}^{-2} + \text{higher order terms } \nonumber\\ 
        = & \E^{\top} \R^* \mathbf V^*{ \mathbf \Lambda^*}^{-2} + \E^\top\E \mathbf V^*{ \mathbf \Lambda^*}^{-2} + \text{negligible terms} + \text{higher order terms} \label{eq: expansion of V}.
    \end{align}
The two terms of the upper bound in \eqref{eq: simplified V perturbation bound} correspond to the decomposition above as follows:  
\begin{align*}
    \norm{\E^\top \R^* \V^* { \mathbf \Lambda^*}^{-2}}_{2,\infty} & = \norm{\mathbf E^\top \U^* {\mathbf\Lambda^*}^{-1}}_{2,\infty} \lesssim \frac{\sigma \sqrt{J\log (N \vee J)}}{{\sigma^*_{K}}} \sqrt{\frac{\mu_2 K}{J}}, \\
    \norm{\E^{\top}\E \V^* {\mathbf \Lambda^*}^{-2} }_{2,\infty} & \lesssim  \frac{\sigma\sqrt{K \log (N \vee J)}}{\sigma_K^*} + \frac{\sigma^2 MN}{{\sigma_K^*}^2} \sqrt{\frac{\mu_2 K}{J}}\\ 
    &\lesssim \frac{\sigma \sqrt{J\log(N\vee J)}}{{\sigma_K^*}} \sqrt{\frac{\mu_2 K}{J}} + \frac{\kappa^* \sigma^2 MN}{{\sigma^*_K}^2} \sqrt{\frac{\mu_2 K }{J}} ,
\end{align*}
hold with probability at least $1  - O((N\vee J)^{-10})$.
The first line follows from inequality (S.8) in Lemma S.3 and the second line holds by Lemma S.7 in the Supplementary Material. 
\end{remark}

Allowing for arbitrarily flexible locally dependent noise is a key distinguishing feature of our theory.
In recent years, there has been significant progress in $\ell\ti$ eigenspace perturbation theory for random matrices \citep{fan2018formula, cape2019two, mao2021mm, chen2021spectral, cai2021subspace, abbe2022, yan2024inference}. 
However, in this literature, most studies consider independent entries in the noise matrix, with only a few exceptions including \cite{lei2019unified}, \cite{modell2024intensity}, and \cite{agterberg2022entrywise}. To facilitate the interpretation, we first delve further into the simplified forms presented in Remark~\ref{remark:1}. 
On the one hand, our upper bound \eqref{eq: simplified U perturbation bound} for the left singular subspace $\U$ involves $\tilde\sigma$ as a critical quantity to measure the noise level, and provides a sharp characterization of the SVD-based method's performance (in light of a similar decomposition to \eqref{eq: expansion of V} for $\U$).
On the other hand, it is inevitable to involve the block size $M$ in our leave-one-block-out argument for deriving the bound for the right singular subspace $\V$.
Actually, we highlight that the dependence on $M$ in our bounds \eqref{eq: simplified V perturbation bound} and \eqref{eq: simplified inf bound} is \emph{sharp}, by considering the following special case. 
Imagine the scenario where each block size is exactly $M$, and the entries $R_{i,j},~ j \in S_l$ in each block $S_l$ are duplicates of a single variable $\check R_{i, l}\sim \mathcal N(0, \sigma^2)$ for each $i\in[N]$. Invoking \eqref{eq: expansion of V}, the fluctuation $\V_{j,:}\V^{\top}\V^* - \V^*_{j,:}$ is partially attributed to the component $\E_{:,j}^{\top}\E_{:,S_l} \V_{S_l,:}^* {\mathbf \Lambda^*}^{-2}$, given $j\in S_l$. Applying elementary concentration inequalities (e.g.~Hoeffding's inequality for sub-Gaussian entries), we find that $(\E^{\top} \E)_{S_l,S_l}$ is concentrated around $\sum_{i\in [N]} \text{Var}(\check R_{i,l}) \mathbf 1_{M} \mathbf 1_{M}^\top$. This leads to the result that $\max_{j\in[J]}\norm{\E_{:,j}^{\top}\E_{:,S_l} \V_{S_l,:}^* {\mathbf \Lambda^*}^{-2}}_2$ is lower bounded by $\frac{\sigma^2 MN }{{\sigma_{1}^*}^2} \norm{\V^*}_{2,\infty}$ with high probability. This confirms that the dependence of \eqref{eq: simplified V perturbation bound} on $M$ is sharp and cannot be improved. 
Moreover, under the same setting, the minimax lower bound for the entrywise error, as informed by \cite{koltchinskii2011nuclear}, is $\sigma \sqrt{K({M}/{J}+ {1}/{N})}$. Provided that $\mu_1, \mu_2 , \kappa^* ,K = O(1)$, our upper bound \eqref{eq: simplified inf bound} is optimal up to a logarithmic term.

We further clarify how our bounds differ from previous results. 
In the literature,
\cite{lei2019unified} and \cite{modell2024intensity} studied eigenspace perturbation to accommodate certain dependence structures. 
To adapt these results from symmetric random matrices to asymmetric matrices, a common approach is to embed the asymmetric matrix as two blocks within a larger symmetric matrix, a technique known as the ``symmetric dilation'' trick.
However, this symmetrization technique often falls short of delivering a sharp dependence on the ratio $J/N$ and the incoherence degrees $\mu_1$ and $\mu_2$, because the corresponding singular subspace perturbation bounds for asymmetric random matrices resulting from symmetrization typically involve $J\vee N$ and $\mu_1 \vee \mu_2$. 
However, distinguishing $N$ from $J$ and distinguishing $\mu_1$ from $\mu_2$ become particularly important when encountering unbalanced observation matrices ($N=o(J)$ or $J=o(N)$) with unbalanced incoherence degrees ($\mu_1\ll \mu_2$ or $\mu_1\gg \mu_2$). To see the importance of distinguishing $\mu_1$ between $\mu_2$, we note that $\mu_1 \le \kappa^2(\bPi^*) / K$ when Assumption~\ref{assumption:pure} holds by Lemmas \ref{lemma:simplex} and S.7, thus $\mu_1 = O(1)$ if $\kappa(\bPi^*) = O(\sqrt{K})$. Section \ref{sec:prior} will show that $\kappa(\bPi^*) = O(\sqrt{K})$ holds with high probability if the membership scores are generated from a Dirichlet distribution, a common assumption in Bayesian inference for GoM models. In contrast, $\mu_2$ often exhibits much larger values due to the prevalence of sparse patterns and heterogeneity among the items. 
This distinction is supported by numerous real-world datasets, including the single-cell data in Section \ref{sec:real data} where the empirical value of $\mu_1$ is around $2.9$ while for $\mu_2$ it is $55.7$. We address this issue by establishing perturbation bounds for the left and right singular vectors \emph{separately}. Our technique is conceptually similar to that of the arXiv version of 
\cite{yan2024inference}, but with a nuanced treatment of local dependence. 
    
Lastly, we remark on the application of matrix concentration universality in our proof to overcome potential dependence issues. 
Specifically, given an $N$-by-$J$ noise matrix $\mathbf E$ with block dependence in Assumption~\ref{assumption:block}, we employ Corollary~2.15 in \cite{brailovskaya2024universality} to derive that $\norm{\mathbf E} \lesssim \tilde\sigma\sqrt{N}+ \sigma \sqrt{J}$ with high probability. To compare this with the preceding approaches, consider the bound provided by the matrix Bernstein inequality, which often yields $\norm{\mathbf E} \lesssim (\tilde\sigma\sqrt{N}+ \sigma \sqrt{J})\sqrt{\log(N\vee J)}$ with high probability. This result is $\sqrt{\log(N\vee J) }$-looser than our universality-based concentration. Moreover, this elimination of the $\sqrt{\log (N\vee J)}$ factor also extends to the bounds on $\norm{\mathbf E\mathbf V^*}$ and $\norm{{\mathbf U^*}^{\top} \mathbf  E}$.

\subsection{Uniform Consistency of Parameter Estimation}\label{sec:consistency}

We establish the error bounds for the estimators $\hat\bPi = \U\U_{\hat{\S},:}^{-1}, \ \hat \bT = \V\Lam\U_{\hat\S,:}^\top$ in Theorem \ref{theorem: estimation error}.

\begin{theorem}[Uniform Consistency of Parameter Estimation]
\label{theorem: estimation error}
    Suppose Assumptions \ref{assumption:block}, \ref{assumption:pure}, \ref{assumption: signal strength} hold, and additional assume that $\sqrt{K} \kappa^2(\mathbf \Pi^*)\sigma_1(\mathbf \Pi^*) \xi_1 \ll 1$. Then there exists a permutation matrix $\mathbf P$ such that with probability at least $1- O\big((N\vee J)^{-10})$, 
    \begin{equation}
    % \label{eq:pi_bound}
        \norm{\hat{\mathbf \Pi} - \mathbf \Pi^* \mathbf P}_{2, \infty} = \ O(\kappa^2(\mathbf \Pi^*) \sigma_1(\mathbf \Pi^*) \xi_1), \quad
    % \label{eq:theta_bound}
        \|\hat{\bT}-\bT^*\bP\|_{\infty} = \ O\left( \kappa^2(\bPi^*) \xi_3\right).
    \end{equation} 
\end{theorem}
The $\ell\ti$ bound of $\hat\bPi$ primarily depends on the left singular subspace perturbation level $\xi_1$, as the estimate is derived from $\hat\U$. The additional term $\kappa^2(\bPi^*) \sigma_1(\bPi^*)$ in the bound of $\hat{\mathbf \Pi}$ arises from the vertex hunting step via the successive projection. 
Similarly, the $\ell_{\infty}$ error bound of $\hat\bT$ involves the upper bound of $\norm{\R^* - \U\Lam\V^\top}_\infty$, which is reasonable given that $\hat\bT^\top$ is a submatrix of $\U\Lam\V^\top$, and $\kappa^2(\bPi^*)$ again stems from estimating $\hat\S$ via vertex hunting.

\begin{corollary}\label{corollary 2}
    Suppose the assumptions in Theorem \ref{theorem: estimation error} hold and 
    $\max\{\kappa(\mathbf \Pi^*), \kappa(\mathbf \Theta^*)\} = O(1)$. 
    Then there exists a permutation matrix $\mathbf P$ such that with probability at least $1 - O\big( (N\vee J)^{-10}\big)$,
    \begin{align}
        & \norm{\hat{\mathbf \Pi} - \mathbf \Pi^* \mathbf P}_{2,\infty} \notag \\ 
        \lesssim & \frac{\tilde\sigma \sqrt{\mu_1 K \log (N\vee J)}}{\sigma_{K}(\mathbf \Theta^*)} + \frac{\sigma^2 \sqrt{JK}}{\sigma_{K}(\mathbf \Pi^*)\sigma_{K}^2(\mathbf \Theta^*)} \left( \sqrt{\frac{J}{N}} \sqrt{\mu_1} + \sqrt{1+\frac{MN}{J}} \frac{MB \log (N\vee J)}{\sigma \sqrt{J}} \sqrt{\mu_2} \right), \label{eq: Pi estimation error}\\[3mm] 
        & \norm{\hat{\mathbf \Theta} - \mathbf \Theta^* \mathbf P}_\infty \notag \\
        \lesssim & K\sigma \log (N \vee J) \sqrt{\mu_1\mu_2} \sqrt{\frac{M}{J} + \frac{1}{N}} + MB\log (N\vee J) \left(\frac{\mu_1 K}{MN} + \frac{\mu_2 K}{J} \right). \label{eq: Theta estimation error}
    \end{align}
\end{corollary}

To see the rates of convergence with {Corollary \ref{corollary 2}}, 
consider the case where $\sigma^2_K(\bPi^*)\gtrsim N/K, \sigma^2_K(\bT^*) \gtrsim J/(MK)$.
Under the mild assumptions for \eqref{eq: simplified U perturbation bound} and \eqref{eq: simplified inf bound} in Remark \ref{remark:1}, $\hat\bpi_i$ converges to $\bpi^*_i$ with a rate of $\tilde{O}\left(\sigma K^2 M \sqrt{\mu_1} \big( \frac{1}{\sqrt{J}} \vee \frac{\sigma K}{N} \big) \right)$,
and $\hat\bt_j$ converges to $\bt^*_j$ with a rate of $\tilde{O}\left( \sigma K^2 \sqrt{\mu_1\mu_2} \big( \sqrt{\frac{M}{J}} \vee \frac{1}{\sqrt{N}} \big) \right)$.

\subsection{Application to GoM Models with Dirichlet Prior}
\label{sec:prior}
Dirichlet prior on the membership scores is commonly adopted in Bayesian inference for GoM models \citep{erosheva2002grade, erosheva2007describing, gu2023dimension}.
To illustrate the theoretical results, we discuss a Dirichlet prior application on the GoM model for polytomous data estimated using the flattening procedure. 
Assume that $\bo\pi_i^*$ is independently sampled from a Dirichlet distribution with parameters $\bo\alpha = (\alpha_1,\ldots,\alpha_K)$. For simplicity, we assume $C_l= M$ for all $l\in[L]$.
Similarly, we assume that for each $l\in[L], k\in[K]$, vector $(\theta^*_{M(l-1)+c,k})_{c=1}^{M}$ independently follows a Dirichlet distribution with parameters $\bo\beta=(\beta_{1},\dots, \beta_{M})$.

\begin{assumption}\label{assumption:dirichlet}
    Assume that $\max_k \alpha_k \le C_1 \min_k \alpha_k$ for some $C_1>0$, $\max_c \beta_c \le C_2 \min_c \beta_c$ for some $C_2>0$, $\max_k \alpha_k = O(1), \max_c \beta_c = O(1)$. %$\alpha_0=O(1)$, and $\beta_0=O(1)$
    In addition, $N\gtrsim M^2K^5 \log^4(N\vee J), J \gtrsim M^4 K^6 \log^4(N\vee J)$ and
    \begin{align}
        \nu \le \frac{\sqrt{N / \log N}}{6\sqrt{3}(1+\alpha_0)} , \quad \ c_1 \ge 4 K \sqrt{15\log L / L} \label{eq:assumption-c_2}.
    \end{align}
\end{assumption}
Note that \eqref{eq:assumption-c_2} is similar to Assumption 3.3 in \cite{mao2021mm}.

\begin{corollary}\label{cor:dirichlet}
    Let $\bo\pi_i^*\stackrel{i.i.d}{\sim}\operatorname{Dirichlet}(\boldsymbol{\alpha})$ and $(\theta^*_{M(l-1)+c,k})_{c=1}^{M}\stackrel{i.i.d}{\sim} \operatorname{Dirichlet}(\bo\beta)$. Suppose Assumption~\ref{assumption:dirichlet} holds. Then with probability at least $1-O(K(N\wedge L)^{-3})$, one has
    \begin{align*}
        & \norm{\hat{\mathbf \Pi} - \mathbf \Pi^* \mathbf P}_{2, \infty} \lesssim \frac{M^{3/2} K^{5/2} \sqrt{\log(N \vee J)}}{\sqrt{J}} + \frac{\sqrt{M^2 K^{9/2}}}{N}, \\
        & \norm{\hat{\bT}-\bT^*\bP}_{\infty} \lesssim \frac{M^{5/2} K^{9/2} \log(N\vee J)}{\sqrt{J}} + \frac{M^2 K^{9/2} \log(N \vee J)}{\sqrt{N}}.
    \end{align*}
\end{corollary}

Corollary \ref{cor:dirichlet} gives strong entrywise guarantees for our spectral estimator for the GoM model for categorical data. To our best knowledge, this is the first uniform estimation error control for any method applied to the GoM model in high-dimensional settings.

\section{Simulation Studies}\label{sec:simulation}

We conduct three simulation studies to assess the efficiency and accuracy of the proposed method. Section \ref{subsec-simu-flatten} presents results for the GoM model for multivariate polytomous data, where the item responses are polytomous and local dependence is introduced by the flattening technique.  Section \ref{subsec-simu-intrinsic} presents results for the GoM model for multivariate binary data, where the binary item responses are designed to have intrinsic blockwise dependent entries. 
In the third study in Section \ref{subsec-simu-poisson}, we consider the generalized-GoM model for Poisson data and compare our method with the popular NMF algorithm \citep{lee2000algorithms}. 
For all three studies, we consider $K=3$ and generate $100$ independent replications for each simulation setting. 
Within each setting and each replication, the rows of $\bPi$ are independently simulated from the Dirichlet$(\mathbf{1}_K)$ distribution, and the first $K$ rows of $\bPi$ are set to the identity matrix $\mathbf{I}_K$ to satisfy Assumption~\ref{assumption:pure}. For both the simulation studies and real-data applications, we employ a pruning step on $\U$ before SPA in Algorithm \ref{algorithm2} to de-noise \citep{mao2021mm}, where we set the tuning parameters as $r=10, q=0.4, e=0.2$ as suggested in \cite{chen2024spectral}.

\subsection{Local Dependence from Flattening Polytomous Responses}\label{subsec-simu-flatten}
We first consider local dependence introduced by the flattening technique for polytomous data, with the number of response categories set to $C_l\equiv C= 3$ for all $l\in[L]$. Consider sample size $N\in\{200, 1000, 2000, 3000, 4000, 5000\}$, and $J=N/5$.
For each $j\in[J]$ and $k\in[K]$, $\bT^*_{(C(j-1)+1):(Cj),k}$ is independently simulated from Dirichlet$(0.2\cdot \mathbf{1}_C)$. 
For each pair $(i,j)$, the non-zero index in $\R_{i, (C(j-1)+1):(Cj)}$ is independently generated from the categorical distribution with the probability vector being $\bT^*_{(C(j-1)+1):(Cj),:} \bPi_{i,:}^{*\top}$.

\begin{figure}[hbt!]
    \centering
    \includegraphics[width=0.75\textwidth]{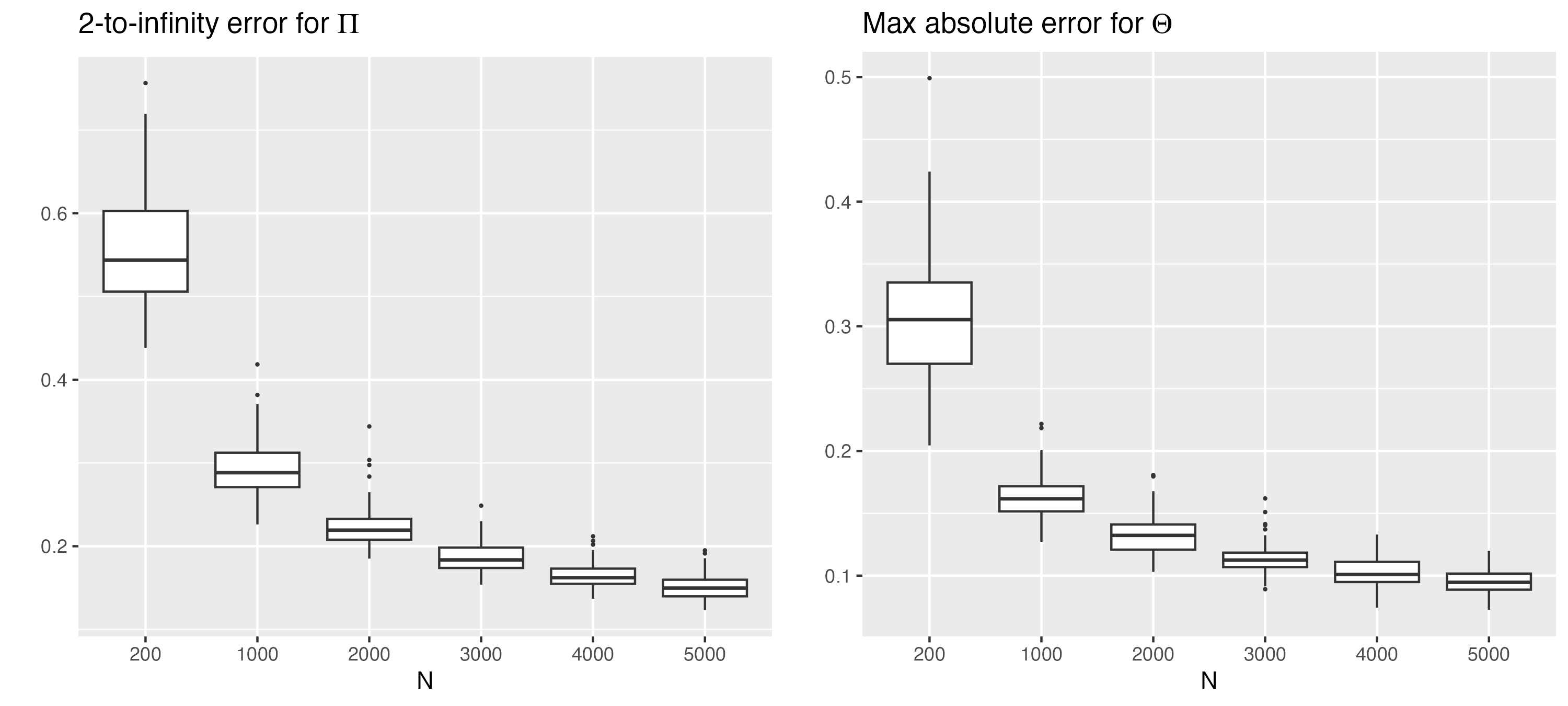}
    \caption{Simulation study I, estimation error for the proposed method using the flattening technique. The left panel presents the two-to-infinity error for $\hat\bPi$ and the right presents the maximum absolute error for $\hat\bT$.}
    \label{fig:error_flatten}
\end{figure}

Figure \ref{fig:error_flatten} displays the boxplots of the two-to-infinity estimation error for $\hat{\bPi}$ and the maximum absolute estimation error for $\hat{\bT}$ obtained from the proposed method. We observe that the error decreases and approaches $0$ as $N, J$ grow for both estimates, which verifies our uniform consistency results for parameter estimation.

We compare our method with the Gibbs sampling method and the variational inference method for the polytomous GoM model. The parameters in the Gibbs sampler are initialized with their generating distributions. The derivation of each updating step in the Gibbs sampler can be found in Section S.3.1 in the Supplementary Material. We set the first $5000$ samples as burn-in and take the average of $2000$ samples after the burn-in phase as the Gibbs sampling estimates.
We employ the variational inference (VI) algorithm proposed in \cite{wang2015fitting}, implemented in the \texttt{mixedMemModel} function in R package \texttt{mixedMem}. As Gibbs sampling is not scalable to large datasets and would take a very long time to compute when $N, J$ are large, we only consider $N\in\{200, 1000, 2000\}$ to compare the three methods' performances. 
Table \ref{tab:t_compare} presents the average computation time in seconds across the $100$ replications. When $N=2000$, our proposed method is about $10,000$ times faster compared to Gibbs sampling and $30$ times faster compared to VI. 
Table \ref{tab:err_compare} summarizes the average two-to-infinity error of $\hat\bPi$ and the max absolute error of $\hat\bT$.
In comparison to Gibbs sampling, the proposed method delivers comparable results in accuracy. We also find that the VI method in this case does not yield consistent results, and it is very sensitive to initialization. In summary, our proposed method is efficient in computation, especially for large-scale datasets, and is comparably accurate with popular existing methods.

\begin{table}[bt!]
\centering
\scalebox{0.95}{
\begin{tabular}{c c c c} 
 \hline
  & $N=200$ & $N=1000$ & $N=2000$ \\ 
 \hline
 Proposed & 0.07 & 2.25 & 12.84 \\ 
 VI & 6.63 & 85.14 & 264.40 \\
 Gibbs & 709.44 & 16458.22 & 65464.17 \\
 \hline
\end{tabular}
}
\caption{Simulation study I, average computation time in seconds for polytomous responses. The flattening technique is used in the proposed method. VI stands for the variational inference algorithm implemented in the \texttt{mixedMem} R package.}
\label{tab:t_compare}
\end{table}

\begin{table}[bt!]
\centering
\scalebox{0.95}{
\begin{tabular}{c c c c c c c c} 
 \hline
 $\bT$ & $N=200$ & $N=1000$ & $N=2000$ & $\bPi$ & $N=200$ & $N=1000$ & $N=2000$ \\ 
 \hline
 Proposed & 0.057 & 0.030 & 0.024 & Proposed & 0.098 & 0.047 & 0.035 \\
 VI & 0.252 & 0.265 & 0.272 & VI & 0.204 & 0.198 & 0.198 \\
 Gibbs & 0.072 & 0.032 & 0.023 & Gibbs & 0.080 & 0.038 & 0.028 \\
 \hline
\end{tabular}
}
\caption{Simulation study I, average mean absolute error for $\bT$ and $\bPi$ for polytomous responses. The flattening technique is used in the proposed method. VI stands for the variational inference algorithm implemented in the \texttt{mixedMem} R package.}
\label{tab:err_compare}
\end{table}

\subsection{Intrinsic Local Dependence across Different Items}\label{subsec-simu-intrinsic}

\begin{figure}[bt!]
    \centering
    \includegraphics[width=0.75\textwidth]{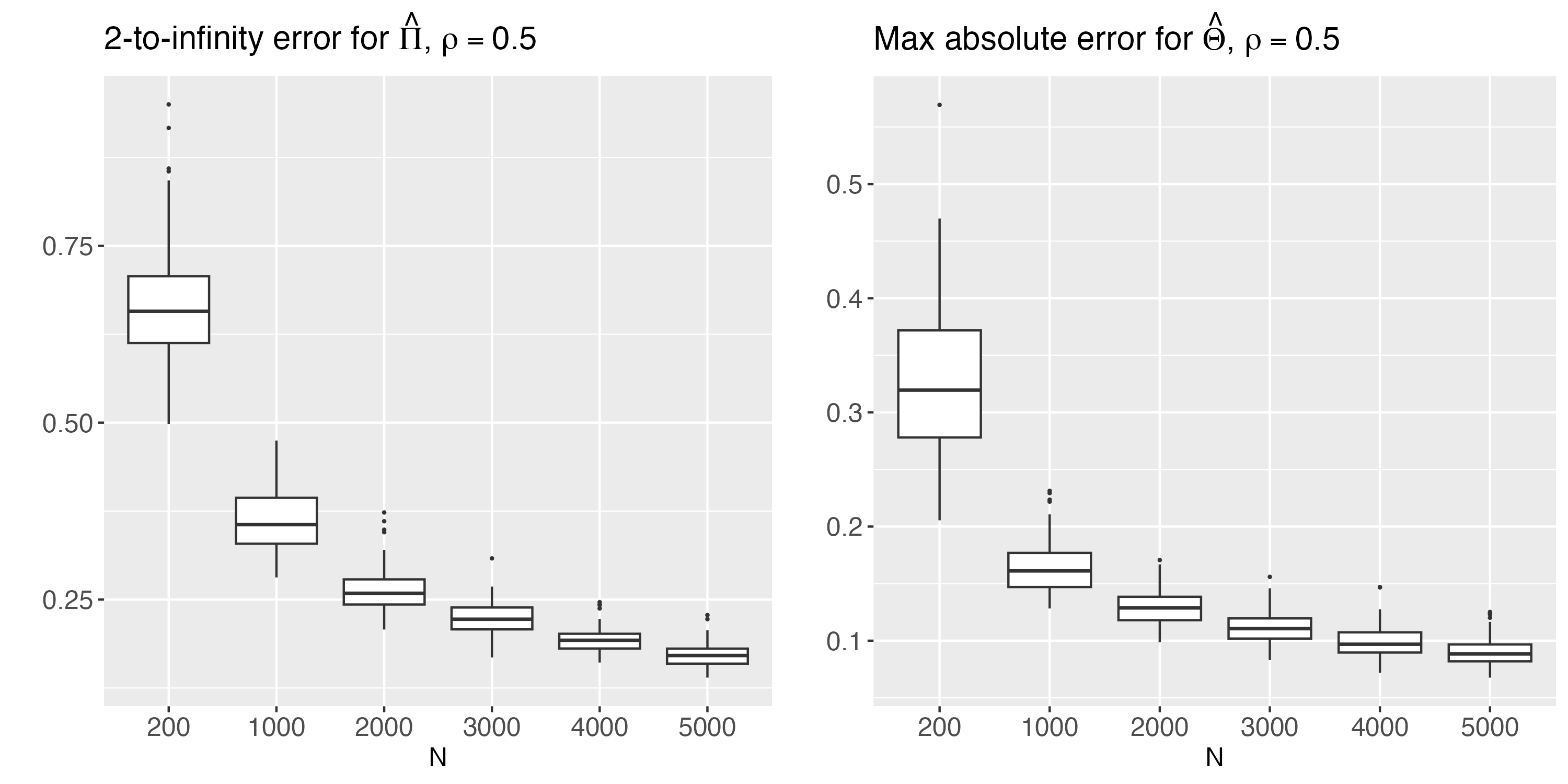}
    \caption{Simulation study II with block-wise dependent data. The left panel presents the two-to-infinity error for $\hat\bPi$ and the right presents the maximum absolute error for $\hat\bT$.}
    \label{fig:error_loc1}
\end{figure}

In this section, we generate binary responses with a more general intrinsic local dependence structure.
Now $\bT^*\in[0,1]^{J\times K}$ is no longer a flattened parameter matrix, but rather contains the item response probabilities as the Bernoulli parameters $\theta_{jk}^*\stackrel{i.i.d}{\sim} \text{Beta}(0.2, 0.2)$.
The membership scores $\bo\pi_i^*$ are generated with the same mechanism as stated in Section \ref{subsec-simu-flatten}.
To generate the locally dependent responses, we first generate latent variables $\bo\eta_i\in\mathbb{R}^J$ independently and identically from the Gaussian distribution $N(\bo0, \bo\Sigma)$, with $\bo\Sigma$ being blockwise diagonal with $J/M$ copies of a $M\times M$ block covariance matrix $\bo\Sigma_0$. Here $\bo\Sigma_0 \in\mathbb{R}^{M\times M}$ has an autoregressive structure with parameter $\rho$, that is, $(\bo\Sigma_0)_{ij}=\rho^{|i-j|}$.
We then define a threshold $d_{ij}=\Phi^{-1}(\bPi_{i,:}^\top \bT_{j,:})$ where $\Phi$ is the cumulative distribution function for the standard Normal distribution. Finally, each binary response is defined as $R_{i,j}=\mathbbm{1}(\eta_{ij} < d_{ij})$. From this construction, we have $\mathbb E[R_{i,j}] = \Pr(R_{i,j}=1)=\bPi_{i,:}^\top \bT_{j,:}$, and the elements in $\R_{i,:}$ are blockwisely dependent with a block size $M$. Specifically, we set $\rho=0.5$ and $M=10$, which induces a relatively strong level of dependence. We still set $N\in\{200, 1000, 2000, 3000, 4000, 5000\}$ and let $J=N/5$.
Figure \ref{fig:error_loc1} presents the error boxplots in this scenario. We can see that the estimates from our method are consistent as $N$ and $J$ grow, as expected from the theory.

\subsection{Generalized-GoM Estimation for Poisson Data}\label{subsec-simu-poisson}
\begin{figure}[ht!]
    \centering
    \includegraphics[width=\linewidth]{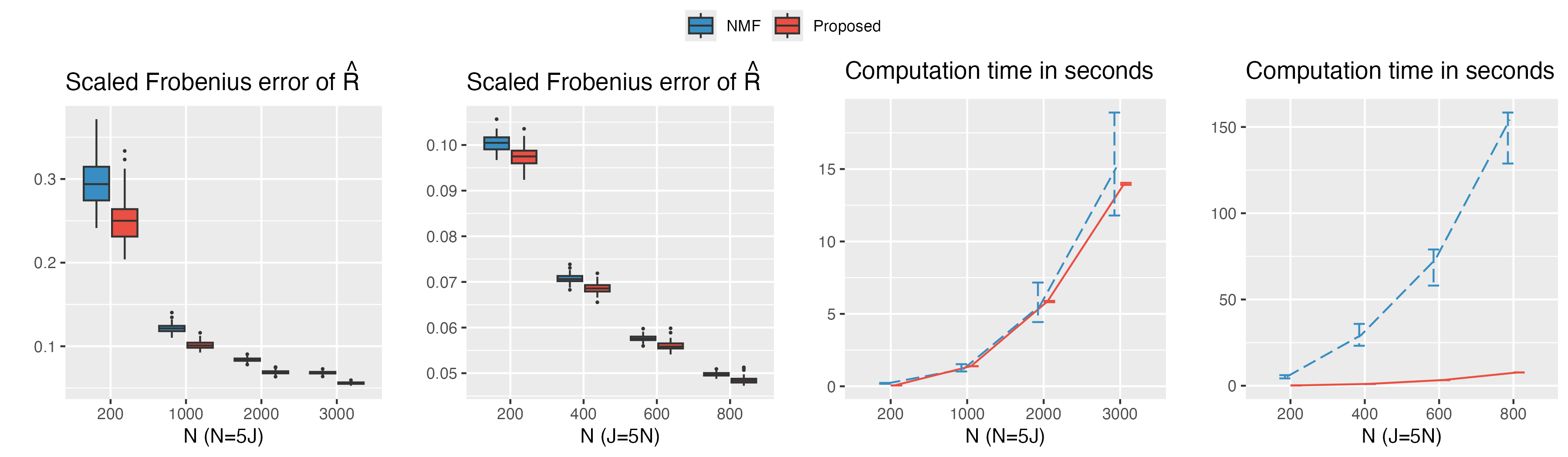}
    \caption{Simulation study III, comparing the estimation error and computation time for the proposed method and NMF with $N=5J$ and $J=5N$.}
    \label{fig:pois_compare}
\end{figure}

We simulate data with independent Poisson entries, and compare our method with the NMF algorithm of multiplicative update proposed in \cite{lee2000algorithms}. The Poisson rate parameters $\theta^*_{j,k}$ are independently sampled from $\text{Gamma}(1, 2)$. 
To see the effect of the ratio between $N$ and $J$, we consider both $N=5J$ with $N\in\{200,1000,2000,3000\}$ and $J=5N$ with $N\in\{200,400,600,800\}$.
We use the R package \texttt{NMF} to implement the NMF algorithm \citep{gaujoux2010flexible}, where most of the built-in algorithms have been optimized in C++. We compare the two methods in terms of computation time and the scaled Frobenius error $\frac{1}{\sqrt{NJ}} \norm{\R^* - \hat\bPi {\hat\bT}^\top}_F$ in Figure \ref{fig:pois_compare}.
The proposed method yields slightly more accurate estimates across different simulation settings. The computation time of the two methods is comparable when $N=5J$, while the proposed method is much more efficient in the high-dimensional setting with $J=5N$.
It is worth noting that the matrix decomposition estimate from NMF requires post-processing to locate each row of $\hat{\bPi}$ on a simplex. Additionally, the NMF algorithm itself does not give consistent estimates of parameters $\bPi$ and $\bT$ due to the lack of identifiability constraints. Therefore, we have only compared the estimation errors of the low-rank matrix itself in Figure \ref{fig:pois_compare}.

\section{Real-data Applications}\label{sec:real data}
In this section, we apply the proposed method to three real-world datasets arising from diverse applications, including political survey data ($N> J$, polytomous data), population genetics data ($N\ll J$, binomial data), and single-cell sequencing data ($N\ll J$, nonnegative count data). These datasets exhibit various different local dependence behaviors, either resulting from the flattening technique, or from intrinsic dependence, or both.

\subsection{Political Survey Data}
The 2022 pilot survey from the American National Election Studies (ANES) provides data about voting and public opinion after the 2022 midterm elections in the United States (publicly available at \url{https://electionstudies.org/}).
We use a subset of $J=145$ items after excluding followup questions, numeric response questions, and questions that directly ask about party or political ideology. A total $N=1,511$ respondents that answered all these items is included in the analysis.
The number of categories for these items ranges from $2$ to $7$, with the majority being $5$. After flattening the polytomous response matrix, the column dimension of the data matrix becomes $J  = 648$. We choose the number of extreme profiles to be $K=3$, which is the convention in the literature for analyzing US political data \citep{gross2012mixed, wang2015fitting}. The computation time for the whole flattening and parameter estimation procedure takes $3.8$ seconds.

\begin{figure}[b!]
    \centering
    \includegraphics[width=\textwidth]{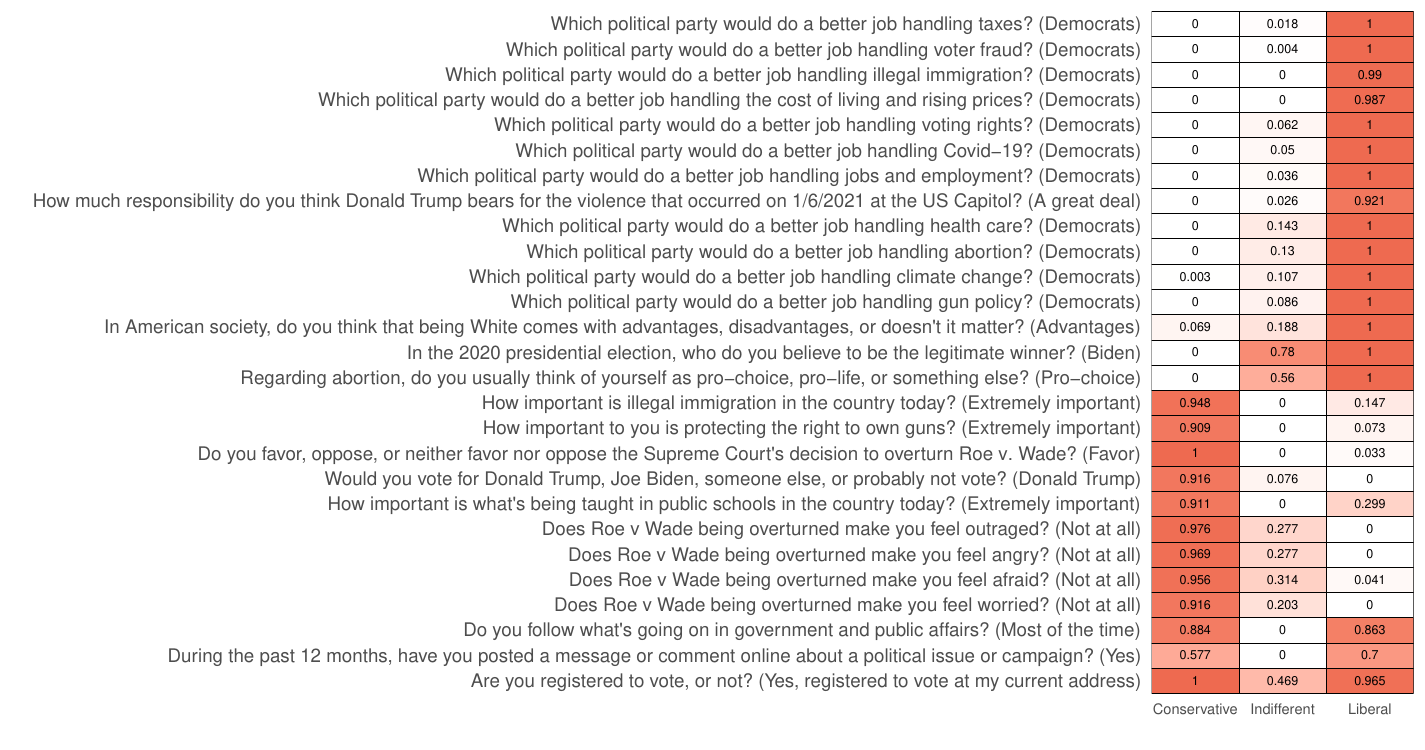}
    \caption{Heatmap of a selected subset of the estimated item response parameters with $K=3$ for the ANES data. Each value corresponds to the estimated probability of choosing the corresponding response in the parenthesis. }
    \label{fig:ANES_heatmap}
\end{figure}

By inspecting the estimated item response probability matrix $\hat\bT$, we can draw very intuitive interpretations on the three extreme profiles.
Figure \ref{fig:ANES_heatmap} gives the heatmap of the estimated response probabilities for a selected subset of items {with a representative category of each item. Note that these items are not necessarily binary and are indeed mostly polytomous with more than two categories.} We can see that the three extreme profiles are well-differentiated by the three columns of $\hat\bT$.
We interpret profile $k=1$ to be conservative in their political ideology. Respondents belonging to this profile are much more likely to vote for Donald Trump, tend to care more about illegal immigration and gun rights, and also much more in favor of the overturn of Roe v. Wade.
On the contrary, profile $k=3$ are the liberals and are much more likely to vote for Joe Biden. They also think the Democrats would do a better job handling a variety of issues such as taxes, voter fraud, Covid-19, jobs and employment, etc.
Profile $k=2$ are interpreted to be indifferent. They tend not to follow what is going on in government and public affairs, are less likely to be registered to vote.

The ANES dataset also contains self-reported party information, {which we do not use for estimating the parameters. To explore how the estimated latent membership scores align with this held-out information, we present the ternary plot of $\hat\bPi$ color coded by the party information in} the right panel in Figure \ref{fig:tern}. In this figure, each dot corresponds to a respondent's $\hat\bpi_i$, whose color represents his or her self-reported party information. We can see that Democrats are closer to the liberal profile, Republicans are closer to the conservative profile, and Independents are closer to the indifferent profile. Since we did not use the party information in the estimation procedure, this result indicates that our method extracts meaningful and interpretable latent representations of people's political ideology.

\begin{figure}[htp]
    \centering
    \begin{minipage}{0.4\textwidth}
        \centering
        \includegraphics[width=\linewidth]{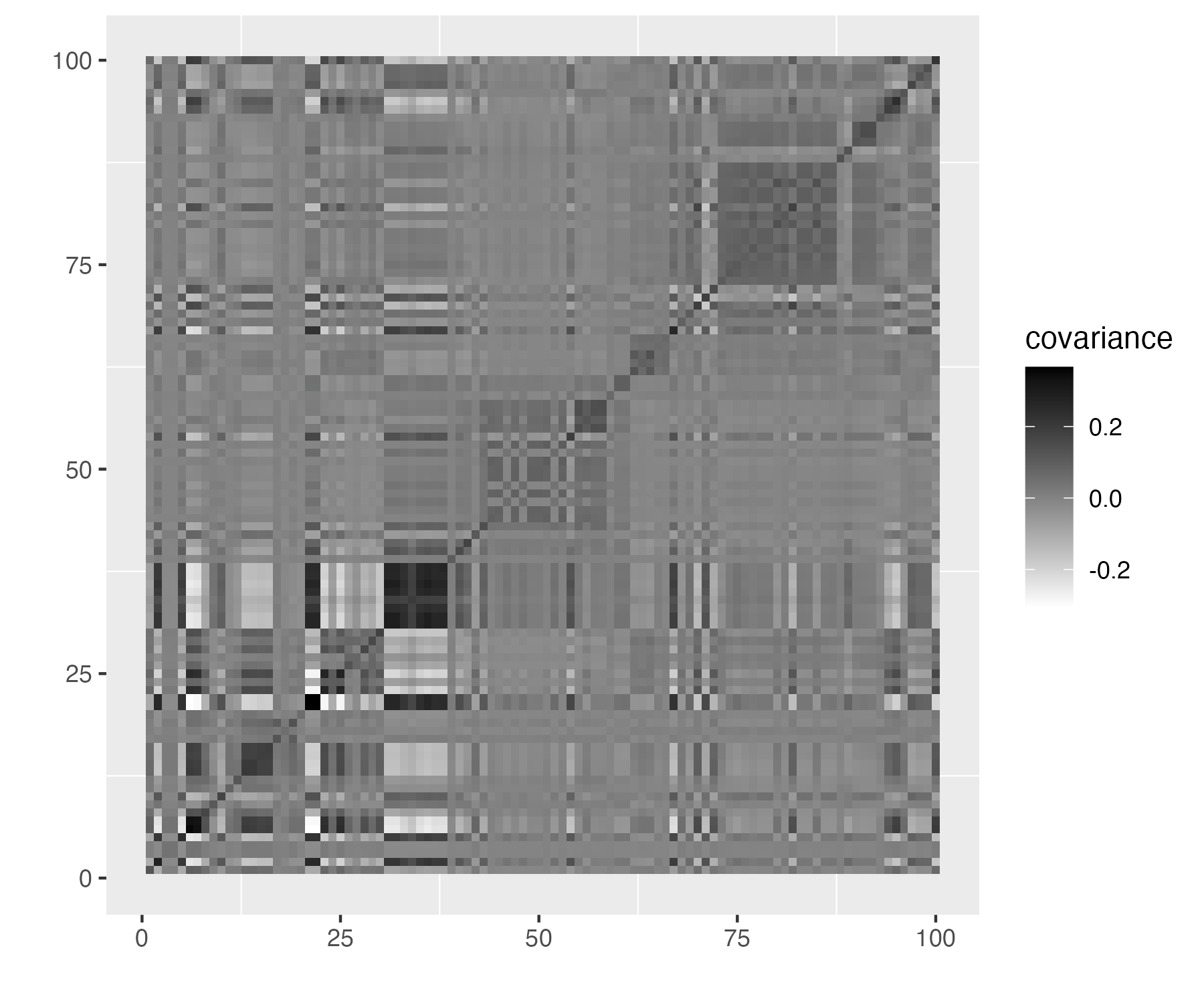}
    \end{minipage}
    \begin{minipage}{0.4\textwidth}
        \centering
        \includegraphics[width=\linewidth]{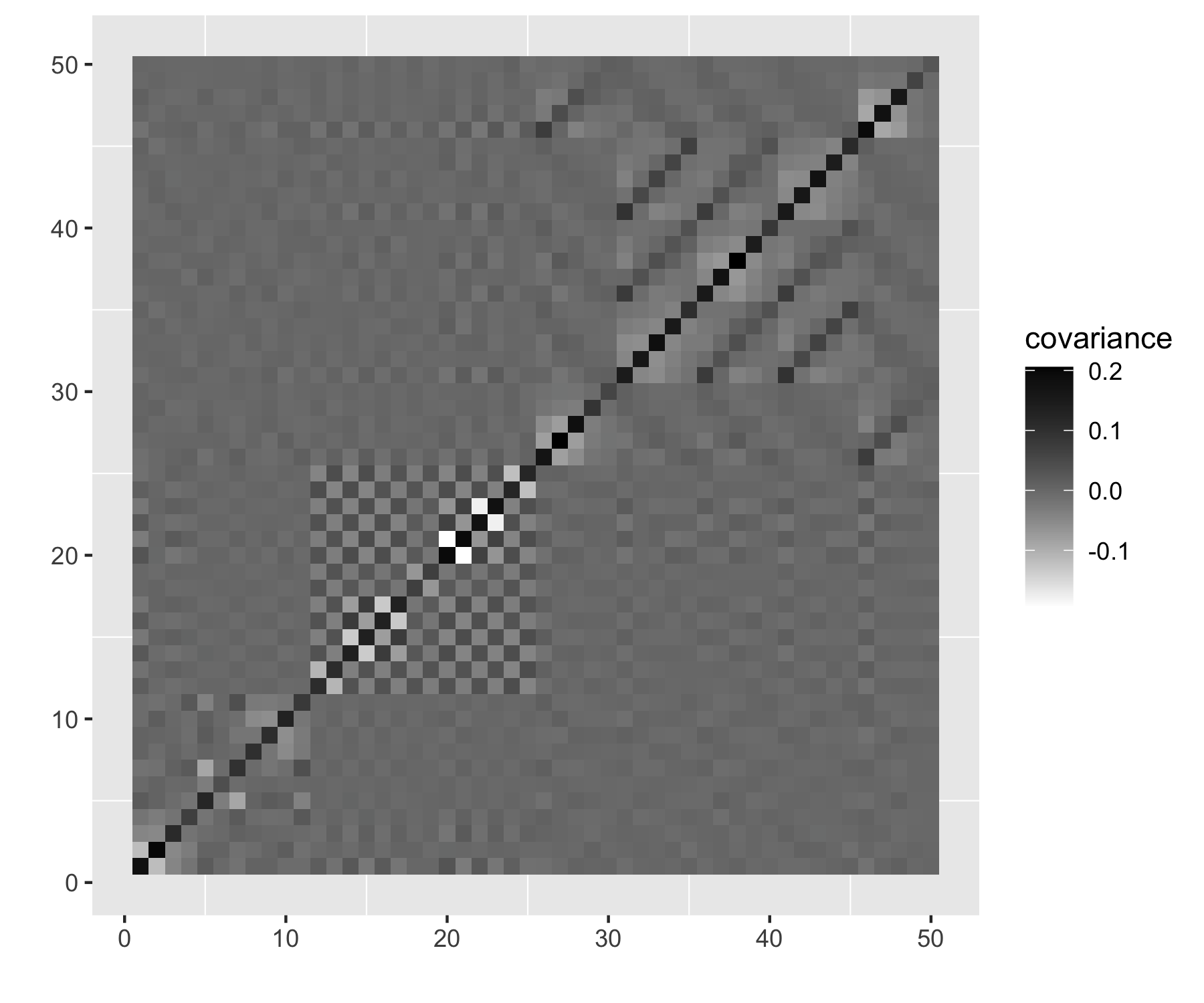}
    \end{minipage}
    \caption{Heatmap of the covariance matrix of the estimated noise matrix $\R-\hat\bPi\hat\bT^\top$. The left panel corresponds to the first $100$ columns of the HapMap3 data and the right panel corresponds to the first $50$ columns of the ANES data.}
    \label{fig:cov_heat}
\end{figure}

To demonstrate the local dependence structure, we provide a covariance matrix heatmap of the first $50$ columns of the estimated error matrix $\hat\E=\R-\hat\bPi\hat\bT^\top$ in the right panel in Figure \ref{fig:cov_heat}. Except for the weak local dependence introduced by the flatting technique (consecutive diagonal blocks of size $3$), we also observe bigger blocks of dependence. It turns out these blocks correspond to different topics of questions in the survey, such as voting participation, global emotion battery, and presidential approval. The answers to the questions within each topic are naturally dependent of each other, giving rise to the bigger blocks in the heatmap.

\subsection{Population Genetics Data}
We consider a genotyping dataset from the HapMap3 project \citep[publicly available at \url{https://www.broadinstitute.org/medical-and-population-genetics/hapmap-3}]{international2010integrating}. A subset of $4$ sub-populations are considered, including the African ancestry in Southwest USA (ASW), Utah residents with Northern and Western European ancestry (CEU), Mexican ancestry in Los Angeles, California (MEX), and Yoruba in Ibadan, Nigeria (YRI). We follow the conventional data preprocessing procedure to remove SNPs with minor allele frequency smaller than 5\% and SNPs with missing data \citep{kranzler2019genome}. The pre-processed dataset consists of $N=467$ individuals and $J=274,128$ SNPs. 

\begin{figure}[bt!]
    \centering
    \begin{minipage}{0.45\textwidth}
        \centering
        \includegraphics[width=\textwidth]{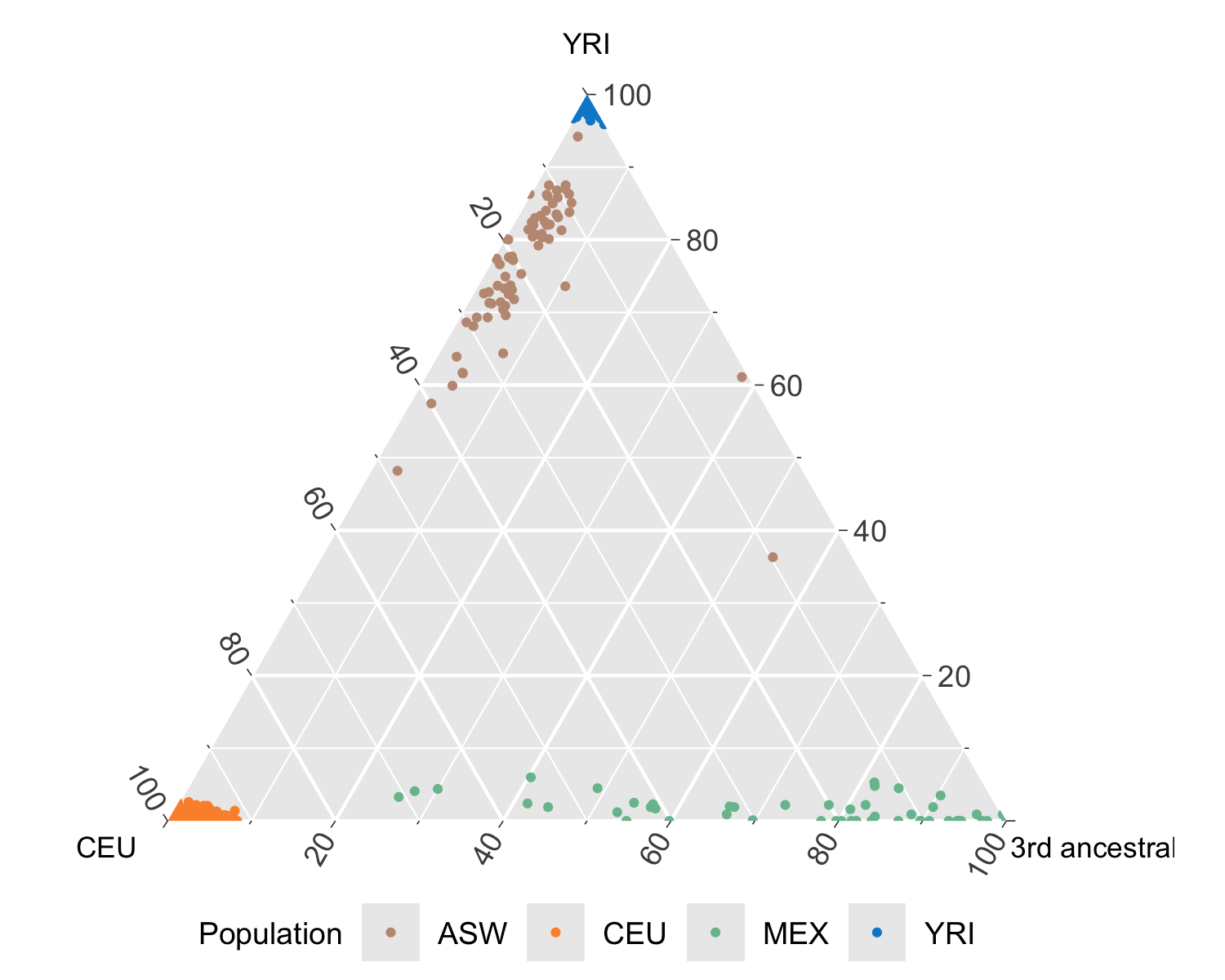}
    \end{minipage} 
    \begin{minipage}{0.45\textwidth}
        \centering
        \includegraphics[width=\textwidth]{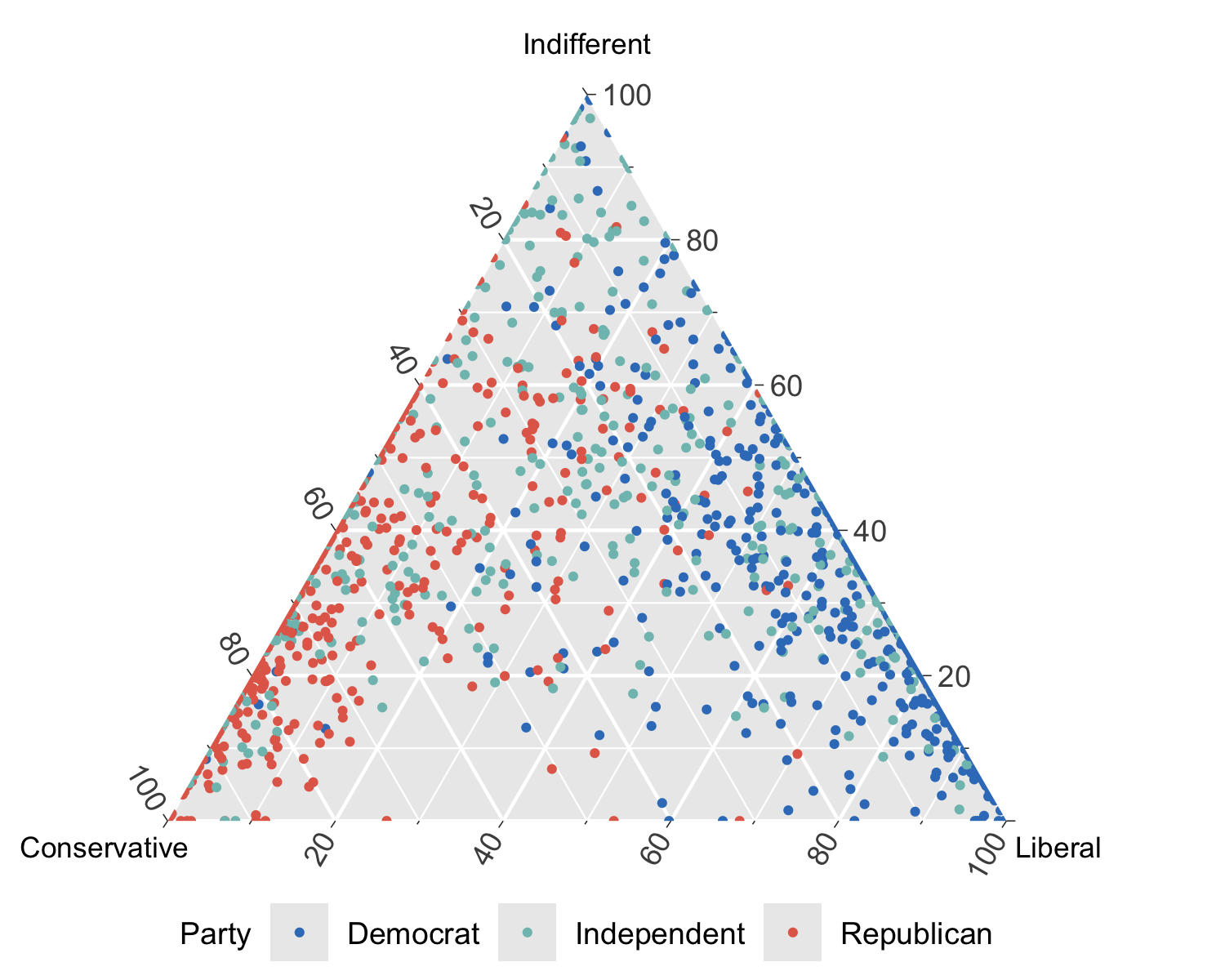}
    \end{minipage}
    \caption{Ternary plots of the estimated membership scores  for the HapMap3 data (left) and ANES data (right). For the Hapmap3 dataset, ASW denotes African ancestry in Southwest USA; CEU denotes Utah residents with Northern and Western European ancestry; MEX denotes Mexican ancestry in Los Angeles, California; YRI denotes Yoruba in Ibadan, Nigeria. For the ANES data, each dot is colored by the subject's self-reported party information.}
    \label{fig:tern}
\end{figure}

The observed genotype $R_{ij}$ for individual $i$ at locus $j$ takes value in $\{0,1,2\}$, which is the count of mutations relative to the reference allele.
The genotype of each SNP is often assumed to follow a binomial distribution \citep{jiang2024tuning}. So, we have $\R^*=\mathbb{E}[\R]=2\bPi^* \bT^{*\top}$ with $\bT^*$ containing the binomial parameters. 
Linkage disequilibrium (LD) is an important and well-studied phenomena for SNP genotype data, which refers to correlations among neighboring alleles \citep{reich2001linkage}. 
Therefore, we consider a binomial model for this dataset, and directly employ our proposed method on $\R/2$ to estimate $\bPi^*$ and $\bT^*$.
The left panel in Figure \ref{fig:tern} is the ternary plot based on the estimated membership scores with our proposed method. Our method is able to reveal that the ASW sample is a mixture of the YRI and CEU samples, and the MEX sample is a mixture between the CEU sample and a third ancestral population. These observations align with the established biological understanding of human population genetics \citep{jakobsson2008genotype}.
% li2008worldwide
Furthermore, the left panel in Figure \ref{fig:cov_heat} confirms the existence of local dependence among the SNPs in the data. As demonstrated in Figure S.1 in the Supplementary Material, the covariance matrix of the estimated error matrix $\hat\E$ is a good estimate of the true error covariance matrix. 

\begin{figure}[bt!]
    \centering
    \includegraphics[width=0.85\textwidth]{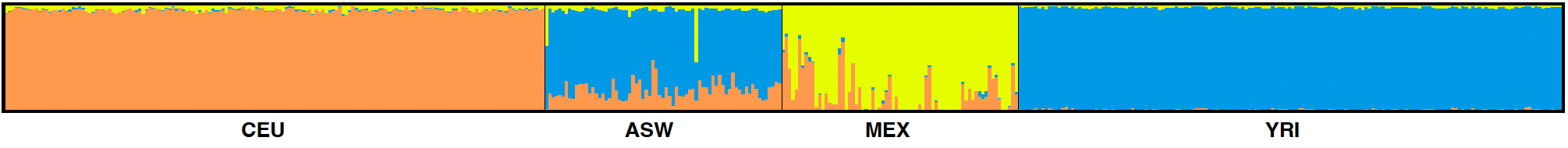}
    \includegraphics[width=0.85\textwidth]{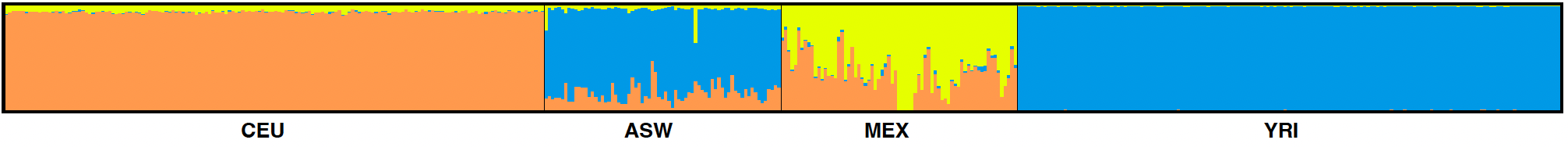}
    \caption{Membership results from STRUCTURE (upper) and the proposed method (lower) with $K=3$ for the HapMap3 dataset. STRUCTURE takes more than $44$ hours to finish while our method takes only $9$ seconds. {Each vertical slice represents an individual, with the lengths of its three color bars proportional to the estimated $\hat\pi_{i,k}$.}}
    \label{fig:distruct}
\end{figure}

We compare the proposed method with a highly influential method called STRUCTURE \citep{pritchard2000inference, falush2003inference}.
STRUCTURE is a method for multilocus genotype data to infer population structure and assign admixed proportions (membership scores) to each individual. Essentially, STRUCTURE falls within the generalized-GoM model, where the admixture proportions are assumed to be sampled from $\text{Dirichlet}(\alpha,\dots, \alpha)$, and the item parameters from $\text{Dirichlet}(\lambda_1,\dots, \lambda_{C_j})$. The hyper-parameter $\alpha$ is assumed to have a uniform hyper-prior $\text{Unif}[0,10]$. STRUCTURE uses an MCMC algorithm to estimate the model parameters and employs Metropolis-Hasting to update $\alpha$ at each iteration. 
Figure \ref{fig:distruct} displays the comparisons of our proposed method with STRUCTURE in terms of the membership score estimates. 
The figures are generated with the \texttt{Distruct} software \citep{rosenberg2004distruct}.
Each vertical slice represents an individual, and each color represents one extreme latent profile.
The lengths of the three color bars for each individual $i$ are proportional to the estimated membership score $\hat\pi_{i,k}$ for $k=1,2,3$.
We can see that both methods capture rather similar admixture  structures in the sample. 
However, STRUCTURE took over $44$ hours and $22$ minutes to run with default settings, yet the proposed method takes only about $9$ seconds. Notably, our method is able to generate similar results with a huge computation advantage. On another note, if we treat the $\{0,1,2\}$ responses as polytomous data and use the flattening technique before estimating parameters, it turns out that it gives very similar estimation results of $\hat\bPi$ to those given by the binomial model reported above.

\vspace{-4mm}
\subsection{Single-cell Sequencing Data}

We finally demonstrate an application to single-cell sequencing data, where the observed features are nonnegative counts.
The dataset contains RNA-seq read counts for single cells of mouse embryo development with $N=259$ cells and $J=22,431$ genes \citep[publicly available at \url{https://github.com/kkdey/CountClust}]{deng2014single}. 
This dataset also provides labels of $10$ pre-implantation development phase categories for each cell.

\begin{figure}[hbt!]
    \centering
    \includegraphics[width=0.85\textwidth]{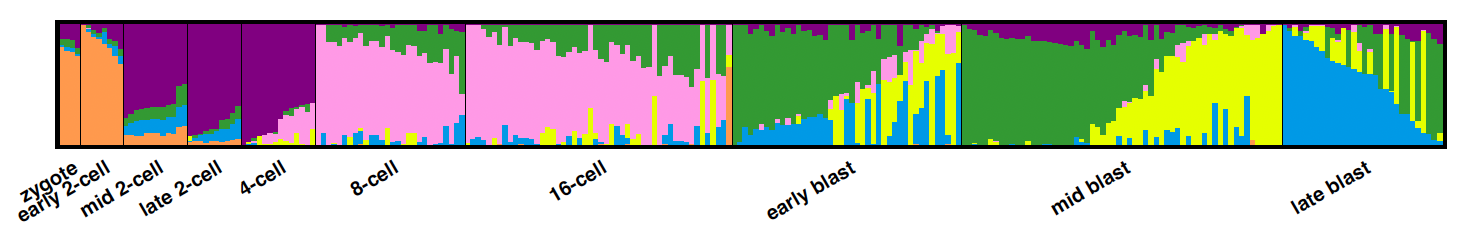}
    \caption{Generalized-GoM estimated memberships for the single cell data with $K=6$. 
    Each color represents an extreme latent profile, and each vertical slice represents a cell. The cells are ordered by their preimplantation development phase, and within each phase sorted by the membership score of the dominating profile in this phase. }
    \label{fig:single_cell_distruct}
\end{figure}

The gene expression counts in our dataset range from $0$ to $693,974$, and we assume a Poisson model. In single-cell sequencing data, the phenomenon of local dependence is commonly referred to as gene co-expression \citep{van2018single}. 
We choose $K=6$ according to the analysis on the same dataset conducted in \cite{dey2017visualizing}. 
Figure \ref{fig:single_cell_distruct} presents our estimated membership scores $\hat\bPi$, with each vertical slice corresponding to a cell. From left (zygote) to right (late blast) are the 10 cell development phases. We can see that cells belonging to the same phase have relatively similar mixed membership allocations, and that the early stage cells are less mixed compared to cells at later stages. Furthermore, we observe a continuous evolution in the membership scores as the cells progress through the developmental stages. These findings align with those reported in \cite{dey2017visualizing}.
Additionally, it takes $0.5$ second to run the proposed method on the count matrix, while the computation time is reported to be $12$ minutes in \cite{dey2017visualizing}, where they use R package \texttt{maptp} \citep{taddy2012estimation} to fit the GoM model. 

\vspace{-10mm}
\section{Discussion}\label{sec:discussion}
This paper proposes a fast spectral method for generalized-GoM models with potential local dependence. For the GoM model for high-dimensional polytomous data, we flatten the three-way quasi-tensor of data into a fat binary matrix and exploit its singular subspace geometry to estimate parameters. We establish fine-grained entrywise estimation error bounds for parameters in a broad family of generalized-GoM models. We also develop a novel $\ell\ti$ singular subspace perturbation theory for arbitrary locally dependent and flexibly distributed noise, which is of independent interest and can be useful for other latent factor models. Our method has a significant computational advantage over popular Bayesian MCMC methods and outputs competitively accurate parameter estimates. Even if one ultimately desires to perform Bayesian posterior inference, we recommend initializing the MCMC algorithm using our spectral method. This sufficiently accurate initialization can help speed up the convergence of the MCMC chain to its stationary distribution.

There are several future directions worth exploring. 
First, Algorithm \ref{algorithm2} requires the input of $K$, which is often unknown in practice. There are various tools to determine the choice of $K$. The simple yet widely-used approach is to inspect the scree plot of the singular values and find the ``elbow point''. Other methods include the gap statistic \citep{tibshirani2001estimating}, parallel analysis \citep{hayton2004factor}, eigen selection \citep{han2023eigen}. It is of special interest to apply the proposed method on an estimated $K$ and still obtain theoretical guarantees, similar to the work in \cite{fan2022simple} and \cite{jin2023optimal}.
Second, missing data can appear in real-world applications especially with big data scenarios. Matrix completion using spectral methods has been studied \citep{cai2021subspace} and could be potentially extended to our setting.
Third, to further increase the modeling power of the generaized-GoM model,
a potential extension is to incorporate the degree-correction parameter similar to the degree-corrected mixed membership model for networks \citep{ke2023special, jin2024mixed, lyu2024degree}. Then one can perform certain normalizations in the spectral domain to correct the degree effects before estimating other parameters. 
Finally, it would be interesting to develop statistical inference results for locally dependent generalized-GoM models, similar to the considerations in \cite{fan2022simple} for network models.

\spacingset{1}

\paragraph{Supplementary Material.}
The Supplementary Material contains all technical proofs of the theoretical results and additional computation details.

\bibliographystyle{apalike}
\bibliography{ref}

\begin{thebibliography}{}

\bibitem[Abbe et~al., 2022]{abbe2022}
Abbe, E., Fan, J., and Wang, K. (2022).
\newblock An {$\ell_p$} theory of {PCA} and spectral clustering.
\newblock {\em The Annals of Statistics}, 50(4):2359--2385.

\bibitem[Agterberg et~al., 2022]{agterberg2022entrywise}
Agterberg, J., Lubberts, Z., and Priebe, C.~E. (2022).
\newblock Entrywise estimation of singular vectors of low-rank matrices with
  heteroskedasticity and dependence.
\newblock {\em IEEE Transactions on Information Theory}, 68(7):4618--4650.

\bibitem[Airoldi et~al., 2015]{airoldi2015handbook}
Airoldi, E.~M., Blei, D.~M., Erosheva, E.~A., and Fienberg, S.~E. (2015).
\newblock {\em Handbook of mixed membership models and their applications}.
\newblock CRC press Boca Raton, FL.

\bibitem[Arroyo et~al., 2021]{arroyo2021inference}
Arroyo, J., Athreya, A., Cape, J., Chen, G., Priebe, C.~E., and Vogelstein,
  J.~T. (2021).
\newblock Inference for multiple heterogeneous networks with a common invariant
  subspace.
\newblock {\em Journal of Machine Learning Research}, 22(142):1--49.

\bibitem[Bhattacharya and Dunson, 2012]{bhattacharya2012simplex}
Bhattacharya, A. and Dunson, D.~B. (2012).
\newblock Simplex factor models for multivariate unordered categorical data.
\newblock {\em Journal of the American Statistical Association},
  107(497):362--377.

\bibitem[Blei et~al., 2003]{blei2003latent}
Blei, D.~M., Ng, A.~Y., and Jordan, M.~I. (2003).
\newblock Latent {D}irichlet allocation.
\newblock {\em Journal of Machine Learning Research}, 3(Jan):993--1022.

\bibitem[Brailovskaya and van Handel, 2024]{brailovskaya2024universality}
Brailovskaya, T. and van Handel, R. (2024).
\newblock Universality and sharp matrix concentration inequalities.
\newblock {\em Geometric and Functional Analysis}, pages 1--105.

\bibitem[Cai et~al., 2021]{cai2021subspace}
Cai, C., Li, G., Chi, Y., Poor, H.~V., and Chen, Y. (2021).
\newblock Subspace estimation from unbalanced and incomplete data matrices:
  $\ell_{2,\infty}$ statistical guarantees.
\newblock {\em Annals of Statistics}, 49(2):944--967.

\bibitem[Cape et~al., 2019]{cape2019two}
Cape, J., Tang, M., and Priebe, C.~E. (2019).
\newblock The two-to-infinity norm and singular subspace geometry with
  applications to high-dimensional statistics.
\newblock {\em The Annals of Statistics}, 47(5):2405--2439.

\bibitem[Chen and Gu, 2024]{chen2024spectral}
Chen, L. and Gu, Y. (2024).
\newblock A spectral method for identifiable grade of membership analysis with
  binary responses.
\newblock {\em Psychometrika}, 89:626–657.

\bibitem[Chen and Thissen, 1997]{chen1997local}
Chen, W.-H. and Thissen, D. (1997).
\newblock Local dependence indexes for item pairs using item response theory.
\newblock {\em Journal of Educational and Behavioral Statistics},
  22(3):265--289.

\bibitem[Chen et~al., 2021]{chen2021spectral}
Chen, Y., Chi, Y., Fan, J., and Ma, C. (2021).
\newblock Spectral methods for data science: A statistical perspective.
\newblock {\em Foundations and Trends{\textregistered} in Machine Learning},
  14(5):566--806.

\bibitem[Chen et~al., 2018]{chen2018robust}
Chen, Y., Li, X., Liu, J., and Ying, Z. (2018).
\newblock Robust measurement via a fused latent and graphical item response
  theory model.
\newblock {\em Psychometrika}, 83:538--562.

\bibitem[Clinton et~al., 2004]{clinton2004statistical}
Clinton, J., Jackman, S., and Rivers, D. (2004).
\newblock The statistical analysis of roll call data.
\newblock {\em American Political Science Review}, 98(2):355--370.

\bibitem[Consortium et~al., 2010]{international2010integrating}
Consortium, I. H.~. et~al. (2010).
\newblock Integrating common and rare genetic variation in diverse human
  populations.
\newblock {\em Nature}, 467(7311):52.

\bibitem[Deng et~al., 2014]{deng2014single}
Deng, Q., Ramsk{\"o}ld, D., Reinius, B., and Sandberg, R. (2014).
\newblock Single-cell {RNA}-seq reveals dynamic, random monoallelic gene
  expression in mammalian cells.
\newblock {\em Science}, 343(6167):193--196.

\bibitem[Dey et~al., 2017]{dey2017visualizing}
Dey, K.~K., Hsiao, C.~J., and Stephens, M. (2017).
\newblock Visualizing the structure of {RNA}-seq expression data using grade of
  membership models.
\newblock {\em PLoS Genetics}, 13(3):e1006599.

\bibitem[Erosheva et~al., 2004]{erosheva2004pnas}
Erosheva, E., Fienberg, S., and Lafferty, J. (2004).
\newblock Mixed-membership models of scientific publications.
\newblock {\em Proceedings of the National Academy of Sciences},
  101(suppl\_1):5220--5227.

\bibitem[Erosheva, 2002]{erosheva2002grade}
Erosheva, E.~A. (2002).
\newblock {\em Grade of membership and latent structure models with application
  to disability survey data}.
\newblock PhD thesis, Carnegie Mellon University.

\bibitem[Erosheva et~al., 2007]{erosheva2007describing}
Erosheva, E.~A., Fienberg, S.~E., and Joutard, C. (2007).
\newblock Describing disability through individual-level mixture models for
  multivariate binary data.
\newblock {\em Annals of Applied Statistics}, 1(2):346.

\bibitem[Falush et~al., 2003]{falush2003inference}
Falush, D., Stephens, M., and Pritchard, J.~K. (2003).
\newblock Inference of population structure using multilocus genotype data:
  linked loci and correlated allele frequencies.
\newblock {\em Genetics}, 164(4):1567--1587.

\bibitem[Fan et~al., 2022]{fan2022simple}
Fan, J., Fan, Y., Han, X., and Lv, J. (2022).
\newblock Simple: Statistical inference on membership profiles in large
  networks.
\newblock {\em Journal of the Royal Statistical Society Series B: Statistical
  Methodology}, 84(2):630--653.

\bibitem[Fan et~al., 2018]{fan2018formula}
Fan, J., Wang, W., and Zhong, Y. (2018).
\newblock An {$\ell_{\infty}$} eigenvector perturbation bound and its
  application to robust covariance estimation.
\newblock {\em Journal of Machine Learning Research}, 18.

\bibitem[Gallagher et~al., 2021]{gallagher2021spectral}
Gallagher, I., Jones, A., and Rubin-Delanchy, P. (2021).
\newblock Spectral embedding for dynamic networks with stability guarantees.
\newblock {\em Advances in Neural Information Processing Systems},
  34:10158--10170.

\bibitem[Gaujoux and Seoighe, 2010]{gaujoux2010flexible}
Gaujoux, R. and Seoighe, C. (2010).
\newblock A flexible {R} package for nonnegative matrix factorization.
\newblock {\em BMC bioinformatics}, 11:1--9.

\bibitem[Gillis and Vavasis, 2013]{Gillis2013}
Gillis, N. and Vavasis, S.~A. (2013).
\newblock Fast and robust recursive algorithmsfor separable nonnegative matrix
  factorization.
\newblock {\em IEEE Transactions on Pattern Analysis and Machine Intelligence},
  36(4):698--714.

\bibitem[Gross and Manrique-Vallier, 2012]{gross2012mixed}
Gross, J.~H. and Manrique-Vallier, D. (2012).
\newblock A mixed-membership approach to the assessment of political ideology
  from survey responses.
\newblock {\em Handbook of mixed membership models and their applications},
  pages 119--140.

\bibitem[Gu et~al., 2023]{gu2023dimension}
Gu, Y., Erosheva, E.~E., Xu, G., and Dunson, D.~B. (2023).
\newblock Dimension-grouped mixed membership models for multivariate
  categorical data.
\newblock {\em Journal of Machine Learning Research}, 24(88):1--49.

\bibitem[Han et~al., 2023]{han2023eigen}
Han, X., Tong, X., and Fan, Y. (2023).
\newblock Eigen selection in spectral clustering: a theory-guided practice.
\newblock {\em Journal of the American Statistical Association},
  118(541):109--121.

\bibitem[Hayton et~al., 2004]{hayton2004factor}
Hayton, J.~C., Allen, D.~G., and Scarpello, V. (2004).
\newblock Factor retention decisions in exploratory factor analysis: A tutorial
  on parallel analysis.
\newblock {\em Organizational Research Methods}, 7(2):191--205.

\bibitem[Jakobsson et~al., 2008]{jakobsson2008genotype}
Jakobsson, M., Scholz, S.~W., Scheet, P., Gibbs, J.~R., VanLiere, J.~M., Fung,
  H.-C., Szpiech, Z.~A., Degnan, J.~H., Wang, K., Guerreiro, R., et~al. (2008).
\newblock Genotype, haplotype and copy-number variation in worldwide human
  populations.
\newblock {\em Nature}, 451(7181):998--1003.

\bibitem[Jiang et~al., 2024]{jiang2024tuning}
Jiang, W., Chen, L., Girgenti, M.~J., and Zhao, H. (2024).
\newblock Tuning parameters for polygenic risk score methods using gwas summary
  statistics from training data.
\newblock {\em Nature Communications}, 15(1):24.

\bibitem[Jin et~al., 2021]{jin2021improvements}
Jin, J., Ke, Z.~T., and Luo, S. (2021).
\newblock Improvements on score, especially for weak signals.
\newblock {\em Sankhya A}, pages 1--36.

\bibitem[Jin et~al., 2024]{jin2024mixed}
Jin, J., Ke, Z.~T., and Luo, S. (2024).
\newblock Mixed membership estimation for social networks.
\newblock {\em Journal of Econometrics}, 239(2):105369.

\bibitem[Jin et~al., 2023]{jin2023optimal}
Jin, J., Ke, Z.~T., Luo, S., and Wang, M. (2023).
\newblock Optimal estimation of the number of network communities.
\newblock {\em Journal of the American Statistical Association},
  118(543):2101--2116.

\bibitem[Jones and Rubin-Delanchy, 2020]{jones2020multilayer}
Jones, A. and Rubin-Delanchy, P. (2020).
\newblock The multilayer random dot product graph.
\newblock {\em arXiv preprint arXiv:2007.10455}.

\bibitem[Kang and Gu, 2024]{kang2024blockwise}
Kang, K. and Gu, Y. (2024).
\newblock A blockwise mixed membership model for multivariate longitudinal
  data: Discovering clinical heterogeneity and identifying parkinson's disease
  subtypes.
\newblock {\em arXiv preprint arXiv:2410.01235}.

\bibitem[Ke and Jin, 2023]{ke2023special}
Ke, Z.~T. and Jin, J. (2023).
\newblock Special invited paper: The score normalization, especially for
  heterogeneous network and text data.
\newblock {\em Stat}, 12(1):e545.

\bibitem[Ke and Wang, 2024]{ke2024using}
Ke, Z.~T. and Wang, M. (2024).
\newblock Using {SVD} for topic modeling.
\newblock {\em Journal of the American Statistical Association},
  119(545):434--449.

\bibitem[Koltchinskii et~al., 2011]{koltchinskii2011nuclear}
Koltchinskii, V., Lounici, K., and Tsybakov, A.~B. (2011).
\newblock Nuclear-norm penalization and optimal rates for noisy low-rank matrix
  completion.
\newblock {\em The Annals of Statistics}, 39(5):2302.

\bibitem[Kranzler et~al., 2019]{kranzler2019genome}
Kranzler, H.~R., Zhou, H., Kember, R.~L., Vickers~Smith, R., Justice, A.~C.,
  Damrauer, S., Tsao, P.~S., Klarin, D., Baras, A., Reid, J., et~al. (2019).
\newblock Genome-wide association study of alcohol consumption and use disorder
  in 274,424 individuals from multiple populations.
\newblock {\em Nature Communications}, 10(1):1499.

\bibitem[Lee and Seung, 2000]{lee2000algorithms}
Lee, D. and Seung, H.~S. (2000).
\newblock Algorithms for non-negative matrix factorization.
\newblock {\em Advances in Neural Information Processing Systems}, 13.

\bibitem[Lee and Seung, 1999]{lee1999learning}
Lee, D.~D. and Seung, H.~S. (1999).
\newblock Learning the parts of objects by non-negative matrix factorization.
\newblock {\em Nature}, 401(6755):788--791.

\bibitem[Lei, 2019]{lei2019unified}
Lei, L. (2019).
\newblock Unified {$\ell_{2,\infty}$} eigenspace perturbation theory for
  symmetric random matrices.
\newblock {\em arXiv preprint arXiv:1909.04798}.

\bibitem[Lyu et~al., 2024]{lyu2024degree}
Lyu, Z., Chen, L., and Gu, Y. (2024).
\newblock Degree-heterogeneous latent class analysis for high-dimensional
  discrete data.
\newblock {\em arXiv preprint arXiv:2402.18745}.

\bibitem[Manton et~al., 1994]{manton1994statistical}
Manton, K.~G., Woodbury, M.~A., and Tolley, H.~D. (1994).
\newblock {\em Statistical Applications Using Fuzzy Sets}.
\newblock Wiley-Interscience, New York.

\bibitem[Mao et~al., 2021]{mao2021mm}
Mao, X., Sarkar, P., and Chakrabarti, D. (2021).
\newblock Estimating mixed memberships with sharp eigenvector deviations.
\newblock {\em Journal of the American Statistical Association},
  116(536):1928--1940.

\bibitem[Modell et~al., 2024]{modell2024intensity}
Modell, A., Gallagher, I., Ceccherini, E., Whiteley, N., and Rubin-Delanchy, P.
  (2024).
\newblock Intensity profile projection: A framework for continuous-time
  representation learning for dynamic networks.
\newblock {\em Advances in Neural Information Processing Systems}, 36.

\bibitem[Pritchard et~al., 2000]{pritchard2000inference}
Pritchard, J.~K., Stephens, M., and Donnelly, P. (2000).
\newblock Inference of population structure using multilocus genotype data.
\newblock {\em Genetics}, 155(2):945--959.

\bibitem[Reich et~al., 2001]{reich2001linkage}
Reich, D.~E., Cargill, M., Bolk, S., Ireland, J., Sabeti, P.~C., Richter,
  D.~J., Lavery, T., Kouyoumjian, R., Farhadian, S.~F., Ward, R., et~al.
  (2001).
\newblock Linkage disequilibrium in the human genome.
\newblock {\em Nature}, 411(6834):199--204.

\bibitem[Robitzsch and Robitzsch, 2017]{robitzsch2017package}
Robitzsch, A. and Robitzsch, M.~A. (2017).
\newblock Package ‘sirt’.
\newblock {\em Computer software]. https://www. maths. bris. ac.
  uk/R/web/packages/sirt/sirt. pdf}.

\bibitem[Rosenberg, 2004]{rosenberg2004distruct}
Rosenberg, N.~A. (2004).
\newblock Distruct: a program for the graphical display of population
  structure.
\newblock {\em Molecular ecology notes}, 4(1):137--138.

\bibitem[Taddy, 2012]{taddy2012estimation}
Taddy, M. (2012).
\newblock On estimation and selection for topic models.
\newblock In {\em Artificial Intelligence and Statistics}, pages 1184--1193.
  PMLR.

\bibitem[Tibshirani et~al., 2001]{tibshirani2001estimating}
Tibshirani, R., Walther, G., and Hastie, T. (2001).
\newblock Estimating the number of clusters in a data set via the gap
  statistic.
\newblock {\em Journal of the Royal Statistical Society: Series B (Statistical
  Methodology)}, 63(2):411--423.

\bibitem[Tropp, 2012]{tropp2012user}
Tropp, J.~A. (2012).
\newblock User-friendly tail bounds for sums of random matrices.
\newblock {\em Foundations of computational mathematics}, 12:389--434.

\bibitem[Tropp, 2015]{tropp2015introduction}
Tropp, J.~A. (2015).
\newblock An introduction to matrix concentration inequalities.

\bibitem[Uffelmann et~al., 2021]{uffelmann2021genome}
Uffelmann, E., Huang, Q.~Q., Munung, N.~S., De~Vries, J., Okada, Y., Martin,
  A.~R., Martin, H.~C., Lappalainen, T., and Posthuma, D. (2021).
\newblock Genome-wide association studies.
\newblock {\em Nature Reviews Methods Primers}, 1(1):59.

\bibitem[Van Der~Wijst et~al., 2018]{van2018single}
Van Der~Wijst, M.~G., Brugge, H., De~Vries, D.~H., Deelen, P., Swertz, M.~A.,
  Study, L.~C., Consortium, B., and Franke, L. (2018).
\newblock Single-cell {RNA} sequencing identifies celltype-specific cis-{eQTLs}
  and co-expression {QTLs}.
\newblock {\em Nature Genetics}, 50(4):493--497.

\bibitem[Wainwright, 2019]{wainwright2019high}
Wainwright, M.~J. (2019).
\newblock {\em High-dimensional statistics: A non-asymptotic viewpoint},
  volume~48.
\newblock Cambridge university press.

\bibitem[Wang and Erosheva, 2015]{wang2015fitting}
Wang, Y.~S. and Erosheva, E.~A. (2015).
\newblock Fitting mixed membership models using mixedmem.

\bibitem[Woodbury et~al., 1978]{woodbury1978gom}
Woodbury, M.~A., Clive, J., and Garson~Jr, A. (1978).
\newblock Mathematical typology: a grade of membership technique for obtaining
  disease definition.
\newblock {\em Computers and Biomedical Research}, 11(3):277--298.

\bibitem[Wu et~al., 2023]{wu2023sparse}
Wu, R., Zhang, L., and Tony~Cai, T. (2023).
\newblock Sparse topic modeling: Computational efficiency, near-optimal
  algorithms, and statistical inference.
\newblock {\em Journal of the American Statistical Association},
  118(543):1849--1861.

\bibitem[Xia, 2021]{xia2021normal}
Xia, D. (2021).
\newblock Normal approximation and confidence region of singular subspaces.
\newblock {\em Electronic Journal of Statistics}, 15(2):3798--3851.

\bibitem[Yan et~al., 2021]{yan2021inference}
Yan, Y., Chen, Y., and Fan, J. (2021).
\newblock Inference for heteroskedastic {PCA} with missing data.
\newblock {\em arXiv preprint arXiv:2107.12365}.

\bibitem[Yan et~al., 2024]{yan2024inference}
Yan, Y., Chen, Y., and Fan, J. (2024).
\newblock Inference for heteroskedastic {PCA} with missing data.
\newblock {\em The Annals of Statistics}, 52(2):729--756.

\bibitem[Zhang et~al., 2020]{zhang2020detecting}
Zhang, Y., Levina, E., and Zhu, J. (2020).
\newblock Detecting overlapping communities in networks using spectral methods.
\newblock {\em SIAM Journal on Mathematics of Data Science}, 2(2):265--283.

\bibitem[Zhao et~al., 2018]{zhao2018dirichlet}
Zhao, S., Engelhardt, B.~E., Mukherjee, S., and Dunson, D.~B. (2018).
\newblock Fast moment estimation for generalized latent {D}irichlet models.
\newblock {\em Journal of the American Statistical Association},
  113(524):1528--1540.

\end{thebibliography}

\clearpage

\renewcommand{\thesection}{S.\arabic{section}}  
\renewcommand{\thetable}{S.\arabic{table}}  
\renewcommand{\thefigure}{S.\arabic{figure}}
\renewcommand{\theequation}{S.\arabic{equation}}
\renewcommand{\theassumption}{S.\arabic{assumption}}
\renewcommand{\thetheorem}{S.\arabic{theorem}}
\renewcommand{\thelemma}{S.\arabic{lemma}}

% set counters to zero
\setcounter{section}{0}
\setcounter{equation}{0}
\setcounter{figure}{0}
\setcounter{assumption}{0}

{\centering\section*{Supplementary Material}}
\addcontentsline{toc}{section}{Supplementary Material}

\spacingset{1.7}
This Supplementary Material is organized as follows. Section \ref{sec-perturbation} includes the theoretical results and proofs for row-wise singular subspace perturbation bounds for asymmetric matrices with local dependence (Theorem~\ref{theorem-bound (general r.v.)}). 
Section \ref{sec-lemma} includes the proof for the parameter estimation error bound (Theorem~\ref{theorem: estimation error}).
Section \ref{sec-computation} contains additional computational details.

\section{Row-wise Singular Subspace Perturbation Bounds for Asymmetric Matrices with Local Dependence}\label{sec-perturbation}

\subsection{Notations and Assumptions}
Recall that we observe the data matrix 
$$\R = \R^* + \E \in \bb R^{N\times J},$$ 
where the deterministic part $\R^*$ has rank $K$ and the noise matrix $\E$ has mean zero.
The rows of the error matrix $\mb E$ are independent. 
The following Assumption \ref{assumption:s_block} (Assumption~\ref{assumption:block} in the main paper) allows local dependence structures on the noise matrix and enables us to treat the error matrix entries as bounded variables with high probability. 
We will verify that this assumption holds for sub-exponential noises in Section \ref{sec:truncation}.

\begin{assumption}[Block Dependent Noise under Flexible Distributions]
\label{assumption:s_block}
Matrix $\E$ satisfies:
\begin{itemize}
\item[(a)] 
There exists a partition $S_1,S_2,\ldots,S_{L}$ of $[J]$ 
with $|S_l| \leq M$ for each $l\in[L]$, such that the vectors $\{\E_{i,S_l}\}_{l=1}^{L}$ are mutually independent %across different pairs of $(i, l)$ 
for different $i\in[N]$ and $l\in [L]$. 
\item[(b)] Either $|E_{i,j}| \leq B$ for all $i\in[N]$, $ j\in[J]$, or there exists a random matrix $\E'=(E_{i,j}') \in\mathbb{R}^{N\times J}$ obeying the same dependence structure in (a), such that for any $i\in[N], j\in[J]$, it holds that $\norm{E'_{i,j}}_\infty\leq B$, $\mathbb E[E'_{i,j}] = 0$, $\norm{ \mathrm{Cov}(\E'_{i,:})} \lesssim \norm{\mathrm{Cov}(\E_{i,:})}$, and $\Pr(E_{i,j} = E'_{i,j}) \ge 1 - O(d^{-22})$.
\end{itemize}
\end{assumption}

\begin{assumption}[Pure Subject]
\label{assumption: s_pure}
    $\bo\Pi^*$ satisfies that every extreme latent profile has at least one pure subject. In addition, rank$(\bT^*)=K$. 
\end{assumption}

\begin{table}[b!]
\centering
\begin{tabular}{ |c|c||c|c| } 
\hline
Notation & Definition & Notation & Definition \\[3mm]
\hline
$\mb \Sigma_{i,S_l}$ & $\cov(\E_{i,S_l})$ & $d$ & $N\vee J$ \\ [3mm]
$\mb \Sigma_i$ & $\cov(\E_{i,:})$ & $r$ & $\dfrac{J}{N}$ \\[3mm] 
$\sigma^2$ & $\max_{i\in[N],j\in[J]}\var(E_{i,j})$ &  $ [B]$ & \{1,2, \dots, B\} \\[3mm]
$\tilde{\sigma}^2$ & $\max_{i\in[N],l\in[L]}\norm{\mb \Sigma_{i,S_l}}$ & $\kappa^*$ & $\dfrac{\sigma_1^*}{\sigma_K^*}$ \\[3mm]
$\mu_1$ & $\dfrac{\norm{\U^*}^2\ti N}{K}$ & $\mu_2$ & $\dfrac{\norm{\V^*}^2\ti J}{K}$  \\
[3mm]
\hline
\end{tabular}
\caption{Notations. 
}
\label{tab:notations}
\end{table}

Consider the SVD of $\R^*$: $\R^*=\U^*\Lam^*\V^{*\top}$ where $\bo\Lam^*=\text{diag}(\sigma_1^*,\dots, \sigma_K^*)$ with $\sigma_1^*\ge\dots \ge \sigma_K^*> 0$ collects the $K$ singular values of $\R^*$. Matrices $\U^*_{N\times K}$, $\V^*_{J\times K}$ respectively consist of the left and right singular vectors and satisfy $\U^{*\top}\U^*=\V^{*\top}\V^*=\mathbf{I}_K$. Similarly, we consider the top-$K$ SVD for the data matrix $\R$: $\R\approx\U\Lam\V^\top$, where $\Lam=\text{diag}(\sigma_1,\dots, \sigma_K)$ with $\sigma_1\ge \dots, \sigma_K\ge0$, $\U\in\mathbb{R}^{N\times K}$, $\V\in\mathbb{R}^{J\times K}$, and $\U^{\top}\U=\V^{\top}\V=\mathbf{I}_K$. Let $(\U^{\perp}, \Lam^{\perp}, \V^{\perp})$ be the SVD of $\R-\U\Lam\V^{\top}$.
We introduce the canonical projection operator $\mc P_{S}:\bb R^{N\times J}\rightarrow \bb R^{N\times J}$, $P_S(\mb E) \mapsto \left(E_{i,j}\mb 1_{\{(i,j)\in S\}}\right)$ where $S\subset [N]\times [J]$. We also denote a tuple returned by the top-$k$ singular value decomposition of a matrix $\mb A \in \bb R^{n_1\times n_2}$ by $\mathsf{svd}(\mb A;k) \coloneqq (\mb U_{\mb A}, \mb \Lambda_{\mb A}, \mb U_{\mb A})$, where $\mb U_{\mb A}\in \bb R^{n_1\times k}$ and $\mb V_{\mb A} \in \bb R^{n_2\times k}$ respectively collect the top-$k$ left and right singular vectors, and $\mb \Lambda_{\mb A}$ is a diagonal matrix with its diagonal entries being the top-$k$ singular values in non-increasing order. 
We define additional notations in Table \ref{tab:notations}, which will be used frequently.

\paragraph{Elementary Facts.} 
One can easily verify that $\sigma^2 \leq \tilde \sigma^2 \leq M \sigma^2$.

\begin{assumption}\label{assumptions}
We assume that $K \ll \min \{N, L\}, ML \asymp J$, and
\begin{align}
    & \sigma_{K}^* \ge \sigma_K(\mathbf \Pi^*) \sigma_K(\mathbf \Theta^*) \gtrsim \kappa^* \sigma \big(M\sqrt{N} + \sqrt{J}), \label{eq: assumption eq3}\\
    & MB(\log d)^2 \max\left\{ \sqrt{\frac{\mu_1 K}{N}},\sqrt{\frac{\mu_2 K}{J}} \right\} \lesssim  \sigma, \label{eq: assumption eq4} 
\end{align}
\end{assumption}

When $M=1$, our setting reduces to the independent noise case.
With Assumption \ref{assumptions}, we assume a low-rank signal matrix and balanced block size in the dependence structure.
Comparing the first line \eqref{eq: assumption eq3} to Theorem 9 in \cite{yan2021inference}, even in the independent noise case, our assumption is weaker than Theorem 9 by removing the log factor in the lower bound of $\sigma^*_{K}$.
The second line \eqref{eq: assumption eq4}  is about the assumption on block size $B$. 
Compared to (6.5) in Assumption 3 in \cite{yan2021inference}, this is slightly stronger in the power of the $\log d$ term, changing from $1/2$ to $2$.

\subsection{Tools for the Proof of Theorem~\ref{theorem-bound (general r.v.)}}\label{sec-parameter-bound}
Note that we treat the entries of $\E$ as bounded variables hereinafter, given that all subsequent high-probability controls have an exceptional probability not greater than $O(d^{-20})$. According to Assumption~\ref{assumption:s_block}, replacing $\E$ with its truncated version does not alter the concentration inequalities.

Intuitively, we can establish concentration inequalities for $\E$ if the local dependence block is adequately small. 
For a vector $\bo a$ of length $C$ and a subset $S\subseteq [C]$, denote $\bo a_{S} = (a_c)_{c\in S}$. 
For example, to use the matrix Bernstein inequality to give an upper bound for $\E_{1,:}\A$ with $\A \in \bb R^{J \times q}$, we can write $\E_{1,:}\A$ as $\sum_{l=1}^L{\tilde{\E}_l}\t \A$ where $\tilde{\E}_l$ is a $J$-dimensional vector that has the $l$th block equal to $\E_{1,S_l}$ and other blocks equal to zero.
Then
\eq{
\norm{\E_{1,:}\A}_2\lesssim \sqrt{(\log J) V} + S\log{J},
}
where 
$$S = \max_{l\in[L]}\norm{\tilde\E_l\A}_2\leq MB \norm{\A}\ti $$ 
and 
$$
V = \max\left\{\norm{\sum_j\bb E\left[{\tilde{\E}_j}\t\A \A\t {\tilde{\E}_j}\right]}, \norm{\sum_j\bb E\left[\A\t {\tilde{\E}_j}{\tilde{\E}_j}\t\A \right]}\right\}\leq \norm{\mb \Sigma} \norm{\A}_F^2\leq J\norm{\mb \Sigma}\norm{\A}\ti^2.
$$

\begin{remark}
The perturbation theories in \cite{abbe2022} also allow dependency on the columns of $\E$, assuming that $\E_{i,:}$ are sub-Gaussian vectors. However, their proofs depend on Hoeffding's inequality instead of Bernstein's inequality, with the help of sub-Gaussianity. 
\end{remark}

%\subsubsection{Main Results}

%\subsection{Auxiliary Lemmas}

\begin{lemma} \label{Lemma-matrix}
    For any matrix $\A\in\mathbb{R}^{m\times n}$ with rank $K$ and $\B\in\mathbb{R}^{n\times k}, \C\in\mathbb{R}^{k\times l}$, we have
    \begin{align}
        & \|\A\|/\sqrt{m} \leq \|\A\|_{2,\infty} \leq \|\A\|\le \|\A\|_F\le \sqrt{m}\|\A\|_{2,\infty}, \label{eq-matrix1} \\ 
        & \|\A\B\|_{2,\infty} \leq \|\A\|_{2,\infty}\|\B\|, \label{eq-matrix2} \\
        & \|\A\B\C\|_{\infty} \le \|\A\|_{2, \infty} \|\B\|\|\C\|_{\infty, 2}. \label{eq-matrix3}
    \end{align}
\end{lemma}

\paragraph{Proof of Lemma~\ref{Lemma-matrix}.}
Refer to \cite{cape2019two} for the first three inequalities in \eqref{eq-matrix1}. The last inequality of \eqref{eq-matrix1} is because $\|\A\|_F = \sqrt{\sum_{i=1}^m\|\A_{i,:}\|^2}\le \sqrt{m\|\A\|_{2,\infty}^2}=\sqrt{m}\|\A\|_{2,\infty}$.
Inequality \eqref{eq-matrix2} holds since
\begin{align*}
    \|\A\B\|_{2,\infty} = \max_i \|\A_{i,:}\B\|\le \max_i \|\A_{i,:}\|\|\B\| = \|\A\|_{2,\infty}\|\B\|.
\end{align*}
And inequality \eqref{eq-matrix3} yields from
\begin{equation*}
    \begin{split}
        |(\A\B\C)_{ij}| = & |\A_{i,:}\B\C_{:,j}| \le \|\A_{i,:}\|\|\B\|\|\C_{:,j}\|\\
        \le &\|\A\|_{2,\infty}\|\|\B\|\|\C\|_{\infty,2}.
    \end{split}
\end{equation*}
\qed

\begin{lemma}\label{lemma:matrix_trace}
    For any matrices $\A\in\mathbb{R}^{d_1\times d_2}, \B\in\mathbb{R}^{d_1\times d_3}$, we have
    \begin{align*}
        \tr(\A\A^\top\B) \le \|\A\|_F^2 \|\B\|.
    \end{align*}
\end{lemma}
\paragraph{Proof of Lemma~\ref{lemma:matrix_trace}.}
It follows directly from 
$$
\tr(\A\A^\top\B) \le \sum_{j\in[d_1]}\sigma_j(\A\A^\top\B) \le \sum_{j\in[d_1]} \sigma_j(\A\A^\top)\|\B\| = \|\A\|^2_F\|\B\|.
$$
\qed
    
Lemmas \ref{lemma: matrix Bernstein} and \ref{lemma: noise matrix concentrations using the universality} below give matrix concentration results for the block dependent noise matrix. %\yg{?}

\begin{lemma}[Bernstein inequalities for block dependent noise matrix]
\label{lemma: matrix Bernstein}
Suppose Assumption~\ref{assumption:s_block} holds. 
Given any deterministic matrix $\A \in \bb R^{J\times q}, \B\in\bb R^{N\times p}$ and a positive constant $\alpha>0$, with probability at least $1- O(d^{-\alpha})$ one has 
\begin{align}
    & \norm{\mb E_{i,:} \mb A} \lesssim \tilde\sigma\sqrt{\alpha \log d }\norm{\mb A}_F  + \alpha M B \log d \norm{\mb A}\ti , \label{eq:bernstein for E_{i,:}A} \\
    & \norm{\E_{:,j}\t \B} \lesssim \sigma\sqrt{\alpha \log d}\norm{\B}_F  + \alpha B \log d  \norm{\B}\ti, \label{eq:bernstein for E_{:,j}B} \\
    & \norm{\mb E \mb A} \lesssim  \tilde \sigma\sqrt{\alpha \log d} \big( \norm{\mb A}_F + \sqrt{N} \norm{\mb A}\big) + \alpha M B \log d \norm{\mb A}\ti , \\
    & \|\E\t \B\| \lesssim \sigma \sqrt{\alpha\log d }(\sqrt{M}\|\B\|_F + \sqrt{J}\|\B\|) + \alpha \sqrt{M} B  \log d\norm{\B}\ti .
\end{align}
\end{lemma}

\paragraph{Proof of Lemma~\ref{lemma: matrix Bernstein}.}
The proof is an application of the matrix Bernstein inequality (for example, Theorem 6.1.1 in \cite{tropp2015introduction}; Corollary 3.3 in \cite{chen2021spectral}). 
We write matrices as a summation of a sequence of independent random matrices: $\E_{i,:}=\sum_{l \in[L]} \mc P_{:,S_l} (\E_{i,:})$, $\E_{:,j} = \sum_{i\in [N]} \mc P_{i,:}(\E_{:,j})$, $\E_{:, S_j}=\sum_{i \in[N]} \mc P_{i, :} (\E_{:,S_j})$, $\E = \sum_{i\in[N]} \sum_{l\in[L]} \mc P_{i,S_l}(\E)$. 
Note that for $\forall i\in[N], l \in [L], j\in[J]$, we can upper bound each independent block matrix:
$\|\E_{i, S_l}\A_{S_l,:}\| = \|\sum_{j\in S_l}\E_{i,j}\A_{j,:}\| \leq M B \|\A\|\ti $, $\|\E_{i,j}\B_{i,:}\| \le B\|\B\|_{2,\infty}$, and $\|\E_{i,S_l}\t \B_{i,:}\|\le \|\E_{i,S_l}\|\|\B\|\ti \le \sqrt{M}B\|\B\|\ti$.

For the first inequality we consider $\E_{i,:}\A$, 
\begin{align*}
    \|\bb E\left[ \A^\top \E_{i,:}^\top \E_{i,:} \A \right]\| = \|\A^\top \bb E[\E_{i,:}^\top \E_{i,:}] \A\| & \le \|\bb E[\E_{i,:}^\top\E_{i,:}]\|\|\A\|^2 \le \max_{l\in[L]} \|\mb\Sigma_{i,S_l}\|\|\A\|^2 \le \tilde\sigma^2 \|\A\|^2, \\
    \bb E[\E_{i,:} \A \A\t \E_{i,:} \t] = \tr\left(\A\A^\top\bb E[\E_{i,:}^\top\E_{i,:}]\right) & \le \|\A\|_F^2\|\bS_i\| \le \|\A\|_F^2 \max_{l\in[L]}\|\bS_{i,S_l}\|^2 \le \tilde\sigma^2\|\A\|_F^2,
\end{align*}
where the second line uses Lemma~\ref{lemma:matrix_trace}. 

For the second inequality, $\E_{:,j}\t\B$ satisfies
\begin{align*}
    \|\bb E[\B\t\E_{:,j} \E_{:,j}\t \B\|] \le \|\bb E[\E_{:,j}\E_{:,j}\t]\| \|\B\|^2 \le \sigma^2 \|\B\|^2, \\
    \bb E[\E_{:,j}\t\B \B\t\E_{:,j}] = \tr\left(\B\B\t \bb E[\E_{:,j}\E_{:,j}\t] \right) \le \sigma^2\|\B\|_F^2,
\end{align*}
where the second line uses Lemma~\ref{lemma:matrix_trace}.

For the third inequality, 
\begin{align*}
    \|\bb E\left[ \A^\top \E^\top \E \A \right]\| \le \sum_{i\in[N]}\|\A^\top\bb E[\E_{i,:}^\top\E_{i,:}]\A \| \le N\tilde\sigma^2 \|\A\|^2, \\
    \|\bb E[\E \A \A\t \E\t]\| = \|\diag(\bb E[\E_{i,:}\A\A^\top\E_{i,:}^\top])\| \le \tilde\sigma^2\|\A\|_F^2.
\end{align*}

For the last inequality,
\begin{align*}
    \norm{\bb E[\E\t\B\B\t\E]} & = \max_{l\in[L]}\|\bb E[\E_{:,S_l}\t\B\B\t\E_{:,S_l}]\| \le M \max_{l\in[L]} \|\mathbb{E} [\E_{:,S_l}^\top \B\B^\top \E_{:,S_l}]\|_{\infty} \\
    & \le M \max_{j\in[J]} \tr(\B\B\t \mathbb{E}[\E_{:,j}^\top\E_{:,j}]) \le M\|\B\|_F^2 \|\mathbb{E}[\E_{:,j}^\top\E_{:,j}]\| \le \sigma^2 M \|\B\|_F^2, \\
    \norm{\bb E[\B\t\E\E\t\B]} & = \|\sum_{i\in[N]} \B_{i,:}\t \bb E[\E_{i,:}\E_{i,:}\t]\B_{i,:}\| \le \sigma^2 J \|\sum_{i\in[N]}\B_{i,:}^\top \B_{i,:}\| \le \sigma^2 J \|\B\|^2.
\end{align*}
\qed

\begin{proposition}[Corollary 2.17 in \cite{brailovskaya2024universality}]
\label{proposition: thm 2.15 in universality paper}
Let $\mb Y = \sum_{i=1}^n \mb Z_i$, where $\mb Z_1, \cdots, \mb Z_n$ are independent (possibly non-self-disjoint) $d\times d$ random matrices with $\bb E[\mb Z_i] = 0$. Then 
\begin{align}
    \bb P\Big[\norm{\mb Y} \geq & \norm{\mb E[\mb Y^* \mb Y]}^{\frac{1}{2}} + \norm{\mb E[\mb Y \mb Y^*]}^{\frac{1}{2}} + \\
    & C\big\{ v(\mb Y)^{\frac{1}{2}}\sigma(\mb Y)^{\frac{1}{2}} (\log d)^{\frac{3}{4}} + \sigma_*(\mb Y) t^{\frac{1}{2}} + R(\mb Y)^{\frac{1}{3}}\sigma(\mb Y)^{\frac{2}{3}}t^{\frac{2}{3}} + R(Y) t   \big\} \Big] \leq 4de^{-t}
\end{align}
for a universal constant $C$ and all $t>0$, where 
\begin{align}
     & \sigma(\mb Y) = \max\big\{\norm{\mb E[\mb Y^* \mb Y]}^{\frac{1}{2}},\norm{\mb E[\mb Y \mb Y^*]}^{\frac{1}{2}} \big\},\\ 
     & v(\mb Y) = \norm{\mathrm{Cov}(\mb Y)}^{\frac{1}{2}},\\ 
     & \sigma_*(\mb Y) = \sup_{\norm{\mb v} = \norm{\mb w} = 1} \left \{\bb E\big[\vert \ip{\mb v, \mb Y  \mb w}\vert^2\big]\right\}^{\frac{1}{2}},\\ 
     & R(\mb Y) = \norm{\max_{1\leq i \leq n}\norm{\mb Z_i}}_\infty,
\end{align}
with $\mathrm{Cov}(\Y)$ denotes the $d^2\times d^2$ covariance matrix of the entries of $\Y$.
\end{proposition}
 
\begin{remark}
    Note that the definitions of $\sigma(\mb Y), v(\mb Y)$, and $\sigma_*(\mb Y)$ do not depend on the decomposition of $\mb Y$. The results in Proposition~\ref{proposition: thm 2.15 in universality paper} also hold for rectangular matrices, as we can consider a square matrix by padding zeros to a rectangular matrix.
\end{remark}

\begin{lemma}[Concentration inequalities for locally dependent noise matrix]\label{lemma: noise matrix concentrations using the universality}
Suppose Assumption~\ref{assumption:s_block} holds. Then we have 
\begin{align}
    & \norm{\mb E} \lesssim \sigma \sqrt{J} + \tilde \sigma \sqrt{N} +  M^{\frac{1}{6}} B^{\frac{1}{3}} \left( \sigma^{\frac{2}{3}}J^{\frac{1}{3}} + \tilde \sigma^{\frac{2}{3}} N^{\frac{1}{3}} \right) \left( \log d\right)^{\frac{2}{3}} + \sqrt{M} B \log d, \\
    & \norm{\mb E_{i,:}} \lesssim  \sigma \sqrt{J} + M^{\frac{1}{6}} B^{\frac{1}{3}} \sigma^{\frac{2}{3}} J^{\frac{1}{3}}(\log d)^{\frac{2}{3}} + \sqrt{M}B \log d, \  \forall i\in[N], \\ 
    & \norm{\mb E_{:,j}} \lesssim \sigma \sqrt{N} + B\log d,\ \forall j\in[J], \\ 
    & \norm{\mb E \mb V^*} \lesssim \tilde \sigma \sqrt{N}  + M^{\frac{1}{6}}B^{\frac{1}{3}}\tilde \sigma^{\frac{2}{3}}N^{\frac{1}{3}}\left(\log d\right)^{\frac{2}{3}} + M B\norm{\mb V^*}\ti \log d, \\
    & \|\E\t\U^*\| \lesssim \sigma\sqrt{J+MK} + \sqrt{M}\sigma \sqrt{\log d} + M^{\frac16} B^{\frac13} \sigma^{\frac23} (J+mK)^{\frac13}\|\U^*\|\ti^{\frac13} (\log d )^{\frac23} \\
    & \quad\quad\quad\quad\quad + \sqrt{M} B \|\U^*\|\ti\log d,
\end{align}		
with probability at least $1 - O(d^{-20})$. 
		
Moreover, if Assumption \ref{assumptions} hold, we have 
\begin{align}
    & \norm{\mb E} \lesssim \sigma \sqrt{J} + \tilde \sigma \sqrt{N}, \label{eq:simplified spectral norm of E} \\
    & \norm{\mb E_{i,:}} \lesssim \sigma\sqrt{J}, \ \forall i\in[N], \label{eq:simplified spectral norm of E_{i,:}} \\ 
    & \norm{\mb E_{:,j}} \lesssim \sigma\sqrt{N}, \ \forall j\in[J], \label{eq:simplified spectral norm of E_{:,j}} \\ 
    & \norm{\mb E \mb V^*} \lesssim \tilde \sigma \sqrt{N}, \label{eq:simplified spectral norm of EV*} \\
    & \|\E\t \U^*\| \lesssim\sigma \sqrt{J}, \label{eq:simplified spectral norm of EtU*}
\end{align}
with probability at least $1 - O(d^{-20})$. 
\end{lemma}

\paragraph{Proof of Lemma~\ref{lemma: noise matrix concentrations using the universality}.}
Using the facts established in the proof of Lemma~\ref{lemma: matrix Bernstein}, we have
\begin{align}
    & \|\bb E[\E^\top\E]\|^{\frac12} \le \tilde \sigma \sqrt{N}, \|\bb E[\E\E^\top]\|^{\frac12} \le \sigma \sqrt{J}, \sigma(\E) \leq \sigma \sqrt{J} + \tilde \sigma \sqrt{N},\\ 
    & v(\E)  \le \max_{i\in[N]}\max_{l\in[L]} \|\bb E[\E_{i,S_l}\t\E_{i,S_l}]\|^{\frac12} \leq \tilde \sigma, \\
    & \sigma_*(\mb E) \le \sup_{\|\mb w\|=1} \| \bb E[\E\mb w \mb w^\top \E^\top]\|^{\frac12} \le \max_{i\in[N]} \sup_{\|\mb w\| = 1} \|\bb E[\E_{i,:}\mb w\mb w^\top\E_{i,:}^\top]\mb\|^{\frac12} \leq \tilde \sigma, \label{eq: upper bound sigma_*(E)}\\ 
    & R(\E) = \left\|\max_{i\in[N],l\in[L]}\|\E_{i,S_l}\|\right\|_{\infty} \leq \sqrt{M}B. 
\end{align}	
Then the first inequality in Lemma~\ref{lemma: noise matrix concentrations using the universality} follows by plugging these conditions into Proposition~\ref{proposition: thm 2.15 in universality paper} with $t = c\log d$ with a sufficiently large constant $c$. 

Similarly, for the second and third inequalities we use 
\begin{align*}
    \sigma(\mb E_{i,:}) = \max\{\|\bb{E}[\E_{i,:}^\top \E_{i,:}]\|^{\frac12}, \|\bb{E}[\E_{i,:}\E_{i,:}^\top]\|^{\frac12}\} \leq \max\{\sigma \sqrt{J}, \tilde\sigma\}=\sigma\sqrt{J}, \\
    v(\mb E_{i,:}) = \norm{\cov(\mb E_{i,:})}^{\frac{1}{2}} \leq \tilde\sigma, \qquad  \sigma_*(\mb E_{i,:}) \leq \tilde\sigma, \qquad R(\mb E_{i,:}) \leq \sqrt{M}B, 
\end{align*}
and
\begin{align*}
    \sigma(\E_{:,j}) = \max\{\|\bb{E}[\E_{:,j}^\top \E_{:,j}]\|^{\frac12}, \|\bb{E}[\E_{:,j}\E_{:,j}^\top]\|^{\frac12}\} \le \sigma\sqrt{N}, \\
    v(\E_{:,j})\le \sigma, \qquad \sigma_*(\E_{:,j})\le \sigma, \qquad R(\E_{:,j})\le B.
\end{align*}

Furthermore, for the fourth and fifth inequalities,
\begin{align}
    \sigma(\mb E\mb V^*) = \max\left\{\|\bb E[\mb E\mb V^*{\mb V^*}\t \mb E\t ] \|^{\frac{1}{2}}, \|\bb E[{\mb V^*}\t \mb E\t \mb E\mb V^*] \|^{\frac{1}{2}} \right\} & \le \tilde \sigma \sqrt{N}, \\
    v(\mb E \mb V^*) \leq \tilde \sigma, \qquad \sigma_*(\mb E\mb V^*) \leq \sigma_*(\mb E) \stackrel{\text{\eqref{eq: upper bound sigma_*(E)}}}{\leq } \tilde\sigma ,
\end{align}
and with $MK \lesssim ML\asymp J$ we have
\begin{align}
    \|\bb E [\mb E\t \mb U^* {\mb U^*}\t \mb E]\| = \max_{l\in[L]} \|\bb E [\mb E_{:,S_l}\t \mb U^* {\mb U^*}\t \mb E_{:,S_l}] \| & \le \max_{l\in[L]} \bb E \| \mb E_{:,S_l}\t \mb U^* {\mb U^*}\t \mb E_{:,S_l} \| \\
    = M \max_{j\in[J]} \bb E [\mb E_{:,j}\t \mb U^* {\mb U^*}\t \mb E_{:,j}] & \le  M \sigma^2\|\U^*\|_F^2= MK\sigma^2,\\
    \|\bb E [\U^{*\top}\E\E^\top \U^*]\| \le \|\bb E[\E\E^\top]\| & \le \sigma^2 J, \\
    \sigma(\mb E\t \mb U^*) \le \sigma\sqrt{J} + \sigma\sqrt{MK} & , \\ 
    R(\E\t\U^*) = \norm{\max_{i\in[N],l\in[L]}\|\E_{i,S_l}\t \U^*_{i,:}\|}_{\infty} & \le \sqrt{M}B\|\U^*\|\ti, \\
    v(\mb E\t\mb U^*) \le \max_{l\in[L]} \norm{\bb E[ \E_{:,S_l}\t\mb U^*\U^{*\top} \E_{:,S_l}]}^{\frac12} & \leq \sqrt{MK}\sigma, \\ 
    \sigma_*({\mb U^*}\t \mb E)^2 \leq \sigma_*(\mb E)^2\stackrel{\text{by \eqref{eq: upper bound sigma_*(E)}}}{\leq } \tilde\sigma^2\leq M\sigma^2.
\end{align}
\qed

Lemma~\ref{lemma:leave-one-out} below utilizes a leave-one-out analysis that decouples the dependency between $\mb R$ and $\mb V \mb V\t \mb V^* - \mb V^*$, and a leave-one-block-out analysis on $\mb R$ and $\mb U \mb U\t \mb U^* - \mb U^*$.
\begin{lemma}
\label{lemma:leave-one-out}
Suppose Assumptions~\ref{assumption:s_block}, \ref{assumptions} hold. Then with probability at least $1- O(d^{-10})$, one has
\begin{align}
    & \norm{\mb R\mb V^* - \mb R \mb V\mb V\t \mb V^*}\ti \\
    \lesssim & \frac{\tilde \sigma \sigma \sqrt{K\log d (M N + J)}}{{\sigma_K^*}} + M B \log d \norm{\mb V \mb V\t \mb V^* - \mb V^*}\ti + \frac{\sigma^2 J}{{\sigma_K^*}} \norm{\mb U \mb U\t \mb U^* - \mb U^*}\ti \\
    & + \frac{\kappa^* \sigma^2 \left(M N+ J\right) }{{\sigma_K^*}}\norm{\mb U^*}\ti + \frac{\sigma M B \log d \sqrt{J}}{{\sigma_K^*}} \norm{\mb V^*}\ti, \label{eq: lemma2_V}\\
    & \norm{\mb R\t \mb U^* - \mb R\t \mb U \mb U\t \mb U^*}\ti \\ 
    \lesssim & \frac{\sigma^2 \sqrt{K\log d (M N + J)}}{{\sigma_K^*}} + B \log d\norm{\mb U \mb U\t \mb U^* - \mb U^* }\ti + \frac{\sigma^2 M N}{{\sigma_K^*}}\norm{\mb V \mb V\t \mb V^* - \mb V^*}\ti \\
    & + \frac{\kappa^* \sigma^2 (M N+J)}{{\sigma_K^*}}\norm{\mb V^*}\ti + \frac{\sigma B \log d \sqrt{M N} }{{\sigma_K^*}}\norm{\mb U^*}\ti. \label{eq: lemma2_U}
\end{align}
\end{lemma}

\paragraph{Proof of Lemma~\ref{lemma:leave-one-out}.}

We first establish some facts using the matrix perturbation theory as a tool for later use. By the perturbation theory under the spectral norm (for example, Theorem 2.9 in \cite{chen2021spectral}, Lemma 2 in \cite{yan2021inference}), \eqref{eq: assumption eq3} in Assumption~\ref{assumptions}, and inequalities \eqref{eq:simplified spectral norm of E}, \eqref{eq:simplified spectral norm of EV*}, \eqref{eq:simplified spectral norm of EtU*} in Lemma~\ref{lemma: noise matrix concentrations using the universality}, we know that 
\begin{align}
    & \max\left\{\norm{\U\U^\top\U^* - \U^*}, \norm{\V\V^\top\V^* - \V^*} \right\}\lesssim \frac{\tilde\sigma \sqrt{N} + \sigma\sqrt{J}}{{\sigma^*_K}}, \label{eq:spectral perturbation} \\
    & \max \left\{\norm{ {\U^*}\t \U \U\t \U^* -\mb I_K }, \norm{ {\mb V^*}\t \mb V \mb V\t \mb V^* -\mb I_K } \right\} \lesssim \frac{\tilde\sigma^2 N + \sigma^2J }{\sigma^{*2}_K}. \label{eq: square of spectral perturbation} 
\end{align}

We first give an overview of the proof. In Part I, we upper bound $\norm{\mb R\mb V^* - \mb R \mb V\mb V\t \mb V^*}\ti$ by separately bounding $\beta_1 \coloneqq \norm{\R^*(\V \V^{\top} \V^* - \V^*)}_{2,\infty}$ and $\beta_2\coloneqq \norm{\E(\V \V^{\top} \V^* - \V^*)}_{2,\infty}$. To bound $\beta_2$ that involves the error matrix $\E$ and the right singular space perturbation, we use the ``leave-one-out'' technique since the rows of $\E$ are independent. 
In Part II, we upper bound $\norm{\mb R\t \mb U^* - \mb R\t \mb U \mb U\t \mb U^*}\ti$ by the summation of $\beta_1'\coloneqq \norm{\R^{*,\top}(\U \U^{\top} \U^* - \U^*)}_{2,\infty}$ and $\beta_2'\coloneqq \norm{\E^{\top}(\U \U^{\top} \U^* - \U^*)}_{2,\infty}$. Similarly, to bound $\beta_2'$ we use the ``leave-one-block-out technique'' since the columns of $\E$ are block dependent.

\paragraph{Part I: upper bounding $\norm{\mb R\mb V^* - \mb R \mb V\mb V\t \mb V^*}\ti$.\\}

We define the following leave-one-row-out matrix for every $i\in[N]$: 
$$
\R^{(-i)} \coloneqq \R^* + \mc P_{-i,:}(\E).
$$
Let $\sigma_1^{(-i)}\ge \dots \ge \sigma^{(-i)}_{N\wedge J}$ by the singular values of $\R^{(-i)}$, and $\mb u_k^{(-i)}$, $\mb v_k^{(-i)}$ be the left and right singular vectors associated with $\sigma_k^{(-i)}$. Write $\U^{(-i)}=(\mb u_1^{(-i)},\dots, \mb u_K^{(-i)})\in\bb R^{N\times K}$ and $\V^{(-i)}=(\mb v_1^{(-i)},\dots, \mb v_K^{(-i)})\in\bb R^{J\times K}$.
We first establish some facts about these quantities. Note that $\|\mc P_{-i,:}(\E)\| \le \|\E\|$. By Weyl's inequality we know $\sigma_K^{(-i)} \ge \sigma_K^* - \|\mc P_{-i,:}(\E)\|$ and $\sigma_{K+1}^{(-i)} \le \|\mc P_{-i,:}(\E)\|$, thus
\begin{equation}\label{ineq: wedin}
    \sigma_K^{(-i)} - \sigma_{K+1}^{(-i)} - \|\mc P_{-i,:}(\E)\| \ge \sigma_K^* - 3\|\mc P_{-i,:}(\E)\|\ge \sigma_K^* - 3\|\E\| \gtrsim \sigma_K^*,
\end{equation}
where the last inequality is from \eqref{eq:simplified spectral norm of E} in Lemma~\ref{lemma: noise matrix concentrations using the universality} and \eqref{eq: assumption eq3} in Assumption~\ref{assumptions}.

It follows by the triangle inequality that 
\begin{align}\label{eq: decomposition of RV* - RVVtV*}
    \norm{\mb R\mb V^* - \mb R \mb V \mb V\t \mb V^*} \ti = & \|(\R^* + \E)(\V^* - \V\V^\top\V^*)\|\ti\\
    \leq & \underbrace{\norm{\mb R^* \left( \mb V\mb V\t \mb V^* - \mb V^*\right) }\ti}_{\beta_1} + \underbrace{\norm{\mb E\left( \mb V\mb V\t \mb V^* - \mb V^* \right) }\ti}_{\beta_2}  . 
\end{align}
We first notice 
\begin{align}
    \beta_1 = \norm{\mb R^*\left( \mb V \mb V\t \mb V^* - \mb V^* \right)}\ti = & \norm{\mb U^* \mb \Lambda^*\left(  {\mb V^*}\t \mb V \mb V\t \mb V^* -\mb I_K \right)}\ti \\
    \leq & \sigma_1^* \norm{\mb U^*}\ti \norm{  {\mb V^*}\t \mb V \mb V\t \mb V^* -\mb I_K } \\
    \lesssim & \frac{\kappa^* \sigma^2\left(M N + J\right)}{\sigma_K^*} \norm{\mb U^*}\ti, \label{eq:9}
\end{align}
where the last inequality yields from \eqref{eq: square of spectral perturbation}.

For $\beta_2$, we use the leave-one-out argument to deal with the dependency between $\mb E_{i,:}$ and $\mb V \mb V\t \mb V^* - \mb V^*$. Applying the triangle inequality gives
\begin{align}\label{eq-gamma1gamma2}
    \beta_2 \leq \underbrace{\max_{i\in[N]}\norm{\mb E_{i,:}\left( \mb V^{(-i)}{\mb V^{(-i)}}\t \mb V^* - \mb V^* \right) }}_{\gamma_1} + \underbrace{\max_{i\in[N]}\norm{\mb E_{i,:}\left( \mb V \mb V\t \mb V^* - \mb V^{(-i)}{\mb V^{(-i)}}\t \mb V^*\right)}}_{\gamma_2}.
\end{align}
Note that $\mb E_{i:}$ is independent of $\mb V^{(-i)}{\mb V^{(-i)}}\t \mb V^* - \mb V^*$, which allows us to apply the matrix Bernstein inequality to control $\gamma_1$ conditional on $\mb V^{(-i)}{\mb V^{(-i)}}\t \mb V^* - \mb V^*$.

\paragraph{Upper bounding $\gamma_1$ in \eqref{eq-gamma1gamma2}.}
We condition on $\mb V^{(-i)}$ and apply Lemma~\ref{lemma: matrix Bernstein} on $\gamma_1$. With probability at least $1 - O(d^{-10})$, one has
\begin{align}
    \gamma_1 \lesssim & \tilde\sigma \sqrt{\log d} \max_{i\in[N]} \norm{\mb V^{(-i)}{\mb V^{(-i)}}\t \mb V^* - \mb V^*}_F + MB \log d \max_{i\in[N]} \norm{\mb V^{(-i)}{\mb V^{(-i)}}\t \mb V^* - \mb V^*}\ti \\
    \le & \tilde\sigma \sqrt{\log d} \norm{\V\V\t\V^* - \mb V^*}_F + \tilde\sigma \sqrt{\log d} \max_{i\in[N]} \norm{\mb V^{(-i)}{\mb V^{(-i)}}\t \mb V^* - \V\V^\top\V^*}_F \\
    & + M B \log d \norm{\V\V\t\V^* - \mb V^*}\ti + M B \log d \max_{i\in[N]} \norm{\mb V^{(-i)}{\mb V^{(-i)}}\t \mb V^* - \V \V^{\top} \mb V^*}\ti \\
    \le & \tilde \sigma \sqrt{\log d} \norm{\mb V \mb V\t \mb V^* - \mb V^*}_F + \tilde \sigma \sqrt{\log d} \max_{i\in[N]} \norm{\mb V^{(-i)} {\mb V^{(-i)}}\t - \mb V \mb V\t }_F \\
    & + M B \log d\norm{\mb V \mb V\t \mb V^* - \mb V^*}\ti + M B \log d \max_{i\in[N]} \norm{\mb V^{(-i)} {\mb V^{(-i)}}\t - \mb V \mb V\t }_F \\ 
    \le & \tilde \sigma \sqrt{\log d} \norm{\mb V \mb V\t \mb V^* - \mb V^*}_F + M B \log d \norm{\mb V \mb V\t \mb V^* - \mb V^*}\ti \\
    & + M B \log d \max_{i\in[N]} \norm{\mb V^{(-i)} {\mb V^{(-i)}}\t - \mb V \mb V\t }_F, \label{eq:gamma1_1}
\end{align}
where the last line follows since $\tilde \sigma \leq \sqrt{M}B $.

We next bound $\max_{i\in[N]} \norm{\mb V^{(-i)} {\mb V^{(-i)}}\t - \mb V \mb V\t }_F$. Combining the Wedin theorem (Theorem 2.9 in \cite{chen2021spectral}) and \eqref{ineq: wedin} gives
\begin{align}
    &~\max\left\{\norm{\mb V \mb V\t - \mb V^{(-i)} {\mb V^{(-i)}}\t }_F, \norm{\mb U \mb U\t - \mb U^{(-i)} {\mb U^{(-i)}}\t}_F\right\} \\
    \lesssim &~\frac{\max\left\{\norm{ \left(\mb R - \mb R^{(-i)} \right) \mb V^{(-i)}}_F, \norm{{\mb U^{(-i)}}\t \left(\mb R - \mb R^{(-i)} \right)}_F\right\}}{{\sigma_K^*}}. \label{eq:2}
\end{align}
For the numerator of the RHS above, we have with probability that at least $1- O(d^{-12})$ that 
\begin{align}
    \norm{\left(\mb R - \mb R^{(-i)} \right)\mb V^{(-i)}}_F = & \norm{\mb E_{i,:}\mb V^{(-i)}} \stackrel{(i)}{\leq} 2 \norm{\mb E_{i,:}\mb V^{(-i)} {\mb V^{(-i)}}^{\top} \mb V^*} \\ 
    \le & 2\norm{\E_{i,:} \V^*} + 2\norm{\E_{i,:}\left( \V^{(-i)}\V^{(-i)\top}\V^* - \V^* \right)} \\
    \stackrel{(ii)}{\lesssim} & \tilde \sigma \sqrt{K \log d} + MB \log d \norm{\mb V^*}\ti + \gamma_1 ,\label{eq:3} \\
    \norm{{\mb U^{(-i)}}\t\left(\mb R - \mb R^{(-i)} \right)}_F = & \norm{\U_{i,:}^{(-i)\top}\E_{i,:}}_F \stackrel{(iii)}{\le} 2 \norm{\left( \mb U^{(-i)}_{i,:}{\mb U^{(-i)}}\t \mb U^* \right)\t\E_{i,:}}_F \\
    \le & 2\norm{\U_{i,:}^{*\top}\E_{i,:}}_F + 2\norm{\left( \mb U^{(-i)}_{i,:}{\mb U^{(-i)}}\t \mb U^* - \mb U^*_{i,:} \right)\t\E_{i,:}}_F \\
    \stackrel{(iv)}{\lesssim} & \sigma\sqrt{J}\norm{\mb U^*}\ti + \sigma \sqrt{J} \norm{\mb U^{(-i)}{\mb U^{(-i)}}\t \mb U^* - \mb U^*}\ti. \label{eq:3_U}
\end{align}
To show that $(i)$ holds, see that with probability at least $1 - O(d^{-20})$ one has
\begin{align}
    \sigma_{K}^2({\mb V^{(-i)}}^{\top} \mb V^*) \geq & 1 -  \norm{\mb V^{(-i)}{\mb V^{(-i)}}^{\top} - \mb V^*{\mb V^*}^{\top}}^2 \\ 
    \stackrel{\text{}}{\geq} & 1 - \big(\sqrt{2}\frac{\norm{\mc P_{-i,:}(\mb E)}}{\sigma_{K}^* - \norm{\mc P_{-i,:}(\mb E)}} \big)^2 \geq 1 - \big(\sqrt{2} \frac{\norm{\mb E}}{\sigma_{K}^* - \norm{\mb E}}\big)^2 \stackrel{}{\gg} \frac{1}{4},
\end{align}
where the first inequality is from Lemma 2.5 in \cite{chen2021spectral}, the second inequality is from Wedin's Theorem, and the last inequality uses \eqref{eq:simplified spectral norm of E}, and \eqref{eq: assumption eq3} in Assumption~\ref{assumptions}. 
In addition, $(ii)$ uses \eqref{eq:bernstein for E_{i,:}A} in Lemma~\ref{lemma: matrix Bernstein}, $(iii)$ uses a similar argument to that of $(i)$, and $(iv)$ uses \eqref{eq:simplified spectral norm of E_{i,:}} in Lemma~\ref{lemma: noise matrix concentrations using the universality}. %Here we recall that $\gamma_1$ is defined in \eqref{eq: decomposition of RV* - RVVtV*}. 
Furthermore, it holds with probability at least $1- O(d^{-12})$ that
\begin{align}
    & \norm{\U^{(-i)} {\U^{(-i)}}\t \U^* - \U^*}\ti \\
    \leq & \norm{\U^{(-i)} {\U^{(-i)}}\t \U^* - \U\U\t \U^*}\ti  + \norm{\U\U\t \U^* - \U^*}\ti \\
    \leq & \norm{\U^{(-i)} {\U^{(-i)}}\t - \U\U\t}\ti  + \norm{\U\U\t \U^* - \U^*}\ti \\
    \stackrel{(v)}{\lesssim} & \frac{1}{{\sigma_K^*}} \left( \tilde \sigma \sqrt{K \log d} + M B \log d \norm{\mb V^*}\ti + \gamma_1 \right) \\
    & + \frac{1}{\sigma_K^*} \left(\sigma\sqrt{J}\norm{\mb U^*}\ti + \sigma \sqrt{J} \norm{\mb U^{(-i)}{\mb U^{(-i)}}\t \mb U^* - \mb U^*}\ti \right) \\
    & + \norm{\mb U \mb U\t \mb U^* - \mb U^*}\ti, \label{eq: U-iU-itU* - U* two to infinity norm}
\end{align}
where $(v)$ uses \eqref{eq:2}, \eqref{eq:3}, \eqref{eq:3_U}.
Rearranging the terms in \eqref{eq: U-iU-itU* - U* two to infinity norm} yields that
\begin{align}
 &\norm{\U^{(-i)} {\U^{(-i)}}\t \U^* - \U^*}\ti \\ 
    \stackrel{(vi)}{\lesssim} & \frac{1}{{\sigma_K^*}} \left(\tilde \sigma \sqrt{K \log d} + M B \log d \norm{\mb V^*}\ti + \sigma\sqrt{J}\norm{\mb U^*}\ti + \gamma_1 \right) \\
    & + \norm{\mb U \mb U\t \mb U^* - \mb U^*}\ti\label{eq:11}
\end{align}
holds with probability at least $1- O(d^{-12})$ where
$(vi)$ uses $\sigma\sqrt{J} = o(\sigma_K^*)$ from Assumption~\ref{assumptions}. 
Plugging \eqref{eq:3}, \eqref{eq:3_U}, \eqref{eq:11} into \eqref{eq:2} leads to
\begin{align}
    & \max\left\{\norm{\mb V \mb V\t - \mb V^{(-i)} {\mb V^{(-i)}}\t }_F, \norm{\mb U \mb U\t - \mb U^{(-i)} {\mb U^{(-i)}}\t}_F\right\}\\
    \lesssim & \frac{1}{{\sigma_K^*}} \left(\tilde \sigma \sqrt{K \log d} + M B \log d \norm{\mb V^*}\ti + \gamma_1 \right)\\
    & + \frac{1}{\sigma_K^*} \left( \sigma\sqrt{J}\norm{\mb U^*}\ti +  \sigma \sqrt{J} \norm{\mb U^{(-i)}{\mb U^{(-i)}}\t \mb U^* - \mb U^*}\ti\right) \\
    \stackrel{(vii)}{\lesssim} & \frac{1}{{\sigma_K^*}} \left(\tilde \sigma \sqrt{K \log d} + M B \log d \norm{\mb V^*}\ti +   \sigma\sqrt{J}\norm{\mb U^*}\ti +  \sigma \sqrt{J} \norm{\mb U \mb U\t \mb U^* - \mb U^*}\ti + \gamma_1 \right) \label{eq:num}
\end{align}
with probability at least $1- O(d^{-12})$, where $(vii)$ again uses $\sigma\sqrt{J} = o(\sigma_K^*)$ from Assumption~\ref{assumptions}.
Plugging \eqref{eq:num} back into \eqref{eq:gamma1_1}, we have with probability at least $1-O(d^{-10})$ 
\begin{align}
    \label{eq:5}
    \gamma_1 \lesssim & \tilde \sigma \sqrt{\log d} \norm{\mb V \mb V\t \mb V^* - \mb V^*}_F + M B \log d \norm{\mb V \mb V\t \mb V^* - \mb V^*}\ti \\
    & + \frac{M B \log d }{{\sigma_K^*}}\left(\tilde \sigma \sqrt{K \log d} + M B \log d \norm{\mb V^*}\ti + \sigma\sqrt{J}\norm{\mb U^*}\ti \right. \\
    & \left. \quad\quad\quad\quad\quad\quad + \sigma \sqrt{J} \norm{\mb U \mb U\t \mb U^* - \mb U^*}\ti + \gamma_1 \right). 
\end{align}
Since $MB\log d \ll \sigma_K^*$, rearranging the terms in \eqref{eq:5} yields that
\begin{align}
    \gamma_1 \lesssim & \tilde \sigma \sqrt{\log d} \norm{\mb V \mb V\t \mb V^* - \mb V^*}_F + M B \log d \norm{\mb V \mb V\t \mb V^* - \mb V^*}\ti \\
    & + \frac{M B \log d }{{\sigma_K^*}}\left(\tilde \sigma \sqrt{K \log d} + M B \log d \norm{\mb V^*}\ti +   \sigma\sqrt{J}\norm{\mb U^*}\ti \right. \\
    & \left. \quad\quad\quad\quad\quad\quad + \sigma \sqrt{J}  \norm{\mb U \mb U\t \mb U^* - \mb U^*}\ti \right), \\
    \stackrel{(xiii)}{\lesssim} & \frac{\tilde \sigma \sigma \sqrt{K\left(M N + J\right) \log d}}{{\sigma_K^*}} + M B \log d \norm{\mb V \mb V\t \mb V^* - \mb V^*}\ti \\
    & + \frac{ M B \log d }{{\sigma_K^*}}\left(\tilde \sigma \sqrt{K \log d} + M B \log d \norm{\mb V^*}\ti + \sigma\sqrt{J}\norm{\mb U^*}\ti \right. \\
    & \left. \quad\quad\quad\quad\quad\quad + \sigma \sqrt{J} \norm{\mb U \mb U\t \mb U^* - \mb U^*}\ti\right),
\end{align}
where $(xiii)$ is from \eqref{eq:spectral perturbation}.

\paragraph{Upper bounding $\gamma_2$ in \eqref{eq-gamma1gamma2}.}
Using \eqref{eq:simplified spectral norm of E_{i,:}} in Lemma~\ref{lemma: noise matrix concentrations using the universality} and \eqref{eq:num}, we have with probability at least $1- O(d^{-10})$ that
\begin{align}
    \gamma_2 & = \max_{i\in[N]}\norm{\mb E_{i,:}\left( \mb V \mb V\t \mb V^* - \mb V^{(-i)}{\mb V^{(-i)}}\t \mb V^*\right)} \\
    &\lesssim \max_{i\in[N]}\norm{\mb E_{i,:}} \norm{ \mb V \mb V\t- \mb V^{(-i)}{\mb V^{(-i)}}\t} \\
    & \lesssim \sigma \sqrt{J} \max_{i\in[N]}\norm{\mb V \mb V\t- \mb V^{(-i)}{\mb V^{(-i)}}\t} \\
    & \lesssim \frac{ \sigma \sqrt{J}}{{\sigma_K^*}} \left(\tilde \sigma \sqrt{K \log d} + M B \log d \norm{\mb V^*}\ti + \sigma\sqrt{J}\norm{\mb U^*}\ti + \right. \\ 
    & \quad\quad\quad\quad \left. \sigma \sqrt{J} \norm{\mb U \mb U\t \mb U^* - \mb U^*}\ti  +  \gamma_1 \right).
\end{align}

Combining \eqref{eq:spectral perturbation} with the upper bounds for $\gamma_1$ and $\gamma_2$, with $\sigma\sqrt{J}=o(\sigma_K^*)$, we have with probability at least $1- O(d^{-10})$ that
\begin{align}
    \beta_2 \lesssim & \frac{\tilde \sigma \sigma \sqrt{K\left(M N + J\right) \log d}}{{\sigma_K^*}} + M B \log d \norm{\mb V \mb V\t \mb V^* - \mb V^*}\ti \\
    & + \frac{ M B \log d }{{\sigma_K^*}}\left(\tilde \sigma \sqrt{K \log d} + M B \log d \norm{\mb V^*}\ti + \sigma\sqrt{J}\norm{\mb U^*}\ti \right. \\
    & \left. \quad\quad\quad\quad\quad\quad + \sigma \sqrt{J} \norm{\mb U \mb U\t \mb U^* - \mb U^*}\ti\right) \\
    & + \frac{ \sigma \sqrt{J}}{{\sigma_K^*}} \left(\tilde \sigma \sqrt{K \log d} + M B\log d \norm{\mb V^*}\ti + \sigma\sqrt{J}\norm{\mb U^*}\ti \right. \\
    & \left. \quad\quad\quad\quad\quad\quad + \sigma \sqrt{J} \norm{\mb U \mb U\t \mb U^* - \mb U^*}\ti  \right) \\ 
    \lesssim & \frac{\tilde \sigma \sqrt{K\log d}\left( \sigma \sqrt{M N+J} + M B \log d  \right) }{{\sigma_K^*}} + M B \log d \norm{\mb V \mb V\t \mb V^* - \mb V^*}\ti \\
    & + \frac{\left( \sigma \sqrt{J	} + MB \log d \right) M B \log d}{{\sigma_K^*}} \norm{\mb V^*}\ti \\
    & + \frac{  \sigma \sqrt{J} \left(M B \log d + \sigma \sqrt{J} \right) }{{\sigma_K^*}} \left( \norm{\mb U^*}\ti + \norm{\mb U \mb U\t \mb U^* - \mb U^*}\ti \right)\\ 
    \lesssim &  \frac{\tilde \sigma \sigma \sqrt{K\log d(M N+J)}}{{\sigma_K^*}} + M B \log d \norm{\mb V \mb V\t \mb V^* - \mb V^*}\ti + \frac{\sigma M B \log d \sqrt{J}}{{\sigma_K^*}} \norm{\mb V^*}\ti \\ 
    & + \frac{\sigma^2 J}{\sigma_K^*} \left( \norm{\mb U^*}\ti + \norm{\mb U \mb U\t \mb U^* - \mb U^*}\ti \right), \label{eq:8}
\end{align}
where the last inequality is from $M B\log d  \lesssim \sigma \sqrt{(M N)\wedge J}$ by \eqref{eq: assumption eq4} in Assumption~\ref{assumptions}.

Now incorporating \eqref{eq:8} and \eqref{eq:9} into \eqref{eq: decomposition of RV* - RVVtV*}, we have with probability at least $1 - O(d^{-10})$ that
\begin{align*}
    \norm{\mb R \mb V^* - \mb R \mb V \mb V\t \mb V^*}\ti \lesssim & \frac{\kappa^* \sigma^2 \left(M N + J\right)}{{\sigma_K^*}} \norm{\mb U^*}\ti + \frac{\tilde \sigma \sigma \sqrt{K\log d (M N + J)}}{{\sigma_K^*}} \\
    & + M B \log d \norm{\mb V \mb V\t \mb V^* - \mb V^*}\ti
    + \frac{\sigma M B \log d  \sqrt{J}}{{\sigma_K^*}} \norm{\mb V^*}\ti \\
    & + \frac{\sigma^2 J}{\sigma_K^*} \left( \norm{\mb U^*}\ti + \norm{\mb U \mb U\t \mb U^* - \mb U^*}\ti \right)\\ 
    \lesssim & \frac{\kappa^* \sigma^2 \left(M N + J\right) }{{\sigma_K^*}}\norm{\mb U^*}\ti + \frac{\tilde \sigma \sigma \sqrt{K\log d (M N + J)}}{{\sigma_K^*}} \\
    & + M B \log d \norm{\mb V \mb V\t \mb V^* - \mb V^*}\ti + \frac{\sigma M B \log d \sqrt{J}}{{\sigma_K^*}} \norm{\mb V^*}\ti \\
    & + \frac{\sigma^2 J}{{\sigma_K^*}} \norm{\mb U \mb U\t \mb U^* - \mb U^*}\ti.
\end{align*}

\paragraph{Part II: upper bounding $\norm{\mb R\mb V^* - \mb R \mb V\mb V\t \mb V^*}\ti$. \\}

For the second part, we will adopt a ``leave-one-block-out'' technique on the noise matrix $\E$. 
We define the following leave-one-block-out matrix for every $j\in[J]$: 
$$
\R^{(j)} \coloneqq \R^* + \mc P_{:,-S_{l_j}}(\E),
$$
where $l_j\in[L]$ such that $j \in S_{l_j}$. 
Let $\sigma_1^{(j)}\ge \dots \ge \sigma^{(j)}_{N\wedge J}$ be the singular values of $\R^{(j)}$, and $\mb u_k^{(j)}$, $\mb v_k^{(j)}$ be the left and right singular vectors associated with $\sigma_k^{(j)}$. Write $\U^{(j)}=(\mb u_1^{(j)},\dots, \mb u_K^{(j)})\in\bb R^{N\times K}$ and $\V^{(j)}=(\mb v_1^{(j)},\dots, \mb v_K^{(j)})\in\bb R^{J\times K}$. With similar arguments that lead to \eqref{ineq: wedin}, we know that 
\begin{equation}\label{ineq: wedin2}
    \sigma_K^{(j)} - \sigma_{K+1}^{(j)} - \|\mc P_{:,-S_{l_j}}(\E)\| \gtrsim \sigma_K^*.
\end{equation}

To begin with, by the triangle inequality we have 
\longeq{\label{eq: decomposition of RU}
    & \norm{\mb R\t \left(\mb U \mb U\t \mb U^*  - \mb U^* \right) }\ti \\
   \leq  & \underbrace{\norm{{\mb R^*}\t \left(\mb U \mb U\t \mb U^* - \mb U^* \right)}\ti}_{\beta_1'} + \underbrace{\norm{\mb E\t \left( \mb U\mb U\t \mb U^* - \mb U^*\right)}\ti}_{\beta_2'}.
}
For $\beta_1'$, with \eqref{eq: square of spectral perturbation} we have with probability at least $1 - O(n^{-10})$
\begin{align}
   \beta_1' & = \norm{{\mb R^*}\t \left(\mb U \mb U\t \mb U^* - \mb U^* \right)}\ti \\
   & \leq \sigma_1^* \norm{\mb V^*}\ti \norm{{\mb U^*}\t  \mb U \mb U\t \mb U^* -\mb I_K} \lesssim \frac{\kappa^*\sigma^2 \left(M N + J\right)}{{\sigma_K^*}} \norm{\mb V^*}\ti . \label{eq:beta1'}
\end{align}
For $\beta_2'$, introducing the "leave-one-block-out" alternative $\U^{(j)}$ yields that
% \longeq{
\begin{align}\notag
    \beta_2' = &~\max_{j\in[J]}\norm{\mb E_{:,j}\t\left(\mb U \mb U\t \mb U^* - \mb U^*  \right) }_2 \\ \label{eq-gamma12prime}
    \leq & ~ \underbrace{\max_{j\in[J]} \norm{\mb E_{:,j}\t \left(\mb U^{(j)} {\mb U^{(j)}}\t \mb U^* - \mb U^*\right)}}_{\gamma'_1}
     + \underbrace{ \max_{j\in[J]} \norm{\mb E_{:,j}\t \left( \mb U \mb U\t \mb U^* - \mb U^{(j)} {\mb U^{(j)}}\t \mb U^*\right) }}_{\gamma_2'}.
\end{align}
% }

\paragraph{Upper bounding $\gamma_1'$ in \eqref{eq-gamma12prime}.}
Utilizing Lemma~\ref{lemma: matrix Bernstein}, one has at least $1 - O(d^{-11})$ that,
\begin{align}
    \gamma_1' \leq & \sigma \sqrt{\log d} \max_{j\in[J]}\norm{\mb U^{(j)} {\mb U^{(j)}}\t \mb U^*  -\mb U^*}_F + B\log d \max_{j\in[J]} \norm{\mb U^{(j)} {\mb U^{(j)}}\t \mb U^*  -\mb U^*}\ti \\ 
    \le & \sigma\sqrt{\log d} \|\U\U\t\U^* - \U^*\|_F + \sigma\sqrt{\log d} \max_{j\in[J]} \|\U^{(j)}\U^{(j)\top}\U^* - \U\U\t\U^*\|_F \\
    & + B\log d \|\U\U\t\U^* - \U^*\|\ti + B\log d \max_{j\in[J]} \|\U^{(j)}\U^{(j)\top}\U^* - \U\U\t\U^*\|\ti \\
    \le & \sigma  \sqrt{\log d}\norm{\mb U \mb U\t  \mb U^*  -\mb U^*}_F + B \log d \norm{\mb U \mb U\t \mb U^*  -\mb U^*}\ti \\
    & + \left( \sigma \sqrt{\log d} + B \log d \right) \max_{j\in[J]} \norm{\mb U^{(j)} {\mb U^{(j)}}\t - \mb U \mb U\t }_F \\ 
    \lesssim & \sigma \sqrt{\log d} \norm{\mb U \mb U\t  \mb U^*  -\mb U^*}_F + B \log d \norm{\mb U \mb U\t \mb U^*  -\mb U^*}\ti \\
    & + B \log d  \norm{\mb U^{(j)} {\mb U^{(j)}}\t - \mb U \mb U\t }_F, \label{eq:18}
\end{align}
where the last line holds since $\sigma \sqrt{\log d} \ll B\log d$. %Note that now $\mb E_{:,j}$ an entry-wisely independent vector. 

To bound $\max_{j\in[J]}\norm{\mb U^{(j)} {\mb U^{(j)}}\t - \mb U \mb U\t }_F$ in the RHS above, we apply Wedin's theorem with \eqref{ineq: wedin2} and obtain
\longeq{\label{eq:leave-one-out perturbation of U}
    & \max\left\{\norm{\mb U \mb U\t - \mb U^{(j)}{\mb U^{(j)}}\t }_F, \norm{\mb V \mb V\t - \mb V^{(j)}{\mb V^{(j)}}\t }_F\right\} \\
    \lesssim & \frac{\max\left\{\norm{\mc P_{:,S_{l_j}}(\mb E)\t \mb U^{(j)}}_F, \norm{\mc P_{:,S_{l_j}}(\mb E) {\mb V^{(j)}}}_F \right\}}{{\sigma_K^*}}.
}
Following similar arguments above, with probability at least $1 - O(d^{-10})$ we have
\begin{align}
    \max_{j\in[J]} \norm{\mc P_{:,S_{l_j}}(\mb E)\t \mb U^{(j)}}_F \le & \sqrt{M}\max_{j\in[J]}\norm{\mb E_{:,j}\t \mb U^{(j)}} \le 2\sqrt{M} \max_{j\in[J]}\norm{\mb E_{:,j}\t \mb U^{(j)\top} \U^{(j)}\U^*} \\
    \le & 2\sqrt{M} \max_{j\in[J]} \left(\norm{\E_{:,j}\t\U^*} + \norm{\E_{:,j}\t\U^* - \E_{:,j}\t\U^{(j)}\U^{(j)\top}\U^*} \right) \\
    \stackrel{(ii)}{\lesssim} & \sigma \sqrt{M K\log d}  +\sqrt{M }B \log d \norm{\mb U^*}\ti + \sqrt{M}\gamma_1', \label{eq:15} \\ 
    \max_{j\in[J]} \norm{\mc P_{:,S_{l_j}}(\mb E) {\mb V^{(j)}}}_F = & \max_{j\in[J]} \norm{\E_{:,S_{l_j}} \V^{(j)}_{S_{l_j}, :}}_F \le 2 \max_{j\in[J]} \norm{\E_{:,S_{l_j}} \V^{(j)}_{S_{l_j}, :} \V^{(j)\top} \V^*}_F \\
    \le & 2\max_{j\in[J]} \norm{\E_{:,S_{l_j}} \V^*_{S_{l_j}, :}}_F + 2 \max_{j\in[J]} \norm{\E_{:, S_{l_j}} \left( \V^{(j)}_{S_{l_j}, :} \V^{(j)\top} \V^* - \V^*_{S_{l_j}, :} \right)}_F \\
    \le & 2M \max_{j\in[J]} \left( \norm{\E_{:,j}} \norm{\V^*}\ti +  \norm{\E_{:, j}} \norm{\V^{(j)} \V^{(j)\top} \V^* - \V^*}\ti \right) \\
    \stackrel{(iii)}{\lesssim} & \sigma M \sqrt{N} \norm{\mb V^*}\ti + \sigma M \sqrt{N} \norm{\mb V^{(j)}{\mb V^{(j)}}\t \mb V^* - \mb V^*}\ti, \label{eq:16}
\end{align}
where $(ii)$ uses \eqref{eq:bernstein for E_{:,j}B} in Lemma~\ref{lemma: matrix Bernstein}, and $(iii)$ uses \eqref{eq:simplified spectral norm of E_{:,j}} in Lemma~\ref{lemma: noise matrix concentrations using the universality}.
Further, we note that
\begin{align}
    & \norm{\mb V^{(j)} {\mb V^{(j)}}\t \mb V^* - \mb V^*} \ti \le \norm{\left(\mb V\mb V\t  - \mb V^{(j)}{\mb V^{(j)}}\t \right) \V^*}\ti + \norm{\mb V \mb V\t \mb V^* - \mb V^* }\ti \\ 
    \le & \norm{\mb V\mb V\t  - \mb V^{(j)}{\mb V^{(j)}}\t }\ti + \norm{\mb V \mb V\t \mb V^* - \mb V^* }\ti \\ 
    \stackrel{(iv)}{\lesssim} & \frac{1}{\sigma_K^*} \left(\sigma \sqrt{M K\log d} + \sqrt{M} B\log d \norm{\mb U^*}\ti + \sqrt{M} \gamma_1' \right) \\
    & + \frac{1}{\sigma_K^*} \left( M \sigma \sqrt{N} \norm{\mb V^*}\ti + \sigma M \sqrt{N} \norm{\mb V^{(j)}{\mb V^{(j)}}\t \mb V^* - \mb V^*}\ti \right) + \norm{\mb V \mb V\t \mb V^* - \mb V^*}\ti\\
    \stackrel{(v)}{\lesssim} & \frac{1}{\sigma_K^*} \left(\sigma \sqrt{M K\log d} + \sqrt{M} B\log d \norm{\mb U^*}\ti + \sqrt{M}\gamma	_1' + \sigma M \sqrt{N} \norm{\mb V^*}\ti \right) \\
    & + \norm{\mb V \mb V\t \mb V^* - \mb V^*}\ti, \label{eq:14}
\end{align}
where $(iv)$ uses \eqref{eq:leave-one-out perturbation of U}, \eqref{eq:15}, \eqref{eq:16}, and $(v)$ uses $M \sigma \sqrt{N} \ll {\sigma_K^*}$ from \eqref{eq: assumption eq3} in Assumption~\ref{assumptions}.
	
Plugging \eqref{eq:15}, \eqref{eq:16}, \eqref{eq:14} into \eqref{eq:leave-one-out perturbation of U}, we have with probability at least $1 - O(d^{-10})$,
\begin{align}
    &\norm{\mb U \mb U\t - \mb U^{(j)}{\mb U^{(j)}}\t}_F \\
    \lesssim & \frac{1}{\sigma_K^*} \left(\sigma \sqrt{M K\log d} + \sqrt{M}B \log d \norm{\mb U^*}\ti + \sqrt{M} \gamma_1' \right) \\
    & + \frac{1}{\sigma_K^*} \left( \sigma M \sqrt{N} \norm{\mb V^*}\ti +\sigma M \sqrt{N} \norm{\mb V^{(j)}{\mb V^{(j)}}\t \mb V^* - \mb V^*}\ti \right) \\ 
    \lesssim & \frac{1}{\sigma_K^*} \left( \sigma \sqrt{M K\log d} + \sqrt{M}B \log d \norm{\mb U^*}\ti + \sqrt{M}\gamma_1' +\sigma M \sqrt{N} \norm{\mb V^*}\ti \right) \\
    & + \frac{\sigma M \sqrt{N}}{{\sigma_K^*}^2}\left(\sigma \sqrt{M K\log d} + \sqrt{M} B\log d \norm{\mb U^*}\ti + \sqrt{M}\gamma_1' + \sigma M \sqrt{N} \norm{\mb V^*}\ti \right) \\
    & + \frac{\sigma M \sqrt{N}}{{\sigma_K^*}} \norm{\mb V \mb V\t \mb V^* - \mb V^*}\ti \\
    \stackrel{(vi)}{\lesssim} & \frac{1}{\sigma_K^*} \left( \sigma \sqrt{M K\log d} + \sqrt{M}B \log d \norm{\mb U^*}\ti + \sqrt{M}\gamma_1' +\sigma M \sqrt{N} \norm{\mb V^*}\ti \right) \\
    & + \frac{\sigma M \sqrt{N}}{{\sigma_K^*}} \norm{\mb V \mb V\t \mb V^* - \mb V^*}\ti, \label{eq:leave-one-out perturbation of U 2}
\end{align}
where $(vi)$ is because $\sigma M \sqrt{N} \ll \sigma_K^*$ from \eqref{eq: assumption eq3} in Assumption~\ref{assumptions}.

Combining \eqref{eq:18} and \eqref{eq:leave-one-out perturbation of U 2} gives that with probability at least $1 - O(d^{-10})$ we have
\begin{align} %\label{eq:20}
    \gamma_1' \lesssim & \sigma \sqrt{\log d} \norm{\mb U \mb U\t  \mb U^*  -\mb U^*}_F + B\log d \norm{\mb U \mb U\t \mb U^*  -\mb U^*}\ti \\
    & + B \log d \max_{j]in[J]} \norm{\mb U^{(j)} {\mb U^{(j)}}\t - \mb U \mb U\t }_F \\ 
    \stackrel{(vii)}{\lesssim} & \frac{\sigma \sqrt{(\tilde\sigma^2 N + \sigma^2 J) K \log d}}{{\sigma_K^*}} + B \log d \norm{\mb U \mb U\t \mb U^*  -\mb U^*}\ti \\ 
    & + \frac{B \log d}{\sigma_K^*} \left( \sigma \sqrt{M K\log d} +  \sqrt{M}B \log d \norm{\mb U^*}\ti + \sqrt{M} \gamma_1' \right. \\
    & \quad\quad\quad\quad\quad \left. +\sigma M \sqrt{N} \norm{\mb V^*}\ti + \sigma M \sqrt{N} \norm{\mb V \mb V\t \mb V^* - \mb V^*}\ti \right), 
    \label{eq: gamma1' (original)}
\end{align}
where $(vii)$ uses \eqref{eq:spectral perturbation}. 

Provided $B\sqrt{M} \log d \ll \sigma_K^*$ from \eqref{eq: assumption eq3} and \eqref{eq: assumption eq4} in Assumption~\ref{assumptions}, we rearrange \eqref{eq: gamma1' (original)} and derive that 
\begin{align}
    \gamma_1' & \lesssim \frac{\sigma \sqrt{K\log d} \left(\sigma \sqrt{M N + J}+ \sqrt{M} B \log d \right)}{{\sigma_K^*}} + B \log d\norm{\mb U \mb U\t \mb U^* - \mb U^* }\ti \\
    & + \frac{B \log d}{\sigma_K^*} \left( \sqrt{M}B \log d \norm{\mb U^*}\ti +\sigma M \sqrt{N} \norm{\mb V^*}\ti + \sigma M \sqrt{N} \norm{\mb V \mb V\t \mb V^* - \mb V^*}\ti \right) 
    \label{eq: gamma1'}
\end{align}
with probability at least $1- O(d^{-10})$.  
\paragraph{Upper bounding $\gamma_2'$  in \eqref{eq-gamma12prime}.}
Analogous to the upper bounding of $\gamma_2$, we have with probability at least $1 - O(d^{-10})$
\begin{align}
    \gamma_2' \le & \max_{j\in[J]} \norm{\E_{:,j}} \norm{\mb U \mb U\t - \mb U^{(j)} {\mb U^{(j)}}\t } \\
    \stackrel{(viii)}{\lesssim} & \sigma \sqrt{N} \norm{\mb U \mb U\t - \mb U^{(j)} {\mb U^{(j)}}\t } \\ 
    \stackrel{(ix)}{\lesssim} & \frac{\sigma \sqrt{N}}{\sigma_K^*} \left( \sigma \sqrt{M K\log d}  + \sqrt{M}B \log d \norm{\mb U^*}\ti + \sqrt{M} \gamma_1' +\sigma M \sqrt{N} \norm{\mb V^*}\ti \right) \\
    & + \frac{\sigma^2 M N}{{\sigma_K^*}} \norm{\mb V \mb V\t \mb V^* - \mb V^*}\ti, \label{eq:gamma_2' concentration}
\end{align}
where $(viii)$ uses \eqref{eq:simplified spectral norm of E_{:,j}} in Lemma~\ref{lemma: noise matrix concentrations using the universality} and $(ix)$ is by \eqref{eq:leave-one-out perturbation of U 2}.
	
Combine \eqref{eq: gamma1'} and \eqref{eq:gamma_2' concentration} and use the fact that the coefficient $\sigma \sqrt{MN} / \sigma_K^*$ of $\gamma_1'$ in \eqref{eq:gamma_2' concentration} is negligible, then with probability at least $1 - O(d^{-10})$
\begin{align}
    \beta_2' \lesssim & \frac{\sigma \sqrt{K\log d} \left( \sigma \sqrt{MN + J} + \sqrt{M} B \log d \right)}{{\sigma_K^*}} + B \log d\norm{\mb U \mb U\t \mb U^* - \mb U^* }\ti \\
    & + \frac{ \sqrt{M} B\log d \left( B \log d + \sigma \sqrt{N} \right) }{{\sigma_K^*}}\norm{\mb U^*}\ti + \frac{\sigma M \sqrt{N} \left( B \log d + \sigma \sqrt{N} \right) }{{\sigma_K^*}}\norm{\mb V^*}\ti \\
    & + \frac{\sigma M \sqrt{N} \left( B \log d + \sigma \sqrt{N} \right) }{{\sigma_K^*}} \norm{\mb V \mb V\t \mb V^* - \mb V^*}\ti \\
    \lesssim & \frac{\sigma^2 \sqrt{K\log d (M N+J)}}{\sigma_K^*} + B \log d\norm{\mb U \mb U\t \mb U^* - \mb U^* }\ti + \frac{\sigma B \log d \sqrt{M N}}{\sigma_K^*} \norm{\U^*}\ti \\
    & + \frac{\sigma^2 M N}{\sigma_K^*} \norm{\V^*}\ti + \frac{\sigma^2 M N}{\sigma_K^*} \norm{\mb V \mb V\t \mb V^* - \mb V^*}\ti, \label{eq:beta2'}
\end{align}
where the last inequality follow by $\sigma\sqrt{MN+J} \gtrsim \sqrt{M}B\log d$ and $\sigma\sqrt{N} \gtrsim B\log d$, from \eqref{eq: assumption eq4} in Assumption~\ref{assumptions}.

By plugging \eqref{eq:beta1'} and \eqref{eq:beta2'} back into \eqref{eq: decomposition of RU}, we finally get with probability at least $1-O(d^{-10})$ that
\begin{align}
    & \norm{\mb R\t \left(\mb U \mb U\t \mb U^*  - \mb U^* \right) }\ti  \\
    \lesssim & \frac{\sigma^2 \sqrt{K\log d (M N+J)}}{\sigma_K^*} + B \log d\norm{\mb U \mb U\t \mb U^* - \mb U^* }\ti + \frac{\sigma B \log d \sqrt{M N}}{\sigma_K^*} \norm{\U^*}\ti \\
    & + \frac{\kappa^* \sigma^2 (MN+J)}{\sigma_K^*} \norm{\V^*}\ti + \frac{\sigma^2 M N}{\sigma_K^*} \norm{\mb V \mb V\t \mb V^* - \mb V^*}\ti.
\end{align}
\qed

\begin{proposition}
\label{proposition: decomposition equalities}
The following equations hold for $(\U^*, \Lam^*, \V^*)$ and $(\U, \Lam, \V)$
\begin{align}
    \label{eq:decomposition of UUtU* - U*}
    \mb U \mb U\t \mb U^* - \mb U^* & = - \mb U \mb U\t \mb E \mb V^*{ \mb \Lambda^*}^{-1} + \R \left(\mb V\mb V\t \mb V^*{\mb \Lambda^*}^{-1} - \mb V^* {\mb \Lambda^*}^{-1}\right) + \mb E\mb V^*{\mb \Lambda^*}^{-1} \\ 
    \label{eq:decomposition of VVtV* - V*}
    \mb V\mb V\t \mb V^* - \mb V^* & = -\mb V\mb V\t \mb E\t \mb U^*{\mb \Lambda^*}^{-1} + \R\t\left(\mb U \mb U\t \mb U^*{\mb \Lambda^*}^{-1} - \mb U^*{\mb \Lambda^*}^{-1}\right) + \mb E\t \mb U^* {\mb \Lambda^*}^{-1}
\end{align}
\end{proposition}

\paragraph{Proof of Proposition~\ref{proposition: decomposition equalities}.}
First notice that
\begin{align*}
    \mb U \mb U\t \mb U^* - \mb U^* & = \U\U^\top \R^*\V^*\Lam^{*-1} - \R^*\V^*\Lam^{*-1} \\
    & = \U\U^\top (\R-\E)\V^*\Lam^{*-1} - (\R - \E)\V^*\Lam^{*-1} \\
    & = -\U\U^\top\E\V^*\Lam^{*-1} + \U\U^\top \R\V^*\Lam^{*-1} - \R\V^*\Lam^{*-1} + \E\V^*\Lam^{*-1} \\
    & = -\U\U^\top\E\V^*\Lam^{*-1} + \R\V\V^\top\V^*\Lam^{*-1} - \R\V^*\Lam^{*-1} + \E\V^*\Lam^{*-1},
\end{align*}
where the last equality holds because
$\U\U^\top \R = \U\U^\top (\U\Lam\V^\top + \U^\perp\Lam^\perp\V^{\perp\top})=\U\Lam\V^\top=(\U\Lam\V^\top + \U^\perp\Lam^\perp\V^{\perp\top})\V\V^\top=\R\V\V^\top$.
Eq.~\eqref{eq:decomposition of VVtV* - V*} holds following a similar reasoning.
\qed

\subsection{Proof of Theorem~\ref{theorem-bound (general r.v.)}}
We will prove Theorem~\ref{theorem-bound (general r.v.)} with Proposition~\ref{proposition: decomposition equalities} and Lemma~\ref{lemma:leave-one-out}.

\paragraph{Part I: Bounding $\norm{\U\U^\top\U^* - \U^*}\ti$ and $\norm{\V\V^\top\V^* - \V^* }\ti$.}
With inequality \eqref{eq:bernstein for E_{i,:}A} in Lemma~\ref{lemma: matrix Bernstein}, we know with probability at least $1-O(d^{-10})$ that,
\begin{align}\label{eq: lemma1_2}
    \norm{\E \V^*}\ti & \lesssim \tilde\sigma\sqrt{K\log d } + MB\log d\|\V^*\|\ti \lesssim \tilde\sigma \sqrt{K\log d },
\end{align}
where the last inequality follows by \eqref{eq: assumption eq4} in Assumption \eqref{eq: assumption eq4}. 
Using the triangle inequality on \eqref{eq:decomposition of UUtU* - U*} in Proposition~\ref{proposition: decomposition equalities} gives that 
\begin{align}
    & \norm{\mb U \mb U\t \mb U^* - \mb U^*}\ti \\ 
    \lesssim & \norm{\U}\ti \norm{\E\V^*}\norm{\Lam^{*-1}} 
     + \norm{\R\V\V^\top\V^*-\R\V^*}\ti \norm{\Lam^{*-1}} + \norm{\E\V^*}\ti\norm{\Lam^{*-1}} \\
     \stackrel{(i)}{\lesssim} & \norm{\U \U^{\top} \U^*}\ti \norm{\E\V^*}\norm{\Lam^{*-1}} 
     + \norm{\R\V\V^\top\V^*-\R\V^*}\ti \norm{\Lam^{*-1}} + \norm{\E\V^*}\ti\norm{\Lam^{*-1}}  \\ 
    \lesssim & \norm{\U \U^{\top} \U^*- \U^*}\ti \norm{\E\V^*}\norm{\Lam^{*-1}} + \norm{\U^*}\ti \norm{\E\V^*}\norm{\Lam^{*-1}} \\
    & + \norm{\R\V\V^\top\V^*-\R\V^*}\ti \norm{\Lam^{*-1}} + \norm{\E\V^*}\ti\norm{\Lam^{*-1}} \\
    \stackrel{(ii)}{\lesssim} & \underbrace{\frac{\tilde \sigma \sqrt{N}}{{\sigma_K^*}}}_{A_1} \norm{\mb U \mb U\t \mb U^* - \mb U^* }\ti + \frac{\tilde \sigma \sqrt{N}}{{\sigma_K^*}} \norm{\mb U^*}\ti + \frac{\tilde\sigma \sigma \sqrt{K\log d (M N + J)}}{{\sigma_K^*}^2} \\
    & + \frac{M B \log d }{\sigma_K^*}\norm{\mb V \mb V\t \mb V^* - \mb V^*}\ti + \underbrace{\frac{\sigma^2 J}{{\sigma_K^*}^2}}_{A_2} \norm{\mb U \mb U\t \mb U^* - \mb U^*}\ti \\
    & + \frac{\kappa^* \sigma^2 \left(M N+ J\right) }{{\sigma_K^*}^2}\norm{\mb U^*}\ti + \frac{\sigma M B \log d \sqrt{J}}{{\sigma_K^*}^2} \norm{\mb V^*}\ti + \frac{\tilde \sigma \sqrt{K \log d}}{{\sigma_K^*}} \\
    \stackrel{(iii)}{\lesssim} & \frac{\tilde \sigma \sqrt{N} }{{\sigma_K^*}}\norm{\mb U^*}\ti + \frac{\tilde\sigma \sigma \sqrt{K \log d (M N + J)} }{{\sigma_K^*}^2} +  \frac{M B \log d}{{\sigma_K^*}} \norm{\V \V\t \V^* - \V^*}\ti \\
    & + \frac{\kappa^* \sigma^2 \left( M N + J\right) }{{\sigma_K^*}^2}\norm{\mb U^*}\ti + \frac{\sigma M B \log d \sqrt{J}}{{\sigma_K^*}^2} \norm{\mb V^*}\ti + \frac{\tilde \sigma \sqrt{K \log d}}{{\sigma_K^*}} \\
    \stackrel{(iv)}{\lesssim} & \frac{\tilde \sigma \sqrt{N}{{\sigma_K^*}} + \kappa^* \sigma^2 J }{{\sigma^*_K}^2} \norm{\mb U^*}\ti + \frac{M B \log d}{{\sigma_K^*}} \norm{\mb V \mb V\t \mb V^* - \mb V^*}\ti \\
    & + \frac{\sigma M B \log d \sqrt{J}}{{\sigma_K^*}^2} \norm{\mb V^*}\ti + \frac{\tilde \sigma \sqrt{K \log d}}{{\sigma_K^*}}, \label{eq:23}
    \end{align}
holds with probability at least $1 - O(d^{-10})$. Here $(i)$ uses Lemma 2.5 in \cite{chen2021spectral} combined with \eqref{eq:spectral perturbation}, $(ii)$ uses \eqref{eq: lemma1_2}, \eqref{eq:simplified spectral norm of EV*} in Lemma~\ref{lemma: noise matrix concentrations using the universality}, and \eqref{eq: lemma2_V} in Lemma~\ref{lemma:leave-one-out}, $(iii)$ is because $A_1 = o(1), A_2 = o(1)$ by \eqref{eq: assumption eq3} in Assumption~\ref{assumptions}, and $(iv)$ holds by \eqref{eq: assumption eq3} in Assumption~\ref{assumptions}.

We will use similar arguments on the right singular subspace. 
By \eqref{eq:bernstein for E_{:,j}B} in Lemma~\ref{lemma: matrix Bernstein},
\begin{align}\label{eq: lemma1_3}
    \norm{\E^\top \U^*}\ti \lesssim \sigma \sqrt{K\log d} + B \log d \norm{\U^*}\ti \lesssim \sigma \sqrt{K\log d }
\end{align}
holds with probability at least $1-O(d^{-10})$,
where the last inequality follows by \eqref{eq: assumption eq4}. 
Therefore, \eqref{eq:decomposition of VVtV* - V*} in Proposition~\ref{proposition: decomposition equalities} leads to
\begin{align}
    & \norm{\mb V{\mb V}\t \mb V^* - \mb V^* }\ti \\ 
    \lesssim & \norm{\mb V \mb V\t \V ^*- \mb V^* }\ti \norm{\mb E\t \mb U^*}\norm{{\Lam^*}^{-1}} + \norm{\mb V^*}\ti \norm{\mb E\t \mb U^*} \norm{{\mb \Lambda^*}^{-1}} \\
    & + \norm{\mb R\t \mb U^* - \mb R\t \mb U \mb U\t \mb U^*}\ti \norm{\Lam^{*-1}} + \norm{\E^\top\U^*}\ti \norm{\Lam^{*-1}} \\
    \stackrel{(iv)}{\lesssim} & \underbrace{\frac{\sigma\sqrt{J}}{{\sigma_K^*}}}_{B_1} \norm{\mb V\mb V\t \V^* -\mb V^* }\ti + \frac{\sigma\sqrt{J}}{\sigma_K^*} \norm{\mb V^*}\ti + \frac{\sigma^2 \sqrt{K\log d (M N + J)}}{{\sigma_K^*}^2} \\
    & + \frac{B \log d}{{\sigma_K^*}}\norm{\mb U \mb U\t \mb U^* - \mb U^* }\ti + \underbrace{\frac{\sigma^2 M N}{{\sigma^*_K}^2}}_{B_2} \norm{\mb V \mb V\t \mb V^* - \mb V^*}\ti \\
    & + \frac{\kappa^* \sigma^2 (M N + J) }{{\sigma^*_K}^2}\norm{\mb V^*}\ti + \frac{\sigma B \log d \sqrt{M N}}{{\sigma_K^*}^2} \norm{\U^*}\ti + \frac{\sigma \sqrt{K\log d}}{\sigma_K^*}, \\
    \stackrel{(v)}{\lesssim} & \frac{\sigma\sqrt{J}}{\sigma_K^*} \norm{\mb V^*}\ti + \frac{\sigma^2 \sqrt{K\log d (M N + J)}}{{\sigma_K^*}^2} + \frac{B \log d}{{\sigma_K^*}}\norm{\mb U \mb U\t \mb U^* - \mb U^* }\ti \\
    & + \frac{\kappa^* \sigma^2 (M N + J) }{{\sigma^*_K}^2}\norm{\mb V^*}\ti + \frac{\sigma B \log d \sqrt{M N}}{{\sigma_K^*}^2} \norm{\U^*}\ti + \frac{\sigma \sqrt{K\log d }}{\sigma_K^*} \\
    \stackrel{(vi)}{\lesssim} & \frac{ \sigma \sqrt{J}\sigma^*_K + \kappa^* \sigma^2 M N} {{\sigma_K^*}^2}\norm{\mb V^*}\ti + \frac{B \log d}{{\sigma_K^*}} \norm{\U \U\t \U^* - \U^*}\ti \\
    & + \frac{\sigma B\log d \sqrt{M N}}{{\sigma_K^*}^2}\norm{\mb U^*}\ti + \frac{\sigma\sqrt{K\log d }}{\sigma_K^*},
    \label{eq:22}
\end{align}
which holds with probability at least $1 - O(d^{-10})$.
Here $(iv)$ uses \eqref{eq: lemma1_3}, \eqref{eq:simplified spectral norm of EtU*} in Lemma~\ref{lemma: noise matrix concentrations using the universality}, and \eqref{eq: lemma2_U} in Lemma~\ref{lemma:leave-one-out}; $(v)$ holds since $B_1=o(1), B_2=o(1)$ by \eqref{eq: assumption eq3} in Assumption~\ref{assumptions}; and $(vi)$ holds by \eqref{eq: assumption eq3} in Assumption~\ref{assumptions}.

Now substitute \eqref{eq:22} into \eqref{eq:23}, and with probability at least $1- O(d^{-10})$ we have
\begin{align*}
    & \norm{\mb U \mb U\t \mb U^* - \mb U^* }\ti \\
    \lesssim & \frac{\tilde \sigma \sqrt{N}{\sigma_K^*} + \kappa^* \sigma^2 J }{{\sigma^*_K}^2} \norm{\mb U^*}\ti+ \frac{\sigma M B \log d \sqrt{J}}{{\sigma_K^*}^2} \norm{\mb V^*}\ti + \frac{\tilde \sigma \sqrt{K \log d}}{{\sigma_K^*}}\\ 
    & + \frac{M B \log d }{{\sigma_K^*}} \left(\frac{\sigma \sqrt{J} \sigma_K^* + \kappa^* \sigma^2 \left(M N + J \right) }{{\sigma_K^*}^2}\norm{\mb V^*}\ti + \frac{B \log d}{{\sigma_K^*}} \norm{\U \U\t \U^* - \U^*}\ti \right.\\
    & \left. + \frac{\sigma B\log d \sqrt{M N}}{{\sigma_K^*}^2}\norm{\mb U^*}\ti + \frac{\sigma\sqrt{K\log d }}{\sigma_K^*} \right) \\
    \stackrel{(vii)}{\lesssim} & \frac{\tilde \sigma \sqrt{N}{\sigma_K^*} + \kappa^* \sigma^2 J }{{\sigma^*_K}^2} \norm{\mb U^*}\ti+ \frac{\sigma M B \log d \sqrt{J}}{{\sigma_K^*}^2} \norm{\mb V^*}\ti + \frac{\tilde \sigma \sqrt{K \log d}}{{\sigma_K^*}} \\
    & + \frac{M B \log d \left(\sigma\sqrt{J} + \sigma\sqrt{M N + J}\right)}{{\sigma^*_K}^2} \norm{\V}\ti + \underbrace{\frac{M B^2 \log^2 d}{{\sigma_K^*}^2}}_{C_1} \norm{\U\U\t\U^* - \U^*}\ti \\
    & + \frac{\sigma M B^2 \log^2 d \sqrt{M N}}{{\sigma_K^*}^3} \norm{\U^*}\ti + \frac{\sigma M B \sqrt{K \log^3 d}}{{\sigma_K^*}^2} \\
    \stackrel{(viii)}{\lesssim} & \frac{\tilde \sigma \sqrt{N}{\sigma_K^*}+ \kappa^* \sigma^2 J }{{\sigma^*_K}^2} \sqrt{\frac{\mu_1 K}{N}} + \frac{\sigma M B\log d \sqrt{MN+J}}{{\sigma^*_K}^2} \sqrt{\frac{\mu_2 K}{J}} + \frac{\tilde \sigma \sqrt{K \log d}}{{\sigma_K^*}} \eqqcolon \xi_1'\\
    \le & \frac{\tilde\sigma \sqrt{\mu_1 K \log d}}{\sigma^*_K} + \frac{\kappa^* \sigma^2 J }{ {\sigma^*_K}^2 } \sqrt{\frac{\mu_1 K}{N}} + \frac{\sigma MB\log d \sqrt{MN+J}}{{\sigma_K^*}^2} \sqrt{\frac{\mu_2K}{J}} \eqqcolon \xi_1
\end{align*}
where $(vii)$ uses \eqref{eq: assumption eq3} and $(viii)$ holds because $C_1=o(1)$ by \eqref{eq: assumption eq3}, \eqref{eq: assumption eq4} in Assumption~\ref{assumptions}.
	
On the other hand, substituting \eqref{eq:23} into \eqref{eq:22} gives that with probability at least $1 - O(d^{-10})$
\begin{align}
    & \norm{\mb V \mb V\t \mb V^* - \mb V^*}\ti \\
    \lesssim & \frac{\sigma \sqrt{J} \sigma_K^* + \kappa^* \sigma^2 M N }{{\sigma_K^*}^2}\norm{\V^*}\ti + \frac{\sigma B\log d \sqrt{M N}  }{{\sigma_K^*}^2}\norm{\mb U^*}\ti + \frac{\sigma\sqrt{K\log d }}{\sigma_K^*} \\
    & + \frac{B \log d}{{\sigma_K^*}} \left(\frac{\tilde \sigma \sqrt{N}{\sigma_K^*} + \kappa^* \sigma^2 \left(M N+ J\right) }{{\sigma^*_K}^2} \norm{\mb U^*}\ti + \frac{M B \log d}{{\sigma_K^*}} \norm{\mb V \mb V\t \mb V^* - \mb V^*}\ti \right. \\
    & \left. + \frac{\sigma M B \log d \sqrt{J}}{{\sigma_K^*}^2} \norm{\mb V^*}\ti + \frac{\tilde \sigma \sqrt{K \log d}}{{\sigma_K^*}} \right) \\
    \stackrel{(ix)}{\lesssim} & \frac{\sigma \sqrt{J} {\sigma_K^*} + \kappa^* \sigma^2 M N}{{\sigma_K^*}^2}\norm{\mb V^*}\ti+ \frac{\sigma B\log d \sqrt{M N}  }{{\sigma_K^*}^2}\norm{\mb U^*}\ti + \frac{\sigma \sqrt{K\log d}}{{\sigma_K^*}} \\ 
    & + \frac{\tilde\sigma B \log d \sqrt{N} + \sigma B \log d \sqrt{M N + J}}{{\sigma_K^*}^2} \norm{\U^*}\ti + \underbrace{\frac{M B^2\log^2 d}{{\sigma_K^*}^2}}_{C_2} \norm{\V\V\t\V^*-\V^*}\ti \\
    & + \frac{\sigma M B^2 \log^2 d \sqrt{J}}{{\sigma_K^*}^3} \norm{\V^*}\ti + \frac{\tilde \sigma B \log d\sqrt{K \log d}}{{\sigma_K^*}^2} \\
    \stackrel{(x)}{\lesssim} & \frac{\sigma \sqrt{J} {\sigma_K^*} + \kappa^* \sigma^2 M N }{{\sigma_K^*}^2} \sqrt{\frac{\mu_2 K}{J}}
    + \frac{\sigma B\log d \sqrt{M N + J} }{{\sigma_K^*}^2} \sqrt{\frac{\mu_1 K}{N}} + \frac{\sigma \sqrt{K\log d}}{{\sigma_K^*}} \eqqcolon \xi_2'\\
    \le & \frac{\sigma \sqrt{\mu_2 K \log d}}{\sigma_K^*} + \frac{\kappa^* \sigma^2 MN}{{\sigma_K^*}^2} \sqrt{\frac{\mu_2 K}{J}}
    + \frac{\sigma B\log d \sqrt{M N + J} }{{\sigma_K^*}^2} \sqrt{\frac{\mu_1 K}{N}} \eqqcolon \xi_2,
\end{align}
where $(ix)$ uses \eqref{eq: assumption eq3} and $(x)$ holds since $C_2=o(1)$ by \eqref{eq: assumption eq3}, \eqref{eq: assumption eq4} in Assumption~\ref{assumptions}.

\paragraph{Part II: Bounding $\norm{\U\Lam\V^\top - \U^*\Lam^*\V^{*\top}}_{\infty}$.}
For the last part, write $\mb R_U \coloneqq\mb U_L  \mb V_L\t$ with $(\mb U_L, \mb \Lambda_L, \mb V_L) = \mathsf{svd}(\mb U\t \mb U^*; K)$ and $\mb R_V \coloneqq\mb U_R\mb V_R\t$ with $(\mb U_R, \mb \Lambda_R, \mb V_R) = \mathsf{svd}(\mb V\t \mb V^*; K)$. Notice that $\R_{U}^\top\R_{U}=\R_{U}\R_{U}^\top=\R_{V}^\top\R_{V}=\R_{V}\R_{V}^\top=\mathbf{I}_K$. Then we have 
\begin{align*}
    & \U \Lam \V\t - \U^* \Lam^* {\V^*}\t \\
    = & (\U\R_U - \U^*) \R_{U}\t \Lam \V\t + \U^* \R_{U}\t\Lam \R_V(\R_V\t\V\t - \V^{*\top}) + \U^* (\R_{U}\Lam\R_V - \Lam^*) \V^{*\top}.
\end{align*}
Using the triangle inequality and Lemma~\ref{Lemma-matrix} gives that
\begin{align}
    & \norm{\mb U \mb \Lambda \mb V\t - \mb U^*\mb \Lambda^* {\mb V^*}\t }_\infty \\
    \leq & \norm{\mb U\mb R_U -\mb U^*}\ti \norm{\mb V \mb \Lambda \R_U}\ti + \norm{\mb U^*}\ti \norm{\mb \Lambda} \norm{\mb V \mb R_V - \mb V^*}\ti \\
    & + \norm{\mb U^*}\ti \norm{\mb R_U\t \mb \Lambda \mb R_V - \mb \Lambda^*} \norm{\mb V^*}\ti \\ 
    \le & \left(\norm{\mb U\mb U\t \mb U^* -\mb U^*}\ti + \norm{\mb U}\ti \norm{\mb R_U -\mb U\t \mb U^*}\right)\norm{\mb V}\ti \norm{\mb \Lambda} \\
    & + \norm{\mb U^*}\ti \norm{\mb \Lambda} \left(\norm{\mb V \mb V\t \mb V^*- \mb V^*}\ti + \norm{\mb V}\ti \norm{\mb V\t \mb V^* -\mb R_V}\right) \\
    & + \norm{\mb U^*}\ti \norm{\mb R_U\t \mb \Lambda \mb R_V - \mb \Lambda^*} \norm{\mb V^*}\ti. \label{eq: lemma2_4}
\end{align}
We then further break the norms of the empirical terms $\mb V, \mb \Lambda$ in \eqref{eq: lemma2_4} down into 
\begin{align}
    \norm{\mb V}\ti \leq & \norm{\mb V\big(\mb R_V - \mb V\t \mb V^*\big)}\ti + \norm{\mb V\mb V\t \mb V^* - \mb V^*}\ti + \norm{\mb V^*}\ti \\ 
    \leq & \norm{\mb V}\ti \norm{\mb R_V - \mb V\t \mb V^*} + \norm{\mb V\mb V\t \mb V^* - \mb V^*}\ti + \norm{\mb V^*}\ti,\label{eq: Vti simple decomposition}
\end{align}
Additionally, we note that $\norm{\mb R_{U} - \mb U\t \mb U^*} = \norm{\mb I - \mb \Lambda_{L}}\leq \norm{\mb I - \mb \Lambda_L^2} = \norm{\mb U_\perp\t \mb U^*}^2$ and similarly $\norm{\mb R_{V} - \mb V\t \mb V^*} \le \norm{\V^\top_{\perp} \V^*}^2$. Then Lemma 2.5 in \cite{chen2021spectral} combined with \eqref{eq:spectral perturbation} yields that 
\begin{align}
    \label{eq: R_U / R_V bound}
        & \max\left\{\norm{\mb R_U - \mb U\t \mb U^*}, \norm{\mb R_V - \mb V\t \mb V^*}\right\} \lesssim  \frac{\tilde\sigma^2 N+ \sigma^2J}{{\sigma_K^*}^2}
\end{align}
with probability at least $1- O(d^{-10})$. Rearranging the term $\norm{\mb V}\ti $ in \eqref{eq: Vti simple decomposition} with \eqref{eq: assumption eq3} in Assumption~\ref{assumptions}, then with probability at least $1 -O(d^{-10})$ we have  
\begin{align}
    \norm{\mb V}\ti \lesssim  \norm{\mb V\mb V\t \mb V^* - \mb V^*}\ti + \norm{\mb V^*}\ti \label{eq: V rearranged decomposition},
\end{align}
and similarly, 
\begin{align}
    \norm{\U}\ti \lesssim \norm{\U\U^\top\U^* - \U^*}\ti + \norm{\U^*}\ti. \label{eq: U rearranged decomposition}
\end{align}

By substituting \eqref{eq: V rearranged decomposition}, and \eqref{eq: U rearranged decomposition} into \eqref{eq: lemma2_4} and with $\norm{\mb \Lambda} \leq \norm{\mb \Lambda^*}+ \norm{\mb \Lambda - \mb \Lambda^*}$,
we have with probability at least $1 -O(d^{-10})$ that
\begin{align}
    &\norm{\mb U \mb \Lambda \mb V\t - \mb U^* \mb \Lambda^* \mb V^*}\ti \\ 
    \lesssim  & \left(\norm{\mb U\mb U\t \mb U^* -\mb U^*}\ti + (\norm{\mb U^*}\ti + \norm{\U\U\t\U^* - \U^*}\ti) \norm{\mb R_U -\mb U\t \mb U^*}\right) \cdot \\
    & \left(\norm{\mb V^*}\ti + \norm{\V^*-\V\V\t\V^*}\ti \right) \left(\norm{\mb \Lambda^*} + \norm{\Lam-\Lam^*} \right) \\
    & + \left(\norm{\mb V \mb V\t \mb V^*- \mb V^*}\ti + (\norm{\mb V^*}\ti + \norm{\V\V\t\V^*-\V^*}\ti) \norm{\mb V\t \mb V^* -\mb R_V}\right) \cdot \\
    & \norm{\mb U^*}\ti (\norm{\mb \Lambda^*} + \norm{\Lam-\Lam^*}) + \norm{\mb U^*}\ti \norm{\mb R_U\t \mb \Lambda \mb R_V - \mb \Lambda^*} \norm{\mb V^*}\ti. \label{eq: ULV-U*L*V*}
\end{align}
By Weyl's inequality and \eqref{eq:simplified spectral norm of E}, $|{\sigma_K} - \sigma_K^*| \le \norm{\Lam - \Lam^*} \lesssim \tilde\sigma\sqrt{N} + \sigma\sqrt{J}. $
%Invoking Lemma 2.5 and Lemma 2.6 in \cite{chen2021spectral} concludes with probability at least $1- O(d^{-10})$ that
In addition, one has 
\begin{align}
    & \norm{\mb R_U\t \mb \Lambda \mb R_V - \mb \Lambda^*} \\
    \leq & \norm{ \left( \R_U^\top - \U^{*\top}\U \right) \Lam \R_{V}} + \norm{\U^{*\top}\U \Lam \left(\R_{V} - \V^\top \V^*\right)} + \norm{\U^{*\top}\U\Lam\V^\top\V^* - \Lam^* } \\
    \leq & \norm{ \left( \R_U^\top - \U^{*\top}\U \right) \Lam \R_{V}} + \norm{\U^{*\top}\U \Lam \left(\R_{V} - \V^\top \V^*\right)}\\ 
    & + \norm{\U^{*\top}\mb R\V^* - \U^{*\top}\U^*\Lam^* \V^{*\top} \V^*}
    + \norm{\U^{*\top}\U^{\perp} \Lam^{\perp}\U^{\perp\top} \V^*}
    \\
    \le & \norm{\mb \Lambda^*}\left( \norm{\R_U^\top - \U^{*\top}\U} + \norm{\R_V^\top - \V^{*\top}\V} \right) + \norm{{\mb U^*}\t  \mb E\mb V^*} \\
    & + \norm{\U^{*\top}\U^{\perp} }\norm{\E}\norm{ \V^{*\top}\U^{\perp}}, \label{eq: decomposition of R_Ut Lambda R_V}
\end{align}
where we use $\norm{\Lam} \le \norm{\Lam^*} + \norm{\Lam-\Lam^*} \lesssim \norm{\Lam^*}$ and $\norm{\Lam^{\perp}}\leq \norm{\mb E}$. 

We separately look into the terms in \eqref{eq: decomposition of R_Ut Lambda R_V} as follows:
\begin{enumerate}
\item Regarding the first term, \eqref{eq: R_U / R_V bound} immediately implies that 
\begin{align}
    & \norm{\mb \Lambda^*}\left( \norm{\R_U^\top - \U^{*\top}\U} + \norm{\R_V^\top - \V^{*\top}\V} \right) \lesssim \frac{\kappa^*(\tilde \sigma^2 N + \sigma^2 J) }{\sigma^*_K}
\end{align}
holds with probability at least $1 - O(d^{-10})$. 
\item For the second term, we first view $\U^{*\top}\E \V^*$ as the sum of independent random matrices $\U^{*\top}\mathcal P_{i,S_l}(\E) \V^*$ over $i\in[N], l \in[L]$ and calculate the following quantities:
\begin{align}
    & \max\bigg\{\Big\| \bb E\big[\U^{*\top}\E \V^* \V^{*\top} \E^{\top} \U^*\big]\Big\| , \Big\|\bb E\big[ \V^{*\top} \E^{\top} \U^* \U^{*\top}\E \V^*\big]\Big\| \bigg\} \\ 
    = & \max\bigg\{\Big\| \U^{*\top} \bb E\big[\E \V^* \V^{*\top} \E^{\top}\big] \U^*\Big\| , \Big\| \V^{*\top} \bb E\big[\E^{\top} \U^* \U^{*\top}\E\big] \V^*\Big\| \bigg\} \\
    \stackrel{(i)}{\le} & \max\left\{ \sum_{i \in[N]}\norm{\U_{i,:}^{*}}_2^2 \big(\bb E\big[\E \V^* \V^{*\top} \E^{\top} \big] \big)_{i,i}, \sum_{l\in [L]} \norm{\V^*_{S_l,:}}^2 \big(\bb E\big[\E\t \U^* \U^{*\top} \E \big]\big)_{S_l,S_l} \right \}  \\ 
    \le & \max\left\{\sum_{i \in[N]}\norm{\U_{i,:}^{*}}_2^2 \big(\bb E\big[\E \V^* \V^{*\top} \E^{\top} \big] \big)_{i,i}, \right. \\
    & \quad\quad\quad \left. \sum_{l\in[L]} \big(\sum_{j\in S_l} \norm{\V^*_{j,:}}^2 \big) \norm{\big( \bb E\big[\E\t \U^* \U^{*\top} \E \big]\big)_{S_l, S_l}}  \right\} \\
    \stackrel{(ii)}{\le}& M K\sigma^2,
\end{align}
where $(i)$ holds since $\bb E\big[\E \V^* \V^{*\top} \E^{\top} \big]$ is a diagonal matrix and $\bb E\big[\E\t \U^* \U^{*\top} \E \big]$ is a block-diagonal matrix. And for $(ii)$ we use Lemma~\ref{lemma:matrix_trace} and
$$
\big(\bb E\big[\E \V^* \V^{*\top} \E^{\top} \big] \big)_{i,i} = \tr(\V^* \V^{*\top} \mathrm{Cov}(\E_{i,:})) \le \norm{\V^*}_F^2 \norm{\cov(\E_{i,:})}= K\tilde\sigma^2 \leq KM\sigma^2,
$$ 
\begin{align*}
    \norm{\big( \bb E\big[\E\t \U^* \U^{*\top} \E \big]\big)_{S_l, S_l}} \leq & \tr \big(\big( \bb E\big[\E\t \U^* \U^{*\top} \E \big]\big)_{S_l, S_l} \big) = \tr\big( \U^*\U^{*\top} \bb E[\E_{:, S_l} \E_{:,S_l}^\top] \big) \\
    \leq & \norm{\U^*}_F^2 \norm{\bb E[\E_{:, S_l} \E_{:,S_l}^\top]} \le K \tr\big( \bb E[\E_{:,S_l}^\top \E_{:, S_l} ] \big) \leq KM\sigma^2.
\end{align*}
In addition, $\norm{\U^{*\top}\mathcal P_{i,S_l}(\E) \V^*} \le \sqrt{M} B \norm{\U^*}\ti \norm{\V^*}\ti = \sqrt{M}BK\sqrt{\frac{\mu_1\mu_2}{NJ}}.$
With these inequalities, we apply the matrix Bernstein inequality which yields
\begin{align}
    & \norm{\U^{*\top}\E \V^*} \lesssim \sigma \sqrt{KM\log d} + \sqrt{M} B K \sqrt{\frac{\mu_1 \mu_2}{NJ}} \log d \lesssim \sigma \sqrt{KM \log d} 
\end{align}
holds with probability at least $1- O(d^{-10})$, where the last inequality holds since $\frac{B}{\sigma} \sqrt{\frac{K \mu_1\mu_2\log d}{NJ}} \lesssim \frac{\sigma}{\sqrt{K} M^2B (\log d)^{3.5}} \ll 1$ by \eqref{eq: assumption eq4}. 
\item For the third term, with probability at least $1- O(d^{-10})$ we have
\begin{align}
    & \norm{\U^{*\top}\U^{\perp} }\norm{\E}\norm{ \V^{*\top}\U^{\perp}} \lesssim \frac{\big(\tilde\sigma\sqrt{N} + \sigma \sqrt{J}\big)^3}{{\sigma_K^*}^2} 
    \lesssim  \frac{\tilde\sigma^2 N + \sigma^2 J}{\sigma_K^*}, 
\end{align}
since $\max\{\norm{\U^{*\top}\U^{\perp}}, \norm{\V^{*\top}\V^{\perp}}\} \leq \frac{\tilde\sigma\sqrt{N} + \sigma \sqrt{J}}{\sigma_K^*}$ with probability $1 - O(d^{-10})$. 
\end{enumerate}
Putting these pieces together, 
\begin{align}
    & \norm{\mb R_U\t \mb \Lambda \mb R_V - \mb \Lambda^*}  \lesssim \frac{\kappa^*(\tilde \sigma^2 N + \sigma^2 J) }{\sigma^*_K} + \sigma \sqrt{KM\log d} 
\end{align}
holds with probability $1 - O(d^{-10})$. 
Therefore, \eqref{eq: ULV-U*L*V*} becomes
\begin{align}
    & \norm{\mb U\mb \Lambda \mb V\t - \mb U^* \mb \Lambda^* {\mb V^*}\t }_\infty \\
    \lesssim & \Big(\xi_1' + (\sqrt{\frac{\mu_1K}{N}} +\xi_1' )\frac{\tilde\sigma^2N + \sigma^2 J}{{\sigma_K^*}^2 } \Big) \Big(\sqrt{\frac{\mu_2K}{J}} + \xi_2' \Big) (\sigma_1^* + \tilde\sigma\sqrt{N} + \sigma\sqrt{J})\\ 
    & + \Big(\xi_2' + (\sqrt{\frac{\mu_2 K}{J}} + \xi_2' )\frac{\tilde\sigma^2N + \sigma^2 J}{{\sigma_K^*}^2 }\Big) \sqrt{\frac{\mu_1K}{N}} (\sigma_1^* + \tilde\sigma\sqrt{N} + \sigma\sqrt{J}) \\
    & + \Big(\frac{\kappa^*(\tilde \sigma^2 N + \sigma^2 J) }{\sigma^*_K} + \sigma \sqrt{K M \log d} \Big)\sqrt{\frac{\mu_1\mu_2 K^2}{NJ }}
    \\ 
    \stackrel{(i)}{\lesssim} & \xi_1' \Big(\sqrt{\frac{\mu_2K}{J}} + \xi_2' \Big) \sigma_1^* + \xi_2' \sqrt{\frac{\mu_1 K}{N}} \sigma_1^* + \frac{\kappa^* (\tilde\sigma^2N + \sigma^2 J)}{\sigma^*_K} \sqrt{\frac{\mu_1\mu_2 K^2}{NJ }} \\
    & + \sqrt{KM} \sigma\sqrt{\log d} \sqrt{\frac{\mu_1\mu_2 K^2}{NJ }} \\
    \stackrel{(ii)}{\lesssim} & \xi_1' \xi_2' \sigma_1^* + \sqrt{\frac{\mu_2 K}{J}} \sigma_1^* \xi_1' + \sqrt{\frac{\mu_1 K}{N}} \sigma_1^* \xi_2' + \frac{\kappa^* (\tilde\sigma^2N + \sigma^2 J)}{\sigma^*_K} \sqrt{\frac{\mu_1\mu_2 K^2}{NJ}} \\
    & + \sqrt{KM} \sigma\sqrt{\log d} \sqrt{\frac{\mu_1\mu_2 K^2}{NJ }}, \label{eq: lemma2_5}
\end{align}
which holds with probability at least $1- O(d^{-10})$. Here $(i)$ follows from the fact that $\tilde\sigma\sqrt{N} + \sigma \sqrt{J} \lesssim \sigma_K^* \le \sigma^*_1$ by \eqref{eq: assumption eq3} and the definition of $\xi_1'$, and 
$(ii)$ holds by \eqref{eq: assumption eq3} and the definitions of $\xi_1', \xi_2'$.

Plugging in the definitions of $\xi'_1, \xi'_2$, one has
\begin{align}
    & \xi_1'\xi_2'\sigma^*_1 \\
    \stackrel{(iii)}{\lesssim} & \sigma^*_1 \left(\frac{\sigma \sqrt{M N+J}}{{\sigma^*_K}} \sqrt{\frac{\mu_1 K}{N}} + \frac{M B\log d}{\kappa^* {\sigma^*_K}} \sqrt{\frac{\mu_2 K}{J}} + \frac{\tilde \sigma \sqrt{K \log d}}{{\sigma_K^*}} \right) \cdot \\
    & \left( \frac{\sigma \sqrt{M N + J}}{{\sigma_K^*}} \sqrt{\frac{\mu_2 K}{J}} + \frac{B\log d }{\kappa^* {\sigma_K^*}} \sqrt{\frac{\mu_1 K}{N}} + \frac{\sigma \sqrt{K\log d}}{{\sigma_K^*}} \right) \\
    \stackrel{(iv)}{\lesssim} & \frac{\kappa^* \sigma^2 (M N+J)}{\sigma_K^*} \sqrt{\frac{\mu_1\mu_2K^2}{NJ}} + \frac{\mu_1 K}{N}\cdot \frac{\sigma B \log d \sqrt{M N + J}}{{\sigma_K^*}} + \frac{\mu_2 K}{J}\cdot \frac{\sigma M B \log d \sqrt{M N + J}}{{\sigma_K^*}} \\
    & + \frac{\kappa^* {\sigma} \sqrt{K\log d(M N + J)}}{{\sigma_K^*}} \left( \sigma \sqrt{\frac{\mu_1 K}{N}} + \tilde\sigma \sqrt{\frac{\mu_2 K}{J}} \right) + \frac{\kappa^* \tilde\sigma \sigma K \log d}{\sigma_K^*} \\
    \stackrel{(v)}{\lesssim} & \frac{\kappa^* K {\sigma^2}\sqrt{M\mu_1\mu_2} \log d }{\sigma_K^*} \sqrt{\frac{r}{M} + \frac{M}{r} } \\
    & + \frac{K \sigma M B \log d}{\sigma_K^*} \left( \frac{\mu_1}{\sqrt{MN}} \sqrt{1+\frac{r}{M}} + \frac{\mu_2}{\sqrt{J}} \sqrt{1 + \frac{M}{r}} \right). \label{eq: entrywise bound 1}
\end{align}
%Recall that $\zeta$ is the ratio between $M B \log d$ and $\sigma\sqrt{M N \wedge J}$. 
Here to derive $(iii)$, we employ the following simplification to $\xi_1'$ and $\xi_2'$:
\begin{align}
    & \xi_1' = \underbrace{\frac{\tilde \sigma \sqrt{N}{\sigma_K^*}+ \kappa^* \sigma^2 J}{{\sigma^*_K}^2}}_{\lesssim \frac{\sigma\sqrt{M N+ J}}{\sigma_K^*}} \sqrt{\frac{\mu_1 K}{N}} + \frac{\tilde \sigma \sqrt{K \log d}}{{\sigma_K^*}} + \underbrace{\frac{\sigma M B\log d \sqrt{M N+J}}{{\sigma^*_K}^2}}_{\lesssim \frac{M B\log d}{\kappa^* \sigma_K^*}} \sqrt{\frac{\mu_2 K}{J}}, \\ 
    & \xi_2' = \underbrace{\frac{\sigma \sqrt{J} {\sigma_K^*} + \kappa^* \sigma^2 M N }{{\sigma_K^*}^2} }_{\lesssim \frac{\sigma\sqrt{M N + J}}{\sigma_K^*}}\sqrt{\frac{\mu_2 K}{J}} + \frac{\sigma \sqrt{K\log d}}{{\sigma_K^*}}
    + \underbrace{\frac{\sigma B\log d \sqrt{M N + J} }{{\sigma_K^*}^2}}_{\lesssim \frac{B\log d}{\kappa^* \sigma_K^* }} \sqrt{\frac{\mu_1 K}{N}},
\end{align}
and $(iv)$ uses \eqref{eq: assumption eq4}. 
To obtain $(v)$, we invoke the following facts that
\begin{align}
    \frac{\kappa^* \sigma^2 (M N+J)}{\sigma_K^*} \sqrt{\frac{\mu_1\mu_2K^2}{NJ}} = & \frac{\kappa^* \sigma^2 M K}{\sigma_K^*}\sqrt{\frac{\mu_1\mu_2N}{J}} + \frac{\kappa^* \sigma^2 K}{\sigma_K^*}\sqrt{\frac{\mu_1\mu_2J}{N}} \\
    \lesssim & \frac{\kappa^* \sigma^2 \sqrt{M} K}{\sigma_K^*}\sqrt{\big(\frac{r}{M} + \frac{M}{r}\big)\mu_1 \mu_2}, \\ 
    \frac{\mu_1 K}{N}\cdot \frac{\sigma B \log d \sqrt{MN + J}}{\sigma_K^*} = & \frac{\mu_1 K \sigma M B \log d}{\sigma_K^* \sqrt{MN}} \sqrt{1+\frac{r}{M}}, \\
    \frac{\mu_2 K}{J} \cdot \frac{\sigma M B \log d \sqrt{M N + J}}{\sigma_K^*} = & \frac{\mu_2 K \sigma M B \log d}{\sigma_K^* \sqrt{J}} \sqrt{\frac{M}{r} + 1},
\end{align}
and 
\begin{align}
    & \frac{\kappa^* \sigma \sqrt{K\log d(M N + J)}}{{\sigma_K^*}} \left( \sigma \sqrt{\frac{\mu_1 K}{N}} + \tilde\sigma \sqrt{\frac{\mu_2 K}{J}} \right) \\ 
    \le & \frac{\kappa^* K \sigma^2 \sqrt{M\log d} \sqrt{\mu_1 \mu_2 }}{\sigma_K^*}\big( \sqrt{1 + \frac{r}{M}} + \sqrt{1 + \frac{M}{r}}\big)\\ 
    \lesssim & \frac{\kappa^* K \sigma^2 \sqrt{M \log d} \sqrt{\mu_1 \mu_2 }}{\sigma_K^*} \sqrt{ \frac{r}{M}+ \frac{M}{r}}.     
\end{align}
In addition,
\begin{align}
    & \sqrt{\frac{\mu_2 K}{J}}\sigma_1^*\xi_1' + \sqrt{\frac{\mu_1 K}{N}}\sigma_1^*\xi_2' \\
    \lesssim & \sqrt{\frac{\mu_2 K}{J}}\sigma_1^* \left( \frac{\sigma \sqrt{MN+J}}{{\sigma^*_K}} \sqrt{\frac{\mu_1 K}{N}} + \frac{M B\log d}{\kappa^* {\sigma^*_K}} \sqrt{\frac{\mu_2 K}{J}} + \frac{\tilde \sigma \sqrt{K \log d}}{{\sigma_K^*}} \right) \\
    & + \sqrt{\frac{\mu_1 K}{N}}\sigma_1^* \left( \frac{\sigma \sqrt{M N + J}}{{\sigma_K^*}} \sqrt{\frac{\mu_2 K}{J}} + \frac{B\log d }{\kappa^* {\sigma_K^*}} \sqrt{\frac{\mu_1 K}{N}} + \frac{\sigma \sqrt{K\log d}}{{\sigma_K^*}} \right) \\
    \lesssim & \kappa^* K \sigma \sqrt{M N + J} \sqrt{\frac{\mu_1\mu_2}{NJ}} + \frac{\mu_2 K}{J} M B \log d + \frac{\mu_1 K}{N} B \log d \\
    & + \kappa^* \sqrt{K\log d} \left( \tilde \sigma \sqrt{\frac{\mu_2 K}{J}} + \sigma \sqrt{\frac{\mu_1 K}{N}} \right) \\
    = & \kappa^* K \sigma \sqrt{\mu_1\mu_2}\sqrt{\frac{M}{J} + \frac{1}{N}} + \frac{\mu_2 K}{J} M B \log d + \frac{\mu_1 K}{N} B \log d \\
    & + \kappa^* \sqrt{K\log d} \left( \tilde\sigma \sqrt{\frac{\mu_2 K}{J}} + \sigma\sqrt{\frac{\mu_1 K}{N}} \right), \label{eq: entrywise bound 2}
\end{align}
Moreover, 
\begin{align}
     \frac{\kappa^* (\tilde\sigma^2N + \sigma^2 J)}{\sigma^*_K} \sqrt{\frac{\mu_1\mu_2 K^2}{NJ }} \le \frac{\sigma^2 \kappa^* K \sqrt{M \mu_1\mu_2}}{\sigma^*_K} \sqrt{\frac{r}{M} + \frac{M}{r}}. \label{eq: entrywise bound 3}
\end{align}

To the end, plugging \eqref{eq: entrywise bound 1}, \eqref{eq: entrywise bound 2}, and \eqref{eq: entrywise bound 3} into \eqref{eq: lemma2_5} yields that  
\begin{align}
    & \norm{\mb U\mb \Lambda \mb V\t - \mb U^* \mb \Lambda^* {\mb V^*}\t }_\infty \\
    \lesssim & \frac{\kappa^* K \sigma^2 \sqrt{M\mu_1\mu_2} \log d }{\sigma_K^*} \sqrt{\frac{r}{M} + \frac{M}{r} } + \frac{K \sigma M B \log d}{\sigma_K^*} \left( \frac{\mu_1}{\sqrt{MN}} \sqrt{1+\frac{r}{M}} + \frac{\mu_2}{\sqrt{J}} \sqrt{1 + \frac{M}{r}} \right) \\ 
    & + \kappa^* K \sigma \sqrt{\mu_1\mu_2}\sqrt{\frac{M}{J} + \frac{1}{N}} + \frac{\mu_2 K}{J} M B \log d + \frac{\mu_1 K}{N} B \log d \\
    & + \kappa^* \sqrt{K\log d} \left( \tilde\sigma \sqrt{\frac{\mu_2 K}{J}} + \sigma\sqrt{\frac{\mu_1 K}{N}} \right) \\ 
    & + \frac{\sigma^2 \kappa^* K \sqrt{M \mu_1\mu_2}}{\sigma^*_K} \sqrt{\frac{r}{M} + \frac{M}{r}} +\sqrt{KM} \sigma\sqrt{\log d} \sqrt{\frac{\mu_1\mu_2 K^2}{NJ }}
    \\ 
    \stackrel{(vi)}{\lesssim} & \frac{\kappa^* K \sigma^2 \sqrt{M\mu_1\mu_2} \log d }{\sigma_K^*} \sqrt{\frac{r}{M} + \frac{M}{r} } \\
    & + \kappa^* K \sigma \sqrt{\mu_1\mu_2 \log d} \sqrt{\frac{M}{J} + \frac{1}{N}} + \frac{\mu_2 K}{J} M B \log d + \frac{\mu_1 K}{N} B \log d \\
    \stackrel{(vii)}{\lesssim} & \kappa^* K \sigma \log d \sqrt{\mu_1\mu_2} \sqrt{\frac{M}{J} + \frac{1}{N}} + \frac{\mu_2 K}{J} M B \log d + \frac{\mu_1 K}{N} B \log d \eqqcolon \xi_3,
\end{align}
where $(vi)$ follows from
\begin{align*} 
    \frac{\sigma}{\sigma_K^*}\sqrt{1 + \frac{r}{M}} \lesssim \frac{1}{\sqrt{M N}}, \quad \frac{\sigma}{\sigma_K^*}\sqrt{1 + \frac{M}{r}} \lesssim \frac{1}{\sqrt{J}}
\end{align*}
with \eqref{eq: assumption eq3} in Assumption~\ref{assumptions}, and 
\begin{align*}
     \kappa^* \sqrt{K\log d} \left( \tilde\sigma \sqrt{\frac{\mu_2 K}{J}} + \sigma\sqrt{\frac{\mu_1 K}{N}} \right) & \le \kappa^* K \sigma \sqrt{\log d} \left(\sqrt{\frac{M\mu_2}{J}} + \sqrt{\frac{\mu_1}{N}} \right) \\
     & \le \kappa^* K \sigma \sqrt{\mu_1 \mu_2 \log d} \sqrt{\frac{M}{J} + \frac{1}{N}}.
\end{align*}
And $(vii)$ also used \eqref{eq: assumption eq3}. 
\qed

\begin{remark}\label{remark: s_thm1}
    To ensure the $\ell_{2,\infty}$-consistency of the singular subspaces and the $\ell_\infty$-consistency of the full matrix, we can impose further assumptions in addition to those in Theorem~\ref{theorem-bound (general r.v.)}. For $\frac{\norm{\mathbf U\mathbf U^{\top} \mathbf U^* - \mathbf U^*}_{2,\infty}}{\norm{\mathbf U^*}_{2,\infty}} = o(1)$ to hold, an easy-to-verify condition is $ \sigma\sqrt{MN \log d + \kappa^*J} = o(\sigma_K^*)$, $\frac{MB}{\sigma} \lesssim \log d$, and $\frac{\mu_2}{\mu_1}\lesssim \frac{ J}{(\log d)^3}$. Similarly, a sufficient condition for $\frac{\norm{\mathbf V\mathbf V^{\top} \mathbf V^* - \mathbf V^*}_{2,\infty}}{\norm{\mathbf V^*}_{2,\infty}} = o(1)$ to hold is $\sigma\sqrt{J\log d + \kappa^*MN}  = o(\sigma_K^* )$, $\frac{B}{\sigma} \lesssim \log(N\vee J)$, and $\frac{\mu_1}{\mu_2}\lesssim \frac{ N}{(\log d)^3}$. 
\end{remark}

\subsection{Verifying Assumption~\ref{assumption:s_block} for Sub-Exponential Noise}\label{sec:truncation}
It is evident from Assumption~\ref{assumption:s_block} that our theory accommodates a broad class of noise distributions. In what follows we shall spell out the corresponding $B$ satisfying Assumption~\ref{assumption:s_block} for sub-exponential distributions, which is common in statistical scenarios. 
Assume that every entry $E_{i,j}$ is sub-exponential with parameters $(v, \alpha)$, i.e., 
\eq{
\bb E\big[\exp(\lambda E_{i,j}) \big] \leq \exp\left(\frac{v^2\lambda ^2}{2}\right), \text{  for $i\in[N],j\in[J]$, and all $|\lambda | \leq \frac{1}{\alpha}$}.
}

The following lemma allows us to treat the original sub-exponential $\E$ as an entrywise-bounded and centered random matrix $\E'$ with high probability. 
\begin{lemma}\label{lemma: sub_exponential}
    There exists a random matrix $\E'=(E_{i,j}') \in \bb R^{N\times J}$ such that 
    \begin{align}
        \mathbb P( \E = \E') \geq 1 - O(d^{-20}),
    \end{align}
    $|E_{i,j}'| \leq 80 \alpha \log d$ and $\bb E[E'_{i,j}] = 0$. Furthermore, we have 
    \longeq{
    \max_{i\in[N], l\in[L]}\norm{\cov(\mb E'_{i,S_l}) } &\leq \tilde\sigma^2 +O(\alpha d^{-8} \sigma + \alpha^2 d^{-19}).
    }
\end{lemma}

\paragraph{Proof of Lemma~\ref{lemma: sub_exponential}.}
Let $C\coloneqq 80 \alpha \log d$, then for each entry $E_{i,j}$ one has 
\begin{align*}
    \max\{\bb P(E_{i,j} \geq C), \bb P(E_{i,j} \leq  -C)\} \leq \exp(-40\log d) = d^{-40}
\end{align*}
and
\begin{align}
    & \bb E[|E_{i,j}| \mathbbm{1}\{|E_{i,j}| \geq C\}] \\
    = & \int_0^{\infty} \bb P \left(|E_{i,j}|\mathbbm{1}\{|E_{i,j}| \geq C\} \ge x \right) dx \\
    = & \int_0^C \bb P \left(|E_{i,j}|\mathbbm{1}\{|E_{i,j}| \geq C\} \ge x \right) dx + \int_C^{\infty} \bb P \left(|E_{i,j}|\mathbbm{1}\{|E_{i,j}| \geq C\} \ge x \right) dx \\
    \leq & C\cdot \bb P(|E_{i,j}| \geq C) +  \int_C^\infty \bb P(|E_{i,j}| \geq x ) \mathrm d x \\
    \leq& 2 C \exp(-\frac{C}{2\alpha}) +  2 \int_{C}^\infty \exp(-\frac{x}{2 \alpha})\mathrm d x \\
    \leq & 2(C+ 2\alpha) d^{-40} = (160\log d + 4) \alpha d^{-40} \label{eq: expection tail of Eij}
\end{align}
by \cite[Proposition 2.9]{wainwright2019high}. Let $\hat E_{i,j} := E_{i,j} \mathbbm{1}\{|E_{i,j}| < C\}$. Then $\hat E_{i,j}=E_{i,j}$ with probability at least $1-2d^{-40}$. Note that $\hat E_{i,j}$ is not centered, we thus make a subtle adjustment by letting $E'_{i,j} \coloneqq \hat E_{i,j}  -\alpha \cdot \text{sign}(\bb E[\hat E_{i,j}]) \beta_{i,j}$, where $\beta_{i,j}$ are independent of $\hat{E}_{i,j}$ and are independent Bernoulli random variables with 
$$
\bb P\left(\beta_{i,j}=1\right) = \frac{1}{\alpha}|\bb E[\hat E_{i,j}]|
\le \frac{1}{\alpha}\bb E[|E_{i,j}| \mathbbm{1}\{|E_{i,j}| \geq C\}] 
\leq (160\log d+ 4) d^{-40},$$ 
where the first inequality holds since $\bb E[\hat E_{i,j}] + \bb E[E_{i,j} \mathbbm{1}\{|E_{i,j}| \geq C\}] = \bb E[E_{i,j
}] = 0$ and the second inequality follows from \eqref{eq: expection tail of Eij}. 
From the definition of $E'_{i,j}$, it is not hard to see that $\bb E[E'_{i,j}] = 0$, $| E'_{i,j}| \leq 2C$, and $ E'_{i,j} =E_{i,j} $ with probability at least $1 - (160 \log d + 5) d^{-40}$ for $i\in[N], j\in[J]$. Taking a union bound over all possible indices yields that 
$\E = \E'$ with probability at least $1 - O(d^{-20})$.

In the end, we need to control the spectral norm of $\cov(\mb E'_{i,S_l})$ for $l\in [L]$. For $j_1, j_2 \in S_l$, it follows from the Cauchy inequality that 
\begin{align}
    & \left|\text{Cov}(E'_{i,j_1}, E'_{i,j_2}) - \bb E[E_{i,j_1} E_{i,j_2}]\right| \\
    = & \left|\bb E[\hat E_{i,j_1} \hat E_{i,j_2}] - \bb E[\hat E_{i,j_1}] \bb E[ \hat E_{i,j_2}] - \bb E[E_{i,j_1} E_{i,j_2}]\right|  \\
    = & \left | \bb{E}\left[E_{i,j_1} E_{i,j_2} \mathbbm{1}\{|E_{i,j_2}| < C\} \right] - \bb{E}\left[E_{i,j_1} \mathbbm{1}\{|E_{i,j_1}| \ge C\} E_{i,j_2} \mathbbm{1}\{|E_{i,j_2}| < C\} \right] \right. \\
    & \left. - \bb E \left[E_{i,j_1} \mathbbm{1}\{|E_{i,j_1}| \geq C\} \right] \bb E \left[E_{i,j_2} \mathbbm{1}\{|E_{i,j_2}| \geq C\} \right] - \bb E[E_{i,j_1} E_{i,j_2}] \right | \\
    = & \left| \bb{E}\left[E_{i,j_1} \mathbbm{1}\{|E_{i,j_1}| \ge C\} E_{i,j_2} \mathbbm{1}\{|E_{i,j_2}| \ge C\} \right] + \bb{E}\left[E_{i,j_1} \mathbbm{1}\{|E_{i,j_1}| < C\} E_{i,j_2} \mathbbm{1}\{|E_{i,j_2}| \ge C\} \right] \right. \\
    & + \bb{E}\left[E_{i,j_1} \mathbbm{1}\{|E_{i,j_1}| \ge C\} E_{i,j_2} \mathbbm{1}\{|E_{i,j_2}| < C\} \right] \\
    & \left. + \bb E \left[E_{i,j_1} \mathbbm{1}\{|E_{i,j_1}| \geq C\} \right] \bb E \left[E_{i,j_2} \mathbbm{1}\{|E_{i,j_2}| \geq C\} \right] \right | \\
    \leq & \bb E[|E_{i,j_1}|^2 \mathbbm{1}\{|E_{i,j_1}| \geq C\}]^{\frac{1}{2}}\bb E[|E_{i,j_2}|^2 \mathbbm{1}\{|E_{i,j_2}| \geq C\}]^{\frac{1}{2}}\\ 
    & + \bb E[|E_{i,j_1}|^2 \mathbbm{1}\{|E_{i,j_1}| < C\}]^{\frac{1}{2}} \bb E[|E_{i,j_2}|^2 \mathbbm{1}\{|E_{i,j_2}| \ge C\}]^{\frac{1}{2}} \\ 
    & + \bb E[|E_{i,j_2}|^2 \mathbbm{1}\{|E_{i,j_1}| \geq C\}]^{\frac{1}{2}} \bb E[|E_{i,j_2}|^2 \mathbbm{1}\{|E_{i,j_1}| < C\}]^{\frac{1}{2}} \\ 
     & + \max_{j\in S_l}\big( \bb E[|E_{i,j}| \mathbbm{1}\{|E_{i,j}| \geq C\}]\big)^2 , \label{eq: cov analysis}
\end{align}
where we make use of the independence of $\beta_{i,j_1}$ and $\beta_{i,j_2}$, $\bb E[\hat E_{i,j}] + \bb E[ E_{i,j} \mathbbm{1}\{| E_{i,j}| \geq C\}] = 0$,  and the definition of $E_{i,j}$. 

Regarding the second moments of the truncated versions in \eqref{eq: cov analysis}, the tail inequality on a sub-exponential random variable gives that 
\begin{align}
    &  \bb E[|\tilde E_{i,j_1}|^2 \mathbbm{1}\{|\tilde E_{i,j_1}| \geq C\}] \leq 
    2 C^2 \bb P[|\tilde E_{i,j_1}| \geq C] + 2 \int_C^\infty x \bb P[|\tilde E_{i,j_1}| \geq x]\mathrm d x \\ 
    \leq & 4C^2 d^{-40} +  4\int_C^\infty x \exp(-\frac{x}{2\alpha}) \mathrm d x \leq  4C^2 d^{-40} + 4\alpha\int_C^\infty  \exp(-\frac{x}{4\alpha}) \mathrm d x \\
    \leq & 4C^2 d^{-40}+  16\alpha^2 d^{-20},
\end{align}
since $\exp(\frac{x}{4\alpha}) \geq \frac{x}{\alpha}$ for $x \geq C$. Besides, we have $\bb E[| E_{i,j}|^2 \mathbbm{1}\{| E_{i,j}| < L\}] \leq \sigma_0^2 $ for $i\in[N],j \in[J]$. Combining these pieces together with \eqref{eq: expection tail of Eij} gives that 
\eq{
\left|\text{Cov}(E'_{i,j_1}, E'_{i,j_2}) - \bb E[ E_{i,j_1}  E_{i,j_2}]\right| \leq O(\alpha d^{-9} \sigma_0 + \alpha^2 d^{-20} ) . 
}

Now we are ready to upper-bound $\norm{\text{Cov}({\mb E}_{i,S_j})  - \text{Cov}({\mb E}'_{i,S_j})}$ provided the entrywise control above. Invoking the elementary fact that $\norm{\mb A}\leq \norm{\A}_F \leq \norm{\A}_\infty \sqrt{n_1 n_2}$ for a matrix $\A\in \bb R^{n_1\times n_2}$, we have 
\eq{
\norm{\text{Cov}({\mb E}_{i,S_j})  - \text{Cov}({\mb E}'_{i,S_j})} \leq O(\alpha d^{-8} \sigma_0 + \alpha^2 d^{-19}),
}
since $M \leq d$. 
\qed

\subsection{Additional Lemmas}
\begin{lemma}\label{lemma: remark_EE^T}
    Suppose Assumptions~\ref{assumption:s_block}, \ref{assumptions} hold. Then with probability at least $1-O(d^{-10})$,
    \begin{align}
        \norm{\mathbf E\mathbf E^{\top} \U^* {\mathbf \Lambda^*}^{-2} }_{2,\infty} & \lesssim \frac{\sigma^2 J}{{\sigma_K^*}^2} \sqrt{\frac{\mu_1 K}{N}} + \frac{\tilde\sigma\sqrt{K \log d}}{\sigma_K^*}, \\
        \norm{\E^\top \E \V^* {\Lam^*}^{-2} }_{2,\infty} & \lesssim \frac{\sigma^2 MN}{{\sigma_K^*}^2} \sqrt{\frac{\mu_2 K}{J}} + \frac{\sigma\sqrt{K \log d}}{\sigma_K^*}.
    \end{align}
\end{lemma}

\paragraph{Proof of Lemma~\ref{lemma: remark_EE^T}.}
\begin{align*}
    & \norm{\mathbf E\mathbf E^{\top} \U^* {\mathbf \Lambda^*}^{-2} }_{2,\infty} = \max_{i\in[N]} \norm{\E_{i,:} \E^{\top} \U^*\mathbf \Lambda^{*-2}} \\ 
    \leq & \max_{i\in[N]} \Big(\norm{\E_{i,:}}^2 \norm{\U^*}_{2,\infty} / {\sigma_K^*}^2 + \norm{\E_{i,:} \mathcal P_{:,-i}(\E^{\top} )\U^*\mathbf \Lambda^{*-2}}_2 \Big)\\ 
    \stackrel{(i)}{\lesssim} & \frac{\sigma^2 J}{{\sigma_K^*}^2} \sqrt{\frac{\mu_1 K}{N}} + \max_{i\in[N]} \big( \tilde\sigma \sqrt{\log d} \norm{\mathcal P_{:,-i}(\E^{\top} )\U^*\mathbf \Lambda^{*-2}}_F + MB \log d \norm{\mathcal P_{:,-i}(\E^{\top} )\U^*\mathbf \Lambda^{*-2}}_{2,\infty} \big) \\ 
    \lesssim & \frac{\sigma^2 J}{{\sigma_K^*}^2} \sqrt{\frac{\mu_1 K}{N}} + \frac{\tilde \sigma \sqrt{ K \log d} }{{\sigma_K^*}^2} \max_{i\in[N]} \norm{\mathcal P_{:,-i}(\E^{\top} )\U^*} + \frac{MB \log d}{{\sigma_K^*}^2} \max_{i\in[N]} \norm{\mathcal P_{:,-i}(\E^{\top} )\U^*}_{2,\infty} \\ 
    \stackrel{(ii)}{\lesssim} & \frac{\sigma^2 J}{{\sigma_K^*}^2} \sqrt{\frac{\mu_1 K}{N}} + \frac{\tilde\sigma \sqrt{ K \log d}}{{\sigma_K^*}^2} \sigma\sqrt{J} + \frac{MB \log d}{{\sigma_K^*}^2} \big( \sigma\sqrt{K \log d} + B \log d \sqrt{\frac{\mu_1 K }{N}} \big) \\ 
    \stackrel{(iii)}{\lesssim} & \frac{\sigma^2 J}{{\sigma_K^*}^2} \sqrt{\frac{\mu_1 K}{N}} + \frac{\tilde \sigma \sigma \sqrt{ K J \log d} }{{\sigma_K^*}^2} \\
    \stackrel{(iv)}{\lesssim } & \frac{\sigma^2 J}{{\sigma_K^*}^2} \sqrt{\frac{\mu_1 K}{N}} + \frac{\tilde\sigma\sqrt{K \log d}}{\kappa^* \sigma_K^*},
\end{align*}
where $(i)$ used \eqref{eq:simplified spectral norm of E_{i,:}} in Lemma~\ref{lemma: noise matrix concentrations using the universality} and \eqref{eq:bernstein for E_{i,:}A} in Lemma~\ref{lemma: matrix Bernstein} conditional on $\mathcal P_{:,-i}(\E^{\top} )\U^*\mathbf \Lambda^{*-2}$, $(ii)$ used a similar treatment to \eqref{eq:simplified spectral norm of EtU*} in Lemma~\ref{lemma: noise matrix concentrations using the universality} on $\norm{\mathcal P_{:,-i}(\E^{\top} )\U^*}$ and \eqref{eq:bernstein for E_{:,j}B} in Lemma~\ref{lemma: matrix Bernstein} on $\norm{\mathcal P_{:,-i}(\E^{\top} )\U^*}\ti$, $(iii)$ used \eqref{eq: assumption eq4} in Assumption~\ref{assumptions}, and $(iv)$ is by \eqref{eq: assumption eq3} in Assumption~\ref{assumptions}.
Similarly, 
\begin{align*}
    & \norm{\E^{\top} \E \V^* {\Lam^*}^{-2} }_{2,\infty} = \max_{l\in[L]} \norm{\E_{:, S_l}^\top \E \V^* \Lam^{*-2}} \\ 
    \leq & \max_{l\in[S_l]} \Big(\norm{\E_{:, S_l}}^2 \norm{\V^*}\ti / {\sigma_K^*}^2 + \norm{\E_{:, S_l}^\top \mathcal P_{:,-S_l}(\E) \V^* \Lam^{*-2}} \Big)\\ 
    \stackrel{(v)}{\lesssim} & \frac{\sigma^2 MN}{{\sigma_K^*}^2} \sqrt{\frac{\mu_2 K}{J}} + \sigma \sqrt{M \log d} \max_{l\in[L]} \norm{\mathcal P_{:,-S_l}(\E)\V^* \Lam^{*-2}}_F \\
    & + \sqrt{M} B \log d \max_{l\in[L]} \norm{\mathcal P_{:,-S_l}(\E)\V^*\Lam^{*-2}}_{2,\infty} \\ 
    \lesssim & \frac{\sigma^2 MN}{{\sigma_K^*}^2} \sqrt{\frac{\mu_2 K}{J}} + \frac{\sigma \sqrt{M K \log d} }{{\sigma_K^*}^2} \max_{l\in[L]} \norm{\mathcal P_{:,-S_l}(\E)\V^*} + \frac{\sqrt{M} B \log d}{{\sigma_K^*}^2} \max_{l\in[L]} \norm{\mathcal P_{:,-S_l}(\E)\V^*}_{2,\infty} \\ 
    \stackrel{(vi)}{\lesssim} & \frac{\sigma^2 MN}{{\sigma_K^*}^2} \sqrt{\frac{\mu_2 K}{J}} + \frac{\sigma \sqrt{M K \log d}}{{\sigma_K^*}^2} \tilde\sigma \sqrt{N} + \frac{\sqrt{M} B \log d}{{\sigma_K^*}^2} \big( \tilde \sigma\sqrt{K \log d} + MB \log d \sqrt{\frac{\mu_2 K }{J}} \big) \\ 
    \stackrel{(vii)}{\lesssim} & \frac{\sigma^2 MN}{{\sigma_K^*}^2} \sqrt{\frac{\mu_2 K}{J}} + \frac{\tilde \sigma \sigma \sqrt{ MKN \log d} }{{\sigma_K^*}^2} \\
    \stackrel{(viii)}{\lesssim } & \frac{\sigma^2 MN}{{\sigma_K^*}^2} \sqrt{\frac{\mu_2 K}{J}} + \frac{\sigma\sqrt{K \log d}}{\kappa^* \sigma_K^*},
\end{align*}
where $(v)$ used \eqref{eq:simplified spectral norm of E_{:,j}} in Lemma~\ref{lemma: noise matrix concentrations using the universality} and \eqref{eq:bernstein for E_{:,j}B} in Lemma~\ref{lemma: matrix Bernstein} conditional on $\mathcal P_{:,-S_l}(\E )\V^*\mathbf \Lambda^{*-2}$, $(vi)$ used a similar treatment to \eqref{eq:simplified spectral norm of EV*} in Lemma~\ref{lemma: noise matrix concentrations using the universality} on $\norm{\mathcal P_{:,-S_l}(\E)\V^*}$ and \eqref{eq:bernstein for E_{i,:}A} in Lemma~\ref{lemma: matrix Bernstein} on $\norm{\mathcal P_{:,-S_l}(\E)\V^*}\ti$, $(vii)$ used \eqref{eq: assumption eq4} in Assumption~\ref{assumptions}, and $(viii)$ is by \eqref{eq: assumption eq3} in Assumption~\ref{assumptions}.
\qed

\section{Lemmas and Proofs for Theorem~\ref{theorem: estimation error}}\label{sec-lemma}
\begin{assumption}\label{assumption:supp_par_est}
\begin{equation}
    \sqrt{K}\kappa^3(\bPi^*)\max\left\{\frac{\tilde\sigma\sqrt{K\mu_1\log d}}{\sigma_K(\mathbf \Theta^*)}, \frac{\kappa^* \sigma^2 J \sqrt{\mu_1 K}}{\sqrt{N} \sigma_K(\bPi^*)\sigma_K(\mathbf \Theta^*)^2}, \frac{\sigma M B \log d \sqrt{(MN+J)\mu_2 K}}{\sqrt{J} \sigma_K(\bPi^*)\sigma_K(\mathbf \Theta^*)^2}\right\} \ll 1.
\end{equation}
Assumption~\ref{assumption:supp_par_est} is a sufficient condition for $\sqrt{K} \kappa^2(\bPi^*) \sigma_1(\bPi^*) \xi_1 \ll 1$ to hold. To see this,
\begin{align}
    & \sqrt{K} \sigma_1(\mathbf \Pi^*) \cdot \kappa^2(\mathbf \Pi^*)\cdot \xi_1 \\
    = & \sqrt{K}\sigma_1(\bPi^*)\kappa^2(\bPi^*)\frac{\tilde \sigma \sqrt{N \log d}}{{\sigma^*_K}} \sqrt{\frac{\mu_1 K}{N}} + \sqrt{K}\sigma_1(\bPi^*)\kappa^2(\bPi^*) \frac{\kappa^* \sigma^2 J}{{\sigma^*_K}^2}  \sqrt{\frac{\mu_1 K}{N}} \\
    & + \sqrt{K}\sigma_1(\bPi^*)\kappa^2(\bPi^*) \frac{\sigma MB\log d \sqrt{MN+J}}{{\sigma_K^*}^2} \sqrt{\frac{\mu_2 K}{J}} \\ 
    \le & \sqrt{K} \kappa^3(\bPi^*) \frac{\tilde\sigma \sqrt{K \mu_1 \log d}}{\sigma_K(\bT^*)} + \sqrt{K} \kappa^3(\bPi^*) \frac{\kappa^* \sigma^2 J \sqrt{\mu_1 K}}{\sqrt{N} \sigma_K(\bPi^*)\sigma_K(\mathbf \Theta^*)^2} \\
    & + \sqrt{K} \kappa^3(\bPi^*) \frac{\sigma M B \log d \sqrt{(MN+J)\mu_2 K}}{\sqrt{J} \sigma_K(\bPi^*)\sigma_K(\mathbf \Theta^*)^2}.
\end{align}
\end{assumption}

\begin{lemma}\label{lemma-ineqs}
${1}/{\sqrt{K}}\le\|\bPi^*\|_{2,\infty}\le 1$,  $\|\bT^*\|_{2,\infty} \le \sqrt{K}$,
$\sigma_1(\bPi^*)\le \sqrt{N}$,  $\sigma_1(\bT^*)\le \sqrt{J  K}$, and
$\kappa(\R^*) \leq \kappa(\bT^*) \kappa(\bPi^*)$.
\end{lemma}

\paragraph{Proof of Lemma~\ref{lemma-ineqs}.}
The first inequality is because $\frac{1}{\sqrt{K}} \le \|\bpi_i\|\le \|\bpi_i\|_1=1$ by Cauchy-Schwarz inequality.
The second inequality directly follows by the fact that each row of $\bT^*$ is a length $K$ Bernoulli probability vector. 
The third inequality holds because $\sigma_1(\bPi^*) \leq \norm{\bPi^*}_F \leq \sqrt{N} \norm{\bPi^*}_{2,\infty} \leq \sqrt{N}$ with Lemma~\ref{Lemma-matrix}.
% The proof of the first inequality can be found in Lemma B.2 of \cite{mao2021mm}. 
The fourth inequality is because $\sigma_1(\bT^*)\le \|\bT^*\|_F\le \sqrt{J  K}$.
The last inequality is because $\sigma_1(\R^*)\le \sigma_1(\bT^*)\sigma_1(\bPi^*)$ and $\sigma_K(\R^*)\ge \sigma_K(\bT^*)\sigma_K(\bPi^*)$.
\qed

\paragraph{Proof of Lemma~\ref{lemma:simplex}.}
According to the proof of Theorem 2 in \cite{chen2024spectral}, Assumption \ref{assumption:pure} is sufficient for the generalized-GoM to be identifiable.
Take the rows indices $\S$ for both sides of $\R=\U^*\Lam\V^{*\top}$, then
\begin{equation}\label{eq-theta_U}
    \U^*_{\S,:}\Lam^*\V^{*\top}=\R^*_{\S,:}=\bPi^*_{\S,:} \bT^{*\top}=\bT^{*\top}.
\end{equation}
Further note that 
\begin{equation}\label{eq-U}
    \U^*=\R^*\V^*\Lam^{*-1}=\bPi^*\bT^{*\top}\V^*\Lam^{*-1}.
\end{equation}
Plug \eqref{eq-theta_U} into \eqref{eq-U}, and we have
\begin{equation}\label{eq-PiU}
    \U^* = \bPi^*\U^*_{\S,:}\Lam^*\V^{*\top}\V^*\Lam^{*-1} = \bPi^*\U^*_{\S,:}.
\end{equation}
Eq \eqref{eq-PiU} indicates that $\U^*_{\S,:}$ is full rank since both $\U^*$ and $\bPi^*$ have rank $K$. Therefore, $\bo\Pi^*=\U^* \U_{\S,:}^{*-1}$ and we have proved the first equation in the lemma.
Furthermore, since $\U^{*\top}\U^*=\mathbf{I}_K$, Eq \eqref{eq-PiU} also leads to $\U_{\S,:}^{*\top} \bPi^{*\top} \bPi^*\U^*_{\S,:} = \mathbf{I}_K$ and
\begin{equation}\label{eq:U_Pi_singular}
    \sigma_1(\U^*_{\S,:})=\frac{1}{\sigma_K(\bPi^*)}, \quad \sigma_K(\U^*_{\S,:})=\frac{1}{\sigma_1(\bPi^*)},\quad \kappa(\U^*_{\S,:}) = \kappa(\bPi^*).
\end{equation}
\qed

\begin{lemma}[Theorem 3 in \cite{Gillis2013}]\label{lemma-g1}
Let $\M'=\M+\N\in\mathbb{R}^{m\times n}$ where $\M=\W\Z=\W[\mathbf{I}_r|\Z'], \W\in\mathbb{R}^{m\times r}$ has rank $r$, $\Z\in\mathbb{R}_+^{r\times n}$, and $\sum_{k=1}^r\Z'_{kj}\le 1,\forall j$ and $r\ge 2$. Let $J$ be the index set extracted by SPA. If $\|\N\|_{\infty, 2}\le \xi$ with
    \begin{equation}\label{Eq:lemma-g1}
        \left(1+80\frac{\|\W\|_{\infty, 2}^2}{\sigma_r^2(\W)}\right) \xi < \min\left\{\frac{1}{2\sqrt{r-1}}, \frac14\right\}\sigma_r(\W),
    \end{equation}
    then there exists a permutation $P$ of $\{1,\cdots, r\}$ such that
\begin{equation}\label{Eq:lemma-g1-2}
    \max_{1\le j\le r}\|\M'_{:,J(j)} - \W_{:,P(j)}\|\le \bar{\xi}:=\xi\left(1+80\frac{\|\W\|_{\infty, 2}^2}{\sigma_r^2(\W)}\right).
\end{equation}
\end{lemma}

\begin{lemma}[Modified from Theorem G.2 of \cite{mao2021mm}] \label{lemma-g2}
Let $\hat{\S}$ be the estimated simplex vertex index set returned by SPA. Suppose Assumptions~\ref{assumption:s_block}, \ref{assumption: s_pure}, \ref{assumptions} hold, and further assume that $\sqrt{K} \kappa^2(\bPi^*) \sigma_1(\bPi^*) \xi_1 \ll 1$. Then there exists a permutation matrix $\bP\in\mathbb{R}^{K\times K}$ such that 
\begin{equation}
     \|\U_{\hat{\S},:}-\bP^\top \U_{\S,:}^*\U^{*\top}\U\|_{2,\infty} = O(\kappa^2(\bPi^*)\xi_1)
\end{equation}
with probability at least $1-O\left((N\vee J )^{-10}\right)$.
\end{lemma}

\paragraph{Proof of Lemma~\ref{lemma-g2}.}
In order to use Lemma~\ref{lemma-g1}, let $r=K$, $\M' = \U^{*\top}\U \U^\top$, $\M = \U^{*\top}$, $\W = \U^{*\top}_{\S,:}$, and $\Z=\bPi^{*\top}$. Note that $\U^{*\top}=\U_{\S,:}^{*\top}\bPi^{*\top}=\U_{\S,:}^{*\top}(\mathbf{I}_K|\bPi_{(K+1):N, :}^{*\top})$.
According to Theorem~\ref{theorem-bound (general r.v.)}, $\N=\M'-\M$ satisfies $\|\N\|_{\infty, 2} \le \xi_1$ with probability at least $1-O\left((N\vee J )^{-10}\right)$. 
Notice that $\|\U^{*}_{\S,:}\|_{2,\infty}\le \|\U^{*}_{\S,:}\|$ with Lemma~\ref{Lemma-matrix}, thus
\begin{equation}
    \frac{\|\W\|_{\infty, 2}^2}{\sigma_K^2(\W)} 
        \le \frac{\sigma_1^2(\U^{*}_{\S,:})}{\sigma_K^2(\U^{*}_{\S,:})} = \kappa^2(\U^*_{\S,:})=\kappa^2(\bPi^*), 
\end{equation}
where the last equality is from \eqref{eq:U_Pi_singular}.
Therefore, $(1+80\frac{K(\W)^2}{\sigma_K^2(\W)})\xi_1 \lesssim \kappa^2(\bPi^*) \xi_1$.
Furthermore, since $\sigma_K(\W)=\sigma_K(\U^{*}_{\S,:}) = \frac{1}{\sigma_1(\bPi^*)}$ with \eqref{eq:U_Pi_singular}, inequality \eqref{Eq:lemma-g1} is satisfied since $\sqrt{K} \kappa^2(\bPi^*) \sigma_1(\bPi^*) \xi_1 \ll 1$.
With Lemma~\ref{lemma-g1}, we know that there exists a permutation matrix $\bP\in\mathbb{R}^{K\times K}$ such that 
$$
\|\U_{\hat{\S},:}\U^\top\U^* - \bP^\top \U^{*}_{\S,:})\|_{2,\infty} = O(\kappa^2(\bPi^*)\xi_1),
$$
with probability at least $1-O\left((N\vee J )^{-10}\right)$.
\qed

\begin{lemma}\label{lemma-g3}
    (Modified based on Lemma G.3 of \cite{mao2021mm}). Suppose Assumptions~\ref{assumption:s_block}, \ref{assumption: s_pure}, \ref{assumptions} hold. Let $\hat{\S}$ be the estimated simplex vertex index set returned by SPA. Then we have
\begin{equation}
     \left\|\U^*\U^{*\top}\U\left(\U^{-1}_{\hat{\S},:} - (\bP^\top \U^*_{\S,:}\U^{*\top}\U)^{-1}\right)\right\|_{2,\infty} = O \left( \sigma_1(\bPi^*)\kappa^2(\bPi^*)\xi_1 \right)
\end{equation}
with probability at least $1-O\left((N\vee J )^{-10}\right)$.
\end{lemma}

\paragraph{Proof of Lemma~\ref{lemma-g3}.} 
Define $\F=\U^{*\top}\U, \W=\bP^\top \U^*_{\S,:}\F$. For every $i\in[N]$, 
\begin{equation}\label{eq-lemma-g2-1}
    \begin{split}
        & \|\e_i^\top \U^*\U^{*\top}\U\left(\U^{-1}_{\hat{\S},:} - (\bP^\top \U^*_{\S,:}\U^{*\top}\U)^{-1}\right)\|\\
        = & \|\e_i^\top \U^*\F(\U_{\hat{\S},:}^{-1} - \W^{-1})\| = \|\e_i^\top \U^*\F\W^{-1}(\W-\U_{\hat{\S},:})\U_{\hat{\S},:}^{-1}\|\\
        = & \|\e_i^\top \U^*\F\F^{-1}\U^{*-1}_{\S,:}\bP(\W-\U_{\hat{\S},:})\U_{\hat{\S},:}^{-1}\| = \|\e_i^\top \bPi^*\bP(\W-\U_{\hat{\S},:})\U_{\hat{\S},:}^{-1}\| \\
        \stackrel{(i)}{\le} & \|\W-\U_{\hat{\S},:}\|_{2,\infty}\|\U_{\hat{\S},:}^{-1}\| = \|\ \bP^\top \U^*_{\S,:}\U^{*\top}\U-\U_{\hat{\S},:}\|_{2,\infty}\|\U_{\hat{\S},:}^{-1}\| \\
        \stackrel{(ii)}{\le} & O(\kappa^2(\bPi^*)\xi_1')\|\U_{\hat{\S},:}^{-1}\| \text{\ \ with probability at least $1-O\left((N\vee J )^{-10}\right)$}.
    \end{split}
\end{equation}
Here $(i)$ holds because $\bPi^*\bP$ is non-negative and has rows with $l_1$ norm 1; $\bPi^*\bP(\W-\U_{\hat{\S},:})$ is a convex combination of rows of $\W-\U_{\hat{\S},:}$, and the $l_2$ norm of a convex combination can not exceed the maximum norm of the vertices. Inequality $(ii)$ uses Lemma~\ref{lemma-g2}.

Now we bound $\|\U_{\hat{\S},:}^{-1}\|$. With Weyl's inequality and $\sigma_k(\bP^\top\U^*_{\S,:}\U^{*\top}\U)=\sigma_{k}(\U^*_{\S,:}), \forall k$, we have with probability at least $1-O\left((N\vee J )^{-10}\right)$
\begin{align*}
    |\sigma_k(\U_{\hat{\S},:})-\sigma_k(\U^*_{\S,:})| \le & \|\U_{\hat{\S},:}-\bP^\top\U^*_{\S,:}\U^{*\top}\U\| \\
    \stackrel{(iii)}{\le} & \sqrt{K}\|\U_{\hat{\S},:}-\bP^\top\U^*_{\S,:}\U^{*\top}\U\|_{2,\infty} \\
    \stackrel{(iv)}{=} & O(\sqrt{K}\kappa^2(\bPi^*)\xi_1'),
\end{align*}
for $k=1,\dots, K$, where $(iii)$ uses Lemma~\ref{Lemma-matrix} and $(iv)$ uses Lemma~\ref{lemma-g2}. It immediately follows that 
\begin{align*}
    \sigma_K(\U_{\hat{\S},:}) & \ge \sigma_K(\U^*_{\S,:})\left(1 - O\left(\frac{\sqrt{K}\kappa^2(\bPi^*)\xi_1'}{\sigma_K(\U^*_{\S,:})}\right) \right)\\
    & = \frac{1}{\sigma_1(\bPi^*)}\left(1 - O\left(\sqrt{K}\kappa^2(\bPi^*)\sigma_1(\bPi^*)\xi_1'\right) \right)
\end{align*}
with \eqref{eq:U_Pi_singular}.
Therefore, 
\begin{equation}\label{eq-lemma-g3-2}
    \|\U_{\hat{\S},:}^{-1}\| =\frac{1}{\sigma_K(\U_{\hat{\S},:})} \le \frac{\sigma_1(\bPi^*)}{\left(1 - O\left(\sqrt{K}\kappa^2(\bPi^*)\sigma_1(\bPi^*)\xi_1'\right) \right)}  = O(\sigma_1(\bPi^*)),
\end{equation} 
since $\sqrt{K}\kappa^2(\bPi^*)\sigma_1(\bPi^*)\xi_1' \ll 1$.
By plugging Eq \eqref{eq-lemma-g3-2} into Eq \eqref{eq-lemma-g2-1}, we get
\begin{align*}
    \norm{\U^*\U^{*\top}\U\left(\U^{-1}_{\hat{\S},:} - (\bP^\top \U^*_{\S,:}\U^{*\top}\U)^{-1}\right)}\ti = O\left(\kappa^2(\bPi^*)\sigma_1(\bPi^*)\xi_1'\right)
\end{align*}
with probability at least $1-O\left((N\vee J )^{-10}\right)$.
\qed

\paragraph{Proof of Theorem~\ref{theorem: estimation error}.} 
    Recall that $\hat{\bPi}=\U\U_{\hat{\S},:}^{-1}$. We control the estimation error of each row up to a permutation: 
\begin{equation*}
    \begin{split}
        & \max_{i\in [N]}\|\e_i^\top(\hat{\bPi}-\bPi^*\bP)\|_2 \\
        = &  \max_{i\in [N]}\|\e_i^\top(\U\U_{\hat{\S},:}^{-1} - \U^*\U^{*-1}_{\S,:}\bP)\|_2\\
        \le &  \max_{i\in [N]}\Big(\|\e_i^\top(\U-\U^*\U^{*\top}\U)\U_{\hat{\S},:}^{-1}\| + \|\e_i^\top \U^*\U^{*\top}\U(\U_{\hat{\S},:}^{-1} - (\U^{*\top}\U)^{-1}\U^{*-1}_{\S,:}\bP)\|_2\Big) \\
        \le & \|\U-\U^*\U^{*\top}\U\|_{2,\infty}\|\U_{\hat{\S},:}^{-1}\| + \|\U^*\U^{*\top}\U(\U_{\hat{\S},:}^{-1} - (\bP^\top \U^{*}_{\S,:}\U^{*\top}\U)^{-1} )\|_{2,\infty} \\
        \stackrel{(i)}{\lesssim} & \sigma_1(\bPi^*)\xi_1' + \kappa^2(\bPi^*)\sigma_1(\bPi^*)\xi_1' \\
        \lesssim & \kappa^2(\bPi^*)\sigma_1(\bPi^*)\xi_1' \text{\ \ with probability at least $1-O\left((N\vee J )^{-10}\right)$},
    \end{split}
\end{equation*}
where $(i)$ uses Lemma~\ref{lemma-g3} and Eq~\eqref{eq-lemma-g3-2}.

On the other hand, recall $\hat\bT=\V\Lam\U_{\hat\S,:}^\top$ and $\xi_3$ is the upper bound in \eqref{eq: general bound for ULambdaV} in Theorem~\ref{theorem-bound (general r.v.)}. 
\begin{align*}
        & \|\hat{\bT}\bP^\top-\bT^*\|_{\infty} \\
        = & \|\bP\U_{\hat\S,:}\Lam\V^\top - \U_{\S,:}^*\Lam^*\V^{*\top}\|_{\infty}\\
        \le & \| (\bP\U_{\hat\S,:} - \U_{\S,:}^*\U^{*\top}\U)\Lam\V^{\top}\|_{\infty} + \|\U_{\S,:}^*\U^{*\top}\U\Lam\V^\top - \U_{\S,:}\Lam\V^\top\|_{\infty}  \\
        & + \|\U_{\S,:} \Lam \V^\top - \U_{\S,:}^*\Lam^*\V^{*\top}\|_{\infty}\\
        \stackrel{(i)}{\le} & \|\U_{\hat{\S},:}-\bP^\top\U^*_{\S,:}\U^{*\top}\U\|_{2, \infty}\cdot \sigma_1(\R) \cdot\|\V\|_{2, \infty} + \|\U_{\S,:}^*\U^{*\top}\U - \U_{\S,:}\|_{2, \infty}\cdot\sigma_1(\R)\cdot\|\V\|_{2, \infty}\\
        & + \|\U\Lam\V^\top - \R^*\|_{\infty} \\
        \stackrel{(ii)}{\lesssim} &~ \big(\kappa^2(\bPi^*) + 1 \big) \cdot\xi_1' \cdot \sigma_1(\R) \cdot\|\V\|_{2, \infty} + \xi_3 \\
        \stackrel{(iii)}{\lesssim } & \kappa^2(\bPi^*) \cdot\xi_1' \cdot \sigma_1(\R)\cdot \big(\sqrt{\frac{K\mu_2}{J }} + \xi_2'  \big)+ \xi_3 \\
        \stackrel{(iv)}{\lesssim} & \kappa^2(\bPi^*) \sigma_1^* ( \sqrt{\frac{K\mu_2}{J}} + \xi_2' )\xi_1' + \xi_3 \\
        \stackrel{(v)}{\lesssim} & \kappa^2(\bPi^*) \xi_3,
\end{align*}
with probability at least $1-O\left((N\vee J )^{-10}\right)$. 
Here $(i)$ uses Lemma~\ref{Lemma-matrix}, $(ii)$ uses Lemma~\ref{lemma-g2} and the definitions of $\xi_1', \xi_3$, and $(iii)$ is from \eqref{eq: V rearranged decomposition} and the definitions of $\mu_2$ and $\xi_2$. In addition, $(iv)$ holds since $|\sigma_1(\R) - \sigma_1(\R^*)| \leq \norm{\mathbf E}\lesssim \sigma\sqrt{J} + \tilde\sigma\sqrt{N} \le \sigma(\sqrt{J} + \sqrt{MN})\lesssim \sigma_K^*\le\sigma_1^*$ holds with probability at least $1 - O((N\vee J)^{-20})$, where we use Assumption \ref{assumptions} and Lemma~\ref{lemma: noise matrix concentrations using the universality}.
In addition, $(v)$ follows since $\xi_1'\xi_2'\sigma_1^* + \sqrt{\frac{\mu_2 K}{J} }\sigma_1^*\xi_1' \lesssim \xi_3$ from \eqref{eq: lemma2_5} in the proof of Theorem~\ref{theorem-bound (general r.v.)}. 
\qed

\paragraph{Proof of Corollary~\ref{corollary 2}.}
Given the additional assumptions, with probability at least $1 - O((N \vee J)^{-10})$ we have
\begin{align}
    & \norm{\hat{\bPi} - \bPi^* \bP}_{2, \infty} \\ 
    \lesssim & \kappa^2(\mathbf \Pi^*)\cdot \sigma_1(\mathbf \Pi^*) \cdot \xi_1' \\ 
    \lesssim & \sigma_1(\mathbf \Pi^*) \left( \frac{\tilde \sigma \sqrt{\mu_1 K \log d}}{{\sigma^*_K}} +  \frac{\sigma^2 J}{{\sigma^*_K}^2}  \sqrt{\frac{\mu_1 K}{N}} + \frac{\sigma MB\log d \sqrt{MN+J}}{{\sigma_K^*}^2} \sqrt{\frac{\mu_2 K}{J}} \right)\\ 
    \le & \frac{\tilde\sigma \sqrt{\mu_1 K \log d}}{\sigma_{K}(\mathbf \Theta^*)} + \frac{\sigma^2 }{\sigma_{K}(\mathbf \Pi^*)\sigma_{K}^2(\mathbf \Theta^*)} \left(J \sqrt{\frac{\mu_1 K}{N}} + \frac{MB\log d \sqrt{MN+J}}{\sigma} \sqrt{\frac{\mu_2 K}{J}} \right) \\
    \le & \frac{\tilde\sigma \sqrt{\mu_1 K \log d}}{\sigma_{K}(\mathbf \Theta^*)} + \frac{\sigma^2 \sqrt{JK}}{\sigma_{K}(\mathbf \Pi^*)\sigma_{K}^2(\mathbf \Theta^*)} \left( \sqrt{r \mu_1} + \sqrt{1+\frac{M}{r}} \frac{MB \log d}{\sigma \sqrt{J}} \sqrt{\mu_2} \right),
\end{align}
where we use the fact that $\sigma_K^* \geq \sigma_{K}(\mathbf \Pi^*)\sigma_{K}(\mathbf \Theta^*)$. 

For the error bound of $\hat\bT$, 
\begin{align}
    \|\hat{\bT}\bP^\top-\bT^*\|_{\infty} \lesssim K\sigma \log d \sqrt{\mu_1\mu_2} \sqrt{\frac{M}{J} + \frac{1}{N}} + MB\log d\left(\frac{\mu_1 K}{MN} + \frac{\mu_2 K}{J} \right)
\end{align}
with probability at least $1 - O((N \vee J)^{-10})$.
\qed

\begin{lemma}\label{lemma:pi-prior}
(Lemma 3.6 in \cite{mao2021mm}.)
If $\bo\pi_i^*\stackrel{i.i.d}{\sim}\operatorname{Dirichlet}(\boldsymbol{\alpha})$ for each $i\in[N]$ with $\alpha_{\max}=\max_k \alpha_k, \alpha_{\min}=\min_k \alpha_k$, $\alpha_0 = \sum_k\alpha_k$ and $\nu:=\alpha_0 / \alpha_{\min}$, then
\begin{equation}\label{eq:dirichelt-pi}
    \begin{split}
        \mathbb{P}\left(\sigma_K^2\left(\bPi^* \right) \geq \frac{N}{2 \nu\left(1+\alpha_0\right)}\right) 
        &\geq 1-K \exp \left(-\frac{N}{36 \nu^2\left(1+\alpha_0\right)^2}\right), \\
        \mathbb{P}\left(\sigma_1^2\left( \bPi^*\right) 
        \leq \frac{3 N\left(\alpha_{\max }+\|\boldsymbol{\alpha}\|^2\right)}{2 \alpha_0\left(1+\alpha_0\right)}\right) 
        &\geq 1-K \exp \left(-\frac{N}{36 \nu^2\left(1+\alpha_0\right)^2}\right), \\
        \mathbb{P}\left(\kappa^2\left(\bPi^* \right) \leq 3 \frac{\alpha_{\max }+\|\boldsymbol{\alpha}\|^2}{\alpha_{\min }}\right) &\geq 1-2 K \exp 
        \left(-\frac{N}{36 \nu^2\left(1+\alpha_0\right)^2}\right).
    \end{split}
\end{equation}
\end{lemma}

\begin{lemma}\label{lemma:theta-prior}
If $(\theta^*_{M(l-1)+c,k})_{c=1}^{M}\stackrel{i.i.d}{\sim} \operatorname{Dirichlet}(\bo\beta)$ for each $l\in[L]$, $k\in[K]$ with $\beta_0 = \sum_{c=1}^M\beta_c$, $c_1=\frac{\beta_0^2-\|\bo\beta\|^2}{\beta_0^2(1+\beta_0)}$, and $c_2=K\frac{\|\bo\beta\|^2}{\beta_0^2} + c_1$, then
\begin{equation}\label{eq:dirichlet-theta}
\begin{split}
    \mathbb{P}\left(\sigma_K^2\left( \bT^* \right) \geq \frac{c_1 L}{2}\right) 
    & \geq 1 - K \exp\left( - \frac{c_1^2L}{80K^2} \right), \\
    \mathbb{P}\left(\sigma_1^2\left( \bT^*\right) \leq\frac{3c_2 L}{2}  \right) 
    & \geq 1 - K \exp\left( - \frac{c_1^2L}{80K^2} \right),\\
    \mathbb{P}\left(\kappa^2\left(\bT^* \right) \le \frac{3c_2}{c_1} \right) & \ge 1 - 2K \exp\left( - \frac{c_1^2L}{80K^2} \right).
\end{split}
\end{equation}
\end{lemma}

\paragraph{Proof of Lemma~\ref{lemma:theta-prior}} (Modified based on the proof of Lemma 3.6 in \cite{mao2021mm}.)
Write 
$$\bT^*=(\bT_1^{*\top}, \dots, \bT_L^{*\top})^\top,$$ 
where $\bT^*_{l}=(\bt^*_{l1},\dots, \bt^*_{lK})\in[0,1]^{C\times K}$ with $\bt_{lk}^*=(\theta^*_{C(l-1)+c,k})_{c=1}^{C} \stackrel{i.i.d}{\sim} D(\bo\beta)$ for each $l\in[L]$ and $k\in[K]$. Denote $\M:=\bT^{*\top}\bT^*$. Note that $\M-\mathbb{E}[\M]=\sum_l \X_l$, where each $\X_l=\bT_l^{*\top}\bT_l^*-\mathbb{E}[\bT_l^{*\top}\bT_l^*]$ is an independent mean zero random $K\times K$ matrix. Since $\|\bt^*_{lk}\|\le \|\bt^*_{lk}\|_1=1$,
$\|\bT_l^{*\top}\bT_l^*\|\le \|\bT_l^*\|_F^2 \le K$.

Also note that $$\mathbb{E}\left[\bT_l^{*\top}\bT_l^*\right] = \frac{\|\bo\beta\|^2}{\beta_0^2}\mathbf{1}\mathbf{1}^\top + \diag\left( \frac{\beta_0^2 - \|\bo\beta\|^2}{\beta_0^2(1+\beta_0)}\right).$$ 
With Weyl's inequality,
\begin{align*}
    \lambda_K\left(\mathbb{E}[\bT_l^{*\top}\bT_l^*]\right) & =\frac{\beta_0^2-\|\bo\beta\|^2}{\beta_0^2(1+\beta_0)} \eqqcolon c_1 < 1,\\
    \lambda_1\left(\mathbb{E}[\bT_l^{*\top}\bT_l^*]\right) & = K\frac{\|\bo\beta\|^2}{\beta_0^2} + \frac{\beta_0^2-\|\bo\beta\|^2}{\beta_0^2(1+\beta_0)} \eqqcolon c_2 < K+1. 
\end{align*}
Therefore,
\begin{align*}
    \|\X_l\|\le \|\bT_l^{*\top}\bT_l^*\| + \|\mathbb{E}[\bT_l^{*\top}\bT_l^*]\| \le K + c_2 < 2K+1 \le 3K.
\end{align*}
Furthermore, by Jensen's inequality,
\begin{align*}
\|\mathbb{E}[\X_l^2]\| \le \mathbb{E}[\|\X_l^2\|] \le \mathbb{E}[\|\X_l\|^2] \le 9K^2.
\end{align*}
By the matrix Bernstein inequality in \cite{tropp2012user}, for all $\tau>0$, we have
\begin{align*}
    \Pr\left( \sigma_1(\M-\mathbb{E}[\M])\ge \tau \right) & = \Pr\left( \sigma_1(\sum_{l=1}^L\X_{l})\ge \tau \right)\le K \exp\left(-\frac{\tau^2}{18LK^2 + 2K\tau}\right) :=\delta_{\tau}.
\end{align*}
%Note that $\bT^*=\bT+\mathbf{1}_{J\times K}\otimes \bo\beta$. 
By Weyl's inequality, with probability at least $1-\delta_{\tau}$,
\begin{align*}
    |\sigma_1(\M) - \sigma_1(\mathbb{E}[\M])|
    & \le \sigma_1(\M-\mathbb{E}[\M]) \le \tau \\
    |\sigma_K(\M) - \sigma_K(\mathbb{E}[\M])|
    & \le \sigma_1(\M-\mathbb{E}[\M]) \le \tau.
\end{align*}
Also we have 
\begin{align*}
\sigma_K(\mathbb{E}[\M])=\sigma_K\left(\sum_l \mathbb{E}\left[\bT_l^{*\top}\bT_l^*\right]\right) \ge L c_1.\\
\sigma_1(\mathbb{E}[\M])=\sigma_1\left(\sum_l \mathbb{E}\left[\bT_l^{*\top}\bT_l^*\right]\right)\le L c_2 
\end{align*}
For $\sigma_K(\M)$, choose $\tau_1=\frac{c_1L}{2}$. Then with probability at least 
$$
1-\delta_{\tau_1} = 1 - K \exp\left(-\frac{c_1^2L}{72K^2 + 4Kc_1}\right) \ge 1 - K \exp\left( - \frac{c_1^2L}{80K^2} \right),
$$
one has
\begin{align*}
\sigma_K^2(\bT^*)=\sigma_K(\M)\ge \sigma_K(\mathbb{E}[\M]) - |\sigma_K(\M) - \sigma_K(\mathbb{E}[\M])| \ge \tau_2=\frac{c_2L}{2}.
\end{align*}
For $\sigma_1(\M)$, choose $\tau_2=\frac{c_2L}{2}$. Then with probability at least 
$$
1-\delta_{\tau_2} = 1 - K \exp\left(-\frac{c_2^2L}{72K^2 + 4Kc_2}\right) \ge 1 - K \exp \left(- \frac{c_1^2L}{80K^2} \right),
$$
one has
\begin{align*}
    \sigma_1^2(\bT^*)=\sigma_1(\M)\le |\sigma_1(\M) - \sigma_1(\mathbb{E}[\M])| + \sigma_1(\mathbb{E}[\M])\le 3\tau_2=\frac{3c_2 L}{2}.
\end{align*}
With probability at least 
$1 - 2K \exp\left( - \frac{c_1^2L}{80K^2} \right)$,
one has
\begin{align*}
\kappa^2(\bT^*) = \frac{\sigma_1^2(\bT^*)}{\sigma_K^2(\bT^*)} \le \frac{3c_2}{c_1}.
\end{align*}
\qed

\paragraph{Proof of Corollary \ref{cor:dirichlet}.}
First note that $J=LM, B\le 1, \sigma\le 1/2, \tilde\sigma\le \sqrt{M}/2$ for the flattened data matrix with Bernoulli entries. 
Define the event 
\begin{align}
    \Omega:=\left \{(\bPi^*, \bT^*): \sigma_K(\bPi^*) \gtrsim \frac{\sqrt{N}}{K}, \sigma_1(\bPi^*) \lesssim \sqrt{\frac{N}{K}}, \kappa(\bPi^*)\lesssim \sqrt{K}, \right. \\
    \left. \sigma_K(\bT^*) \gtrsim \frac{\sqrt{J}}{M}, \sigma_1(\bT^*) \le \frac{\sqrt{KJ}}{M}, \kappa(\bT^*) \le \sqrt{K} \right\}.
\end{align}
With \eqref{eq:assumption-c_2}, the probabilities on the RHS in \eqref{eq:dirichelt-pi} and \eqref{eq:dirichlet-theta} are in the order of $1- O(KN^{-3})$ and $1 - O(KL^{-3})$, respectively. Along with the assumptions on $\bo\alpha$ and $\bo\beta$, we have 
$$
\Pr(\Omega) \ge 1-O\left(K (N\wedge L)^{-3} \right).
$$

Below we discuss for $(\bPi^*, \bT^*)\in\Omega$. With the lower bounds on $N, J$, we have
\begin{align*}
    \sigma \kappa^* (M\sqrt{N} + \sqrt{J}) \lesssim K (M\sqrt{N} + \sqrt{J}) \lesssim \frac{\sqrt{NJ}}{MK} \lesssim \sigma_K(\bPi^*)\sigma_K(\bT^*), 
\end{align*} 
and thus \eqref{eq: assumption eq3} in Assumption~\ref{assumptions} holds. 
Now we bound $\mu_1, \mu_2$ with Lemmas \ref{Lemma-matrix} and \ref{lemma-ineqs} and \eqref{eq:U_Pi_singular}. Since $\U^*=\bPi^*\U_{\S,:}^*$, we have
\begin{align*}
    \mu_1 = \frac{N}{K}\|\U^*\|^2\ti \le \frac{N}{K} \|\bPi^*\|\ti^2 \|\U_{\S,:}^*\|^2 \le \frac{N}{K \sigma_K^2(\bPi^*)} \lesssim K.
\end{align*}
Similarly, $\V^*=\bT^*\U^*_{\S,:}(\U^{*\top}_{\S,:}\U^*_{\S,:})^{-1}\Lam^{*-1}$. Thus,
\begin{align*}
    \mu_2 = \frac{J}{K}\|\V^*\|^2\ti \le \frac{J}{K} \|\bT^*\|^2\ti \frac{\|\U^*_{\S,:}\|^2}{\sigma_K^4(\U^*_{\S,:}) \sigma_K^2(\bPi^*)\sigma_K^2(\bT^*)} \le J \frac{\kappa^4(\bPi^*)}{\sigma_K^2(\bT^*)} \lesssim K^2M^2.
\end{align*}
One can verify that \eqref{eq: assumption eq4} in Assumption~\ref{assumptions} holds with the lower bounds on $N, J$ in Assumption~\ref{assumption:dirichlet}.
Furthermore, %$\zeta \lesssim \frac{M\log(N\vee J)}{\sqrt{(MN)\wedge J}}$ and 
we can verify that Assumption~\ref{assumption:supp_par_est} is satisfied with the lower bounds on $N, J$, thus $\sqrt{K} \kappa^2(\mathbf \Pi^*)\sigma_1(\mathbf \Pi^*) \xi_1 \ll 1$.
Now we have verified that Assumption~\ref{assumptions} holds. By Remark~\ref{remark:1}, with probability at least $1-O((N\vee J)^{-10})$ one has
\begin{align*}
    \xi_1 & \lesssim \frac{M^{3/2} K^2 \sqrt{\log d}}{\sqrt{NJ}} + \frac{M^2 K^4}{N^{3/2}}, \\
    \xi_3 & \lesssim \frac{M^{5/2} K^{7/2} \log d}{\sqrt{J}} + \frac{M^2 K^{7/2} \log d}{\sqrt{N}}.
\end{align*}
Furthermore, with Theorem~\ref{theorem: estimation error},
\begin{align*}
    \norm{\hat{\mathbf \Pi} - \mathbf \Pi^* \mathbf P}_{2, \infty} & \lesssim \frac{M^{3/2} K^{5/2} \sqrt{\log d}}{\sqrt{J}} + \frac{\sqrt{M^2 K^{9/2}}}{N}, \\
     \|\hat{\bT}-\bT^*\bP\|_{\infty} & \lesssim K \xi_3 \lesssim \frac{M^{5/2} K^{9/2} \log d}{\sqrt{J}} + \frac{M^2 K^{9/2} \log d}{\sqrt{N}}.
\end{align*}
\qed

\section{Additional Computational Details}\label{sec-computation}
\subsection{Gibbs Sampling Updating Steps}\label{sec:gibbs}
For simplicity of presentation, we set the number of possible responses as a constant, i.e. $C_j\equiv C$ for all items $j$; extensions to fully general cases is straightforward. Here we use slightly different notations. Denote $\bt_{jk}\in\mathbb{R}^{C}$ as the probabilities for all $C$ response categories for item $j$ and extreme latent profile $k$. When written in the form of the likelihood function, the GoM model is equivalent to assuming the existence of latent class variables $Z_{ij}\in\{1,\dots, K\}$ for each individual $i$ and each item $j$.
\begin{align*}
    R_{ij} \vert Z_{ij} = k & \stackrel{i.i.d}{\sim} \text{Categorical} \left([C]; \bt_{jk}\right); \\
    Z_{ij} | \bpi_i & \stackrel{i.i.d}{\sim} \text{Categorical} ([K]; \bpi_i); \\
    \bpi_i & \stackrel{i.i.d}{\sim} \text{Dirichlet} (\bo \alpha), \ \bo \alpha \in\mathbb{R}^K; \\
    \bt_{jk} & \stackrel{i.i.d}{\sim} \text{Dirichlet} (\bo \beta), \ \bo\beta \in\mathbb{R}^C.
\end{align*}
We write $Z_{ijk}:=\mathbbm{1}(Z_{ij}=k)$ and $R_{ijc} := \mathbbm{1}(R_{ij}=c)$. The likelihood function of the GoM model is then
\begin{equation}
    L(\R; \bT, \Z, \Pi) = \prod_{i=1}^N \prod_{j=1}^J \prod_{k=1}^K \left(\prod_{c=1}^C \theta_{jkc} ^{R_{ijc}} \right)^{Z_{ijk}}.
\end{equation}
\begin{itemize}
    \item For the conditional distribution of $\bt_{jk}$ we have
    \begin{align*}
        p(\bt_{jk} \vert \R, \Z, \bPi) & \propto \prod_{i=1}^N \left( \prod_{c=1}^C \theta_{jkc}^{R_{ijc}} \right)^{Z_{ijk}} \prod_{c=1}^C \theta_{jkc}^{\beta_c-1} \propto \prod_{c=1}^C \theta_{jkc}^{\sum_{i=1}^N R_{ijc}Z_{ijk} + \beta_c - 1},
    \end{align*}
    which leads to
    \begin{equation}
        \bt_{jk} \vert \R, \Z, \bPi \sim \text{Dirichlet} \left( \left[ \sum_{i=1}^N R_{ijc} Z_{ijk} + \beta_c \right]_c \right).
    \end{equation}
    \item For $\bpi_i$, since 
    $$
    p(\bo Z_i \vert \bpi_i) = \prod_{j=1}^J \prod_{k=1}^K \pi_{ik}^{Z_{ijk}},
    $$
    we have
    \begin{align*}
        p(\bpi_i | \Z) & \propto \prod_{k=1}^K \pi_{ik}^{\sum_{j=1}^J Z_{ijk}} \prod_{k=1}^K \pi_{ik}^{\alpha_k-1}  = \prod_{k=1}^K \pi_{ik}^{\sum_{j=1}^J Z_{ijk} + \alpha_k - 1}.
    \end{align*}
    Thus 
    \begin{equation}
        \bpi_i | \Z \sim \text{Dirichlet}\left( \left[\sum_{j=1}^J Z_{ijk} + \alpha_k \right]_k \right).
    \end{equation}
    \item For $Z_{ij}$, with 
    $$
    p(R_{ij} | Z_{ij}=k) = \prod_{c=1}^C \theta_{jkc}^{R_{ijc}}
    $$
    and $p(Z_{ij}=k)=\pi_{ik}$, 
    we have
    $$
    p(Z_{ij}=k | R_{ij}) \propto \prod_{c=1}^C \theta_{jkc}^{R_{ijc}}\pi_{ik},
    $$
    and thus
    \begin{equation}
        p(Z_{ij}=k | \R, \bT, \bPi) = \frac{\prod_{c=1}^C \theta_{jkc}^{R_{ijc}}\pi_{ik}}{\sum_{k=1}^K \prod_{c=1}^C \theta_{jkc}^{R_{ijc}}\pi_{ik}}.
    \end{equation}
\end{itemize}

\subsection{Additional Algorithms and Figures}

Algorithm \ref{algorithm:poisson} below estimates the GoM model for Poisson data. The difference of Algorithm \ref{algorithm:poisson} from Algorithm \ref{algorithm2} is the truncation for $\hat\bT$.

\spacingset{1}
\setcounter{algorithm}{1}
\begin{algorithm}[ht!]
\centering
\caption{Poisson-GoM Estimation with Potential Local Dependence}\label{algorithm:poisson}
\begin{algorithmic}
    \Require Count matrix $\R\in \mathbb{N}_+^{N\times J}$, number of extreme profiles $K$.
    \vspace{0.3em}
    \Ensure Estimated $\hat{\S}\in \mathbb{R}^K, \hat{\bT}^{(\text{post})}\in \mathbb{R}^{J \times K}, \hat{\bPi}^{(\text{post})}\in\mathbb{R}^{N\times K}$
    \vspace{0.3em}
    \State Get the top $K$ singular value decomposition of $\R$ as $\U\Lam\V^\top$
    \State $\Y = \U$
    \For{$k \in \{1,\dots,K\}$}                    
        \State $\hat{S}_k=\argmax (\{\|\Y_{i,:}\|_2:\; i\in[N] \backslash \hat{\bP}\})$
        \State $\mathbf{u} = \Y_{\hat{S}_k,:} / \|\Y_{\hat{S}_k,:}\|_2$
        \State $\Y = \Y (\I_K - \mathbf{u}\mathbf{u}^\top)$
    \EndFor
    \State $\hat{\bPi}=\U(\U_{\hat{\S},:})^{-1}$
    \State $\hat{\bPi}^{(\text{post})}=\diag(\hat{\bPi}_+ \mathbf{1}_K)^{-1} \hat\bPi_+$
    \State $\hat{\bT}=\V \Lam \U^{\top}_{\hat\S,:}$
    \State Let $\hat{\bT}^{(\text{post})} = (\hat{\bt}_{j,k}^{(\text{post})})$ with $\hat{\theta}^{(\text{post})}_{j,k}=
    \begin{cases}
        \hat{\theta}_{j,k} & \text{if } \hat{\theta}_{j,k} \ge 0 \\
        \epsilon & \text{if } \hat{\theta}_{j,k}<0
    \end{cases}$
\end{algorithmic}
\end{algorithm}

\spacingset{1.7}

\begin{figure}[ht!]
    \centering
    \includegraphics[width=0.85\textwidth]{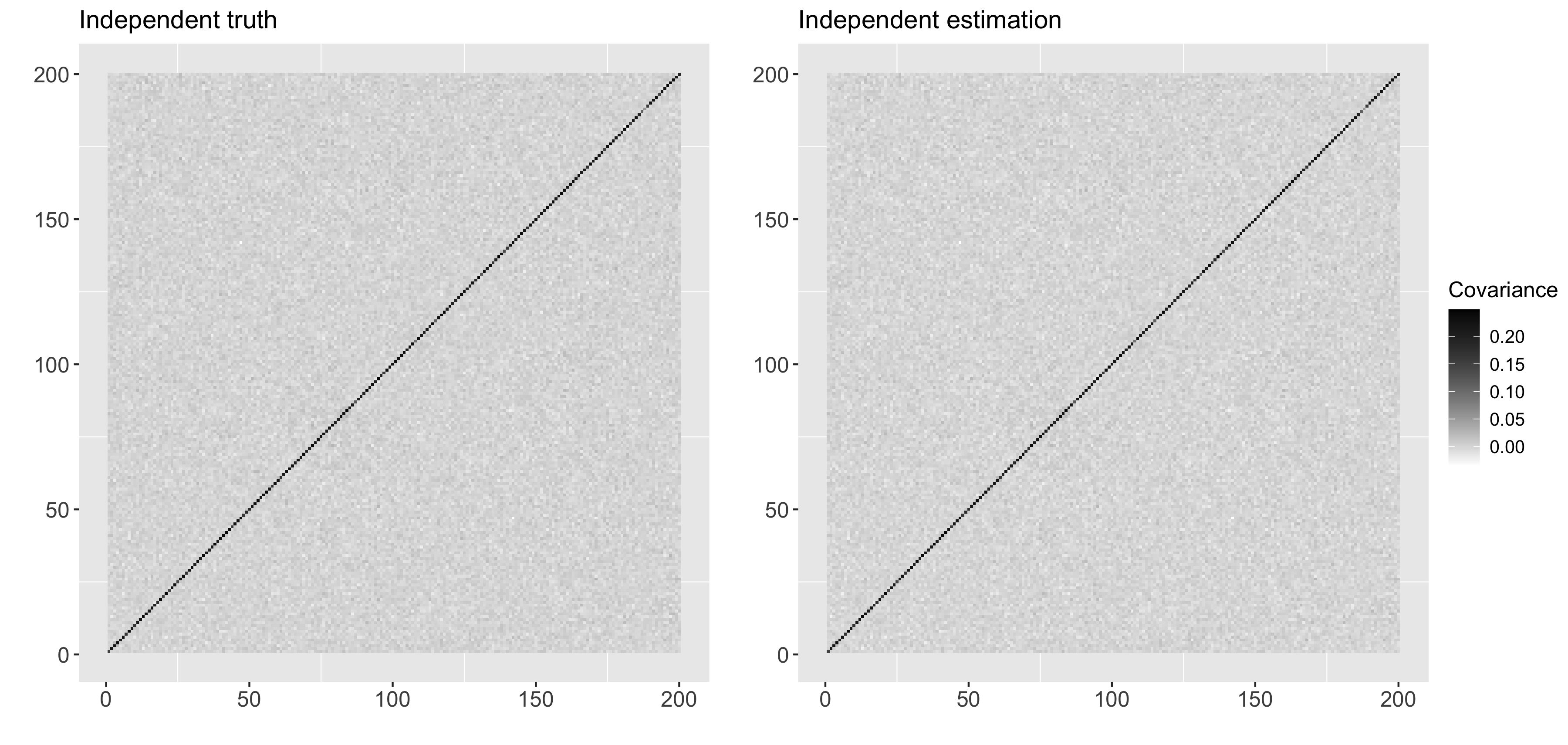}
    \includegraphics[width=0.85\textwidth]{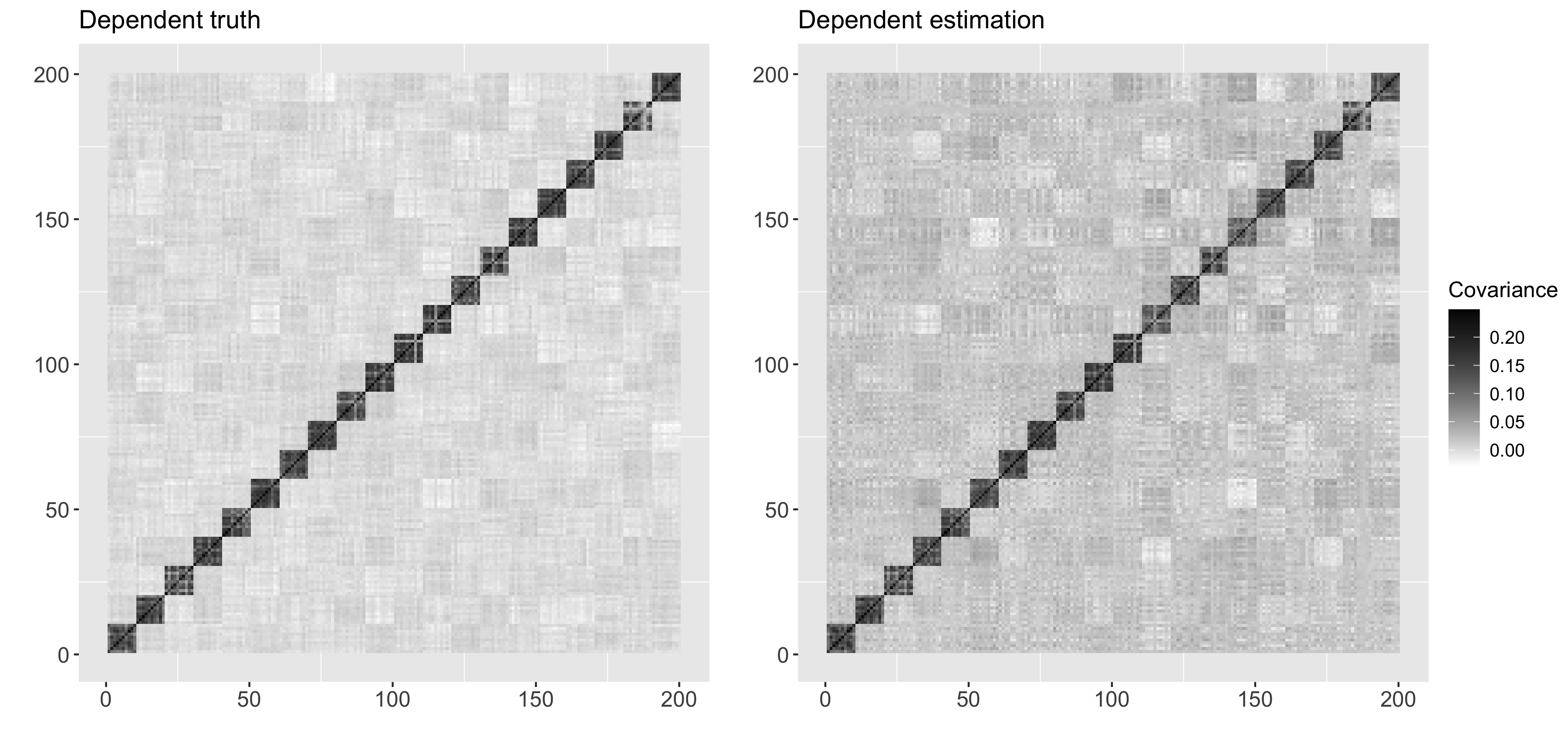}
    \caption{Heatmaps of the covariance matrices of $\R - \R^*$ (left) and $\R-\hat\bPi\hat\bT^{\top}$ (right), without local dependence (upper) and with local dependence (lower).}
    \label{fig:local_dependence}
\end{figure}

\clearpage
Figure \ref{fig:local_dependence} presents the heatmaps of the covariance matrix of the true error matrix $\R - \R^*$ (left) and the estimated error matrix $\R-\hat\bPi\hat\bT^{\top}$ (right) in a simulation study. Whether or not there is local dependence, the covariance matrix of the estimated error matrix empirically approximates that of the ground truth error matrix well. The details of the simulation setting can be found in Section \ref{sec:simulation}.

\end{document}